\documentclass[prd,preprintnumbers,onecolumn,showkeys,floatfix,nofootinbib,notitlepage]{revtex4-1}
\usepackage{graphicx} 
\usepackage{amssymb} 
\usepackage{amsmath}
\usepackage{mathtools}
\usepackage{epsfig}
\usepackage{color}
\usepackage{enumerate}
\newcommand{\be}{\begin{equation}} \newcommand{\ee}{\end{equation}}
\newcommand{\ba}{\begin{eqnarray}} \newcommand{\ea}{\end{eqnarray}}
\newcommand{\bea}{\begin{eqnarray}} \newcommand{\eea}{\end{eqnarray}}
\newcommand{\bean}{\begin{eqnarray*}} \newcommand{\eean}{\end{eqnarray*}}

\newcommand{\RNum}[1]{\uppercase\expandafter{\romannumeral #1\relax}}
\usepackage{slashed} 

\allowdisplaybreaks[1]

\newcommand{\tr}[1]{\ensuremath{{\rm Tr} \left[ #1 \right]}}

\newcommand{\ublob}{\mathcal{U}}
\newcommand{\lblob}{\mathcal{L}}

\begin{document}

\title{Extra Spin Asymmetries From the Breakdown of TMD-Factorization in Hadron-Hadron Collisions}  

\preprint{YITP-SB-13-11}

\author{Ted C. Rogers}
\affiliation{C.N.\ Yang Institute for Theoretical Physics, 
Stony Brook University, Stony Brook, New York 11794--3840, USA}
\email{rogers@insti.physics.sunysb.edu}   
\date{April 15, 2013}

\begin{abstract}
We demonstrate that partonic correlations that would traditionally be identified as subleading on the basis of
a generalized TMD-factorization conjecture can become leading-power because of TMD-factorization breaking 
that arises in hadron-hadron collisions with large transverse momentum back-to-back hadrons produced in the 
final state.  General forms of TMD-factorization fail for such processes because of a previously noted incompatibility 
between the requirements for TMD-factorization and the Ward identities of non-Abelian gauge theories.
We first review the basic steps for factorizing the gluon distribution and then show that a conflict between 
TMD-factorization and the non-Abelian Ward identity arises already at the level of a single extra soft or collinear 
gluon when the partonic subprocess involves a TMD gluon distribution.  Next we show that the resulting 
TMD-factorization violating effects produce leading-power final state spin asymmetries that would be classified as 
subleading in a generalized TMD-factorization framework.  
We argue that similar extra TMD-factorization breaking effects may be necessary to explain a range of open 
phenomenological QCD puzzles.  The potential to observe extra transverse spin or azimuthal asymmetries 
in future experiments is highlighted as their discovery may indicate an influence from novel and unexpected 
large distance parton correlations.  
\end{abstract}
\maketitle

\section{Introduction}
\label{sec:intro}

This paper examines the consequences of factorization breaking in 
inclusive high energy cross sections that are differential in the transverse momentum 
of produced particles, with a focus on the region of small transverse momentum where 
intrinsic motion associated with hadron structure becomes significant.  There is at present a wide range of motivations for
predicting and measuring intrinsic transverse momentum effects experimentally, so much effort continues to be devoted to 
developing methods to account for them in perturbative QCD (pQCD) treatments.  At a phenomenological level, the region 
of very small transverse momentum is most commonly described within a general transverse momentum dependent (TMD)  
parton model wherein the colliding hadrons are treated as collections of nearly free point-like quark and gluon constituents.  
Within the parton model description, the TMD parton distribution functions (PDFs) and fragmentation functions are treated 
much like classical probability densities.  They are conceptually very similar to the PDFs of the more familiar collinear parton 
model except that they describe the distribution of partons in terms of both longitudinal and transverse momentum components.

In pQCD, a parton model description should be replaced by a factorization theorem.
Demanding that factorization be consistent with real pQCD leads to rigid theoretical constraints which, in phenomenology,  
translate into first principles QCD predictions.  When factorization theorems accommodate TMD PDFs and TMD fragmentation 
functions, they are usually called ``TMD-factorization"  or ``$k_T$-factorization" theorems, and the constraints they impose are 
specific to cross sections that are sensitive to intrinsic parton transverse momentum.  A derivation or proof of a TMD-factorization 
theorem for a particular process must show that the transversely differential cross section factorizes into a generalized product 
of a hard part (perturbatively calculable to fixed order in small coupling), and a collection of well-defined non-perturbative TMD 
objects corresponding to TMD PDFs and TMD fragmentation functions for separate external hadrons.  The non-perturbative TMD 
functions contain detailed information about the influence of intrinsic non-perturbative motion of bound state quarks and gluons, so 
measuring them and explaining them theoretically is currently the focus of many efforts to formulate a deeper understanding of 
hadronic bound states in terms of elementary QCD quark-gluon degrees of freedom.  Because of their sensitivity to 3-dimensional 
intrinsic motion, TMD PDFs and fragmentation functions are also important in the study of the spin and orbital angular momentum 
composition of hadrons in terms of fundamental constituents.  For lists of relevant references see, for example, 
Refs.~\cite{Boer:2011fh,Balitsky:2011jrs}.  In addition, for high energy collisions at the LHC, certain TMDs may be useful for 
studying the detailed properties of the Higgs boson~\cite{Boer:2011kf,Schafer:2012yx}.  Generally, TMD-factorization theorems 
are necessary whenever highly accurate calculations of transversely differential cross sections in the region small or zero transverse 
momentum are needed.  This can include calculations used in the study of hadron structure as well as in searches for new physics.  

TMD-factorization theorems have been derived in pQCD with a high degree of rigor for a number of processes,
including Drell-Yan (DY) scattering, semi-inclusive deep inelastic scattering (SIDIS), and the production of back-to-back 
hadrons in $e^+ e^-$ annihilation~\cite{Collins:1981uk,Collins:1981uw,Collins:1984kg,Collins:1987pm,Bodwin:1984hc,Ji:2004wu,Ji:2004xq,collins}.   
For those processes where TMD-factorization theorems are currently understood and known to exist, an important next step 
is to implement unified phenomenological treatments, including global fits with QCD evolution and extractions of the 
non-perturbative TMD functions.  The success or failure of these efforts will be a crucial test of the validity of small-coupling 
techniques in studies of hadronic structure and in new phenomenological regimes beyond what are treatable within ordinary 
collinear factorization.  

However, the focus of this paper is on processes where normal steps for deriving TMD-factorization in pQCD are now understood to fail.  
Recently, standard parton-model-based pictures of TMD-factorization have been found to conflict with pQCD for certain classes of high 
energy processes, most notably in high energy hadron-hadron collisions where a pair of hadrons or jets with large back-to-back 
transverse momentum is produced in the final 
state~\cite{Bomhof:2004aw,Bomhof:2006dp,Bomhof:2007xt,Vogelsang:2007jk,Collins:2007nk,Collins:2007jp,Rogers:2010dm}.  
Interestingly, the conflict with TMD-factorization arises in kinematical regimes where common partonic intuition suggests that factorization 
should be very reliable.  In particular, there is a hard scale, $Q$, set by the transverse momentum of the produced final state hadrons, which 
may be of the same order-of-magnitude as the center-of-mass energy.  Therefore, the TMD-factorization breaking mechanisms discussed in 
Refs.~\cite{Bomhof:2004aw,Bomhof:2006dp,Bomhof:2007xt,Vogelsang:2007jk,Collins:2007nk,Collins:2007jp,Rogers:2010dm} 
are distinct from other well-known kinematical complications with factorization such as those expected in the limit of very small Bjorken-$x$.
They are instead due to an incompatibility between arguments for leading-power TMD-factorization and the non-Abelian gauge invariance of QCD, 
which persists even in the limit of a large hard scale.

For processes and observables where factorization theorems are valid, Ward identities maintain factorization in the large $Q$ limit by ensuring 
that any leading-power factorization breaking contributions that may appear term-by-term in perturbation theory cancel in the inclusive sum.
After all cancellations have occurred, any remaining terms that violate factorization must be shown to be suppressed by powers of $\Lambda_{\rm QCD}/Q$ 
in order for factorization to be said to be valid at leading power.  The details of these steps of a factorization proof constrain the specific form of
factorization for the classic hard QCD processes.  By contrast, the normally anticipated Ward identity cancellations fail in the scenarios discussed in
Refs.~\cite{Bomhof:2004aw,Bomhof:2006dp,Bomhof:2007xt,Vogelsang:2007jk,Collins:2007nk,Collins:2007jp,Rogers:2010dm}, 
leaving leftover leading-power TMD-factorization breaking contributions. The non-cancellations were first noted in 
Refs.~\cite{Bomhof:2004aw,Bomhof:2006dp} for their ability to introduce interesting process dependence in TMD-functions, and they were 
later identified in Refs.~\cite{Collins:2007nk,Collins:2007jp} as constituting a breakdown in the normal steps of a factorization proof.  
Still, it remained common to hypothesize that a more general form of TMD-factorization, called ``generalized TMD-factorization" in 
Refs.~\cite{Vogelsang:2007jk,Rogers:2010dm}, holds for these processes so long as the Wilson lines needed for gauge invariance in 
TMD definitions are allowed to have non-universaland potentially complicated structures.\footnote{We will use the terms ``gauge link" and 
``Wilson line" interchangeably throughout this paper.}  However, it was later found in~\cite{Rogers:2010dm}  that TMD-factorization in the 
hadro-production of back-to-back hadrons fails even in this generalized sense.  In other words, the TMD-factorization
derivations cannot generally be made consistent with gauge invariance even when allowing for non-universal Wilson line 
topologies in the TMD definitions.  Methods for addressing the resulting non-universality within tree-level or parton-model-like
approaches have since been proposed in Refs.~\cite{Gamberg:2010tj,DAlesio:2011mc,Buffing:2011mj,Buffing:2012sz}, and methods 
tailored to the small Bjorken-$x$ limit have been discussed in Ref.~\cite{Dominguez:2011wm,Xiao:2010sp}.  

A basic goal in the study of hadron structure is to establish a unified theoretical framework, rooted in 
elementary QCD quark and gluon concepts, for characterizing the intrinsic partonic correlations associated with QCD bound states and 
relating them to experimental observables.  In processes where TMD-factorization theorems are known to be valid, the information about 
intrinsic hadron structure is contained in well-defined non-perturbative correlation functions like the TMD PDFs and the TMD fragmentation 
functions.  The pattern that emerges is suggestive of a much more general picture of partonic interactions based on descriptions of quark-gluon 
properties for individual and separate external hadrons.  With this broad picture as a foundation, the customary TMD classification schemes 
have usually assumed a form of generalized TMD-factorization for all types of interesting or relevant hard hadronic processes.   The varieties of 
possible spin and angular behavior are then enumerated by first classifying the separate TMD functions according to their individual intrinsic 
properties, and then using them in a large set of both conjectured and derived TMD-factorization formulas.   (We will elaborate in more detail 
on the meaning of ``customary TMD classification schemes" in Sect.~\ref{sec:genfact}.) 

This very general TMD-factorization picture has intuitive appeal because of the straightforward organizational method it implies and because 
it has a very direct connection to parton model intuition. In classifications of the possible spin and angular asymmetries of hard inclusive cross 
sections, it is almost always taken as an assumption.  The only deviations that are usually allowed are those which incorporate non-universal 
normalization factors such as the overall sign reversal expected for the Sivers function in comparisons between the Drell-Yan process and 
SIDIS~\cite{Collins:2002kn} or an overall color factor normalization for more complicated processes~\cite{Vogelsang:2007jk}.

However, even the loosest forms of a TMD-factorization hypothesis impose strong constraints on the possible generalbehavior of transversely 
differential and spin dependent cross sections.  Those constraints constitute valuable first-principles QCD-based predictions where 
TMD-factorization is expected to hold, but it has become common to also apply the standard classification scheme outside of the class of 
processes known to respect TMD-factorization, including in processes like those discussed in 
Refs.~\cite{Bomhof:2004aw,Bomhof:2006dp,Bomhof:2007xt,Vogelsang:2007jk,Collins:2007nk,Collins:2007jp,Rogers:2010dm}. 
The purpose of this paper is to argue by way of example that, for these latter cases, the constraints imposed by a general TMD-factorization 
framework are too restrictive.\footnote{The meaning of ``generalized TMD-factorization," as it is used in this article, is much more general than 
what was called ``generalized TMD-factorization" in Ref.~\cite{Rogers:2010dm}.  The distinction will be explained in detail in 
Sect.~\ref{sec:partonmodel} of the main text.}  It will be shown that factorization breaking partonic correlations can produce unexpected 
patterns of spin and angular dependence in TMD cross sections that would otherwise be forbidden at leading power if a TMD-factorization 
hypothesis is adopted.  Hence, the breakdown in TMD-factorization can modify the general landscape of leading-power spin and angular 
behavior in hadron-hadron scattering rather than simply giving process dependence to already known TMD functions.

The relevance of TMD-factorization breaking extends beyond hadron or nuclear structure studies and is potentially important whenever 
perturbative QCD calculations are sensitive to the details of final states and are intended to have high precision point-by-point over a wide range 
of kinematics.  For example, because of the detailed account of final state kinematics in TMD-factorization, it is a potentially valuable tool in the 
construction of Monte Carlo event generators~\cite{Hautmann:2012pf,Dooling:2012uw}.  Also, it has recently even been found that factorization 
breaking arises in certain collinear cases~\cite{Forshaw:2012bi,Catani:2011st}.  In principle, it should be possible to connect the breakdown 
of TMD-factorization to the treatment of large higher order logarithms in collinear factorization and thus relate the two phenomena.

Very generally, factorization theorems for inclusive processes rely on cancellations in the inclusive sums over final states.  As such, the 
validity of any factorization theorem is placed in danger whenever extra final state constraints or conditions are imposed.  A recent 
discussion in the context of collinear factorization for top-antitop pair production can be found in Ref.~\cite{Mitov:2012gt}.  For the 
TMD-factorization breaking scenarios of 
Refs.~\cite{Bomhof:2004aw,Bomhof:2006dp,Bomhof:2007xt,Vogelsang:2007jk,Collins:2007nk,Collins:2007jp,Rogers:2010dm}, 
it is the specification of a small total transverse momentum for the final state back-to-back hadron or jet pair that is responsible 
for breaking TMD-factorization.  Violations of standard factorization have been observed to have quite large phenomenological consequences, 
such as in measurements of hard diffraction in hadron-hadron collisions and in dijet projection, particularly when compared with measurements of hard deep 
inelastic diffraction~\cite{Ahmed:1995ns,Derrick:1995tw,Affolder:2000vb,Khoze:2000dj,Kaidalov:2001iz,Klasen:2008ah,Kaidalov:2009fp}.  

The concept of a TMD PDF, and especially of an unintegrated gluon PDF, also appears in extensions of small coupling perturbation theory to 
the limit of small-$x$ where other novel phenomena such as parton saturation become relevant.  In this context, however, there are varying 
methods for using TMD gluon PDFs in calculations, and work is still needed to fully reconcile the different approaches with one 
another~\cite{Xiao:2010sp,Dominguez:2011wm,Avsar:2012hj,Avsar:2012tz}.

A collection of noteworthy QCD-related phenomenological puzzles has developed over the past several decades.  
(See also Ref.~\cite{Broadsky:2012rw}.) This includes famously large and non-energy-suppressed transverse single spin asymmetries (SSAs) 
observed in hadron-hadron collision experiments at Argonne National Laboratory (ANL)~\cite{Klem:1976ui,Dragoset:1978gg}, 
Fermilab~\cite{Adams:1991rw,Adams:1991cs}, CERN~\cite{Antille:1980th}, Serpukhov~\cite{Apokin:1990ik}, and 
Brookhaven National Laboratory (BNL)~\cite{Saroff:1989gn,Allgower:2002qi,Adams:2003fx}.  For a recent review of the current experimental status 
of TMDs and spin asymmetries, see Ref.~\cite{Aidala:2012mv}.  Very early, large final state $\Lambda$-hyperon transverse polarizations 
were also observed in \emph{unpolarized} hadron-nucleus collisions at Fermilab~\cite{Bunce:1976yb}.  Similar transverse polarizations have not 
been detected in the inclusive $e^+ e^-$ production of $\Lambda$-hyperons in the ALEPH experiment at LEP~\cite{Buskulic:1996vb}, or 
in semi-inclusive deep inelastic scattering measurements by ZEUS~\cite{Chekanov:2006wz}.  More recently, an apparent sign discrepancy~\cite{Kang:2011hk} 
has been identified in comparisons of Sivers asymmetries in SIDIS measurements~\cite{Airapetian:2004tw,Airapetian:2009ae} with Sivers asymmetries extracted 
from $h^\uparrow h \to \pi + X$ collisions at the BNL Relativistic Heavy Ion Collider (RHIC)~\cite{Adams:2003fx}.  Proposed explanations include
a node in the Sivers function~\cite{Kang:2011hk,Boer:2011fx}, though recent fits disfavor this~\cite{Kang:2012xf}.   It is also possible that the apparent 
$h^\uparrow h \to \pi + X$/SIDIS discrepancy is due to contributions beyond the Sivers effect~\cite{Metz:2012ui,Anselmino:2012rq}.  At present, the full explanation 
remains unclear and more analysis is needed.  At very high energies, interesting phenomena have also emerged, including the CMS ridge structure seen in 
proton-proton collisions at the LHC~\cite{Khachatryan:2010gv}.   Explanations have been proposed in terms of Wilson line interactions~\cite{Cherednikov:2010tr} 
and the color glass condensate formalism~\cite{Dusling:2012cg,Dusling:2012wy,Dusling:2013oia}. In addition, a forward-backward asymmetry in the production 
of top-antitop quark pairs has been observed at the Tevatron~\cite{Abazov:2011rq,Aaltonen:2012it},  generating speculation about effects from 
physics beyond the standard model.  

Given the collection of observations in last few paragraphs, it is a natural stage now to investigate
the possibility that TMD-factorization breaking effects lead to unexpected general patterns of behavior at leading power in the hard scale.
This paper will demonstrate, by applying power counting to a specific example, that such an extra TMD-factorization breaking spin 
dependence can be induced by TMD-factorization breaking initial and final state soft gluon interactions.  As our illustrative example, we 
will use the production of a final state large invariant-mass back-to-back photon-hadron pair in collisions of unpolarized protons.  We will 
show that a single extra initial state soft gluon, of the type more commonly associated with gauge link effects, is sufficient to produce a  
correlation between the helicity of the final state photon and the transverse spin of the exiting final state hadron.  We choose this specific 
example because it provides a relatively simple demonstration of an anomalous spin effect made possible by the breakdown of 
TMD-factorization while requiring only one extra soft/collinear gluon.   However, the discussion is intended to illustrate the more general 
observation that spin correlations that would not ordinarily be expected to be large from the perspective a generalized TMD parton model
can become leading-power in TMD-factorization breaking scenarios.  

The earlier examples of TMD-factorization breaking in 
Refs.~\cite{Bomhof:2004aw,Bomhof:2006dp,Bomhof:2007xt,Vogelsang:2007jk,Collins:2007nk,Collins:2007jp,Rogers:2010dm} 
focused on the appearance of anomalous color factors in low order graphs.  In this paper, the TMD-factorization breaking will also involve 
an anomalous contraction of intrinsic transverse momentum components with Dirac matrices, and so it will appear to come from a different source.  
At the root of all these examples, however, is the breakdown in compatibility between non-Abelian gauge invariance cancellations and the 
factorizability of the transverse momentum dependent cross section at leading power in the region of small transverse momentum.

The paper is organized as follows:  In Sect.~\ref{sec:setup}, we describe the process to be used throughout the paper to illustrate 
TMD-factorization breaking and the appearance of an extra TMD-factorization breaking spin asymmetry.   Section~\ref{sec:setup} also 
provides an overview of notation and conventions.  In Sect.~\ref{sec:superleading}, we give a brief summary of topics needed for discussions 
of TMD-factorization breaking, particularly in cases involving a TMD-gluon distribution in the initial state.  In Sect.~\ref{sec:partonmodel}, we 
precisely state the criteria for what we will call the maximally general form of a TMD-factorization hypothesis.  Also in Sect.~\ref{sec:partonmodel}, 
we summarize the constraints that are placed on the spin and azimuthal dependence of differential cross sections by this maximally generalized 
TMD-factorization hypothesis.  In Sect.~\ref{sec:onegluon} we present the detailed steps necessary to verify TMD-factorization at tree-level, and 
in Sect.~\ref{sec:spec} we deal with spectator-spectator interactions.  In Sect.~\ref{sec:twogluons} we apply direct power counting to demonstrate a 
failure of TMD-factorization for the case of a single extra soft gluon attaching to the initial and final state, and we isolate the leftover 
TMD-factorization breaking contributions.  We show in Sect.~\ref{sec:extra} that power counting implies an extra TMD-factorization violating spin 
correlation in the final state.  In Sect.~\ref{sec:conclusion} we summarize our main observations and make concluding remarks.

\section{Set Up and Notation}
\label{sec:setup}

\begin{figure*}
\centering
\includegraphics[scale=.45]{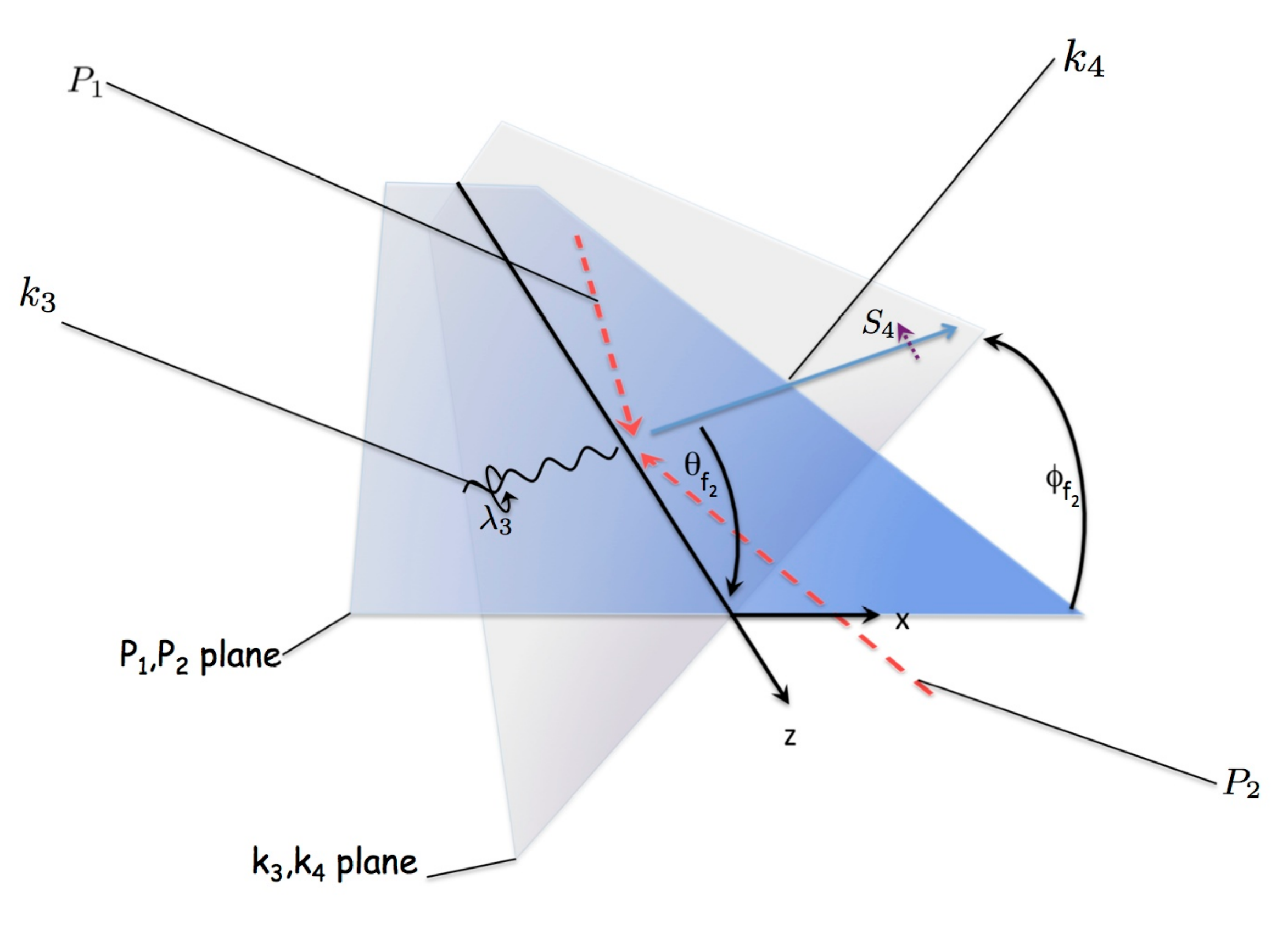}
\caption{Diagram of the process in Eq.~\eqref{eq:theprocess} in the coordinate system of frame-2.  
The final state hadronizing quark has transverse polarization $S_4$ and the final state prompt photon has helicity $\lambda_3$.}
\label{fig:processdiagram}
\end{figure*}

\subsection{The Process}
\label{sec:theprocess}
For concreteness, we will focus on TMD-factorization breaking in the specific example of inclusive 
production of a back-to-back hadron-photon or jet-photon pair (see Fig.~\ref{fig:processdiagram}):
\begin{equation}
\label{eq:theprocess}
{\rm Hadron \, 1} + {\rm Hadron \, 2}  \to {\rm Jet}(S_4) + \gamma(\lambda_3) + X \, .
\end{equation}  
The large transverse momentum of the final $\gamma$ and jet/hadron fixes the hard scale.   
The jet/hadron and the real photon are nearly back-to-back, so 
the cross section, which is differential in the total transverse momentum of the final state 
hadron-photon pair, is sensitive to the intrinsic transverse momentum of the 
constituent partons in the colliding hadrons.  
Hence, TMD-factorization is a natural candidate description.

The outgoing quark in Fig.~\ref{fig:processdiagram} is meant to be representative of a
final state hadron or jet with a transverse polarization $S_4$.  

Let $P_1$ and $P_2$ be the momenta of the incoming colliding hadrons and let $k_4$ and $k_3$ be the 
momenta of the final state jet and real photon respectively.  The total momentum of the final state pair is defined
to be
\begin{equation}
\label{eq:totfinal}
q \equiv k_3 +k_4,
\end{equation}
and $Q^2 \equiv q^2$ is the hard scale.  The kinematical region of interest is characterized by final state 
transverse momenta of order the total center-of-mass energy of the colliding hadrons:
\begin{equation}
(P_1 + P_2)^2 \sim (P_1 + k_3)^2 \sim (P_1 + k_4)^2 = \mathcal{O}(Q^2)\, .
\end{equation} 
That is, the final state center-of-mass transverse momenta of $k_3$ and $k_4$ are large.
\begin{figure}
\centering
\includegraphics[scale=.4]{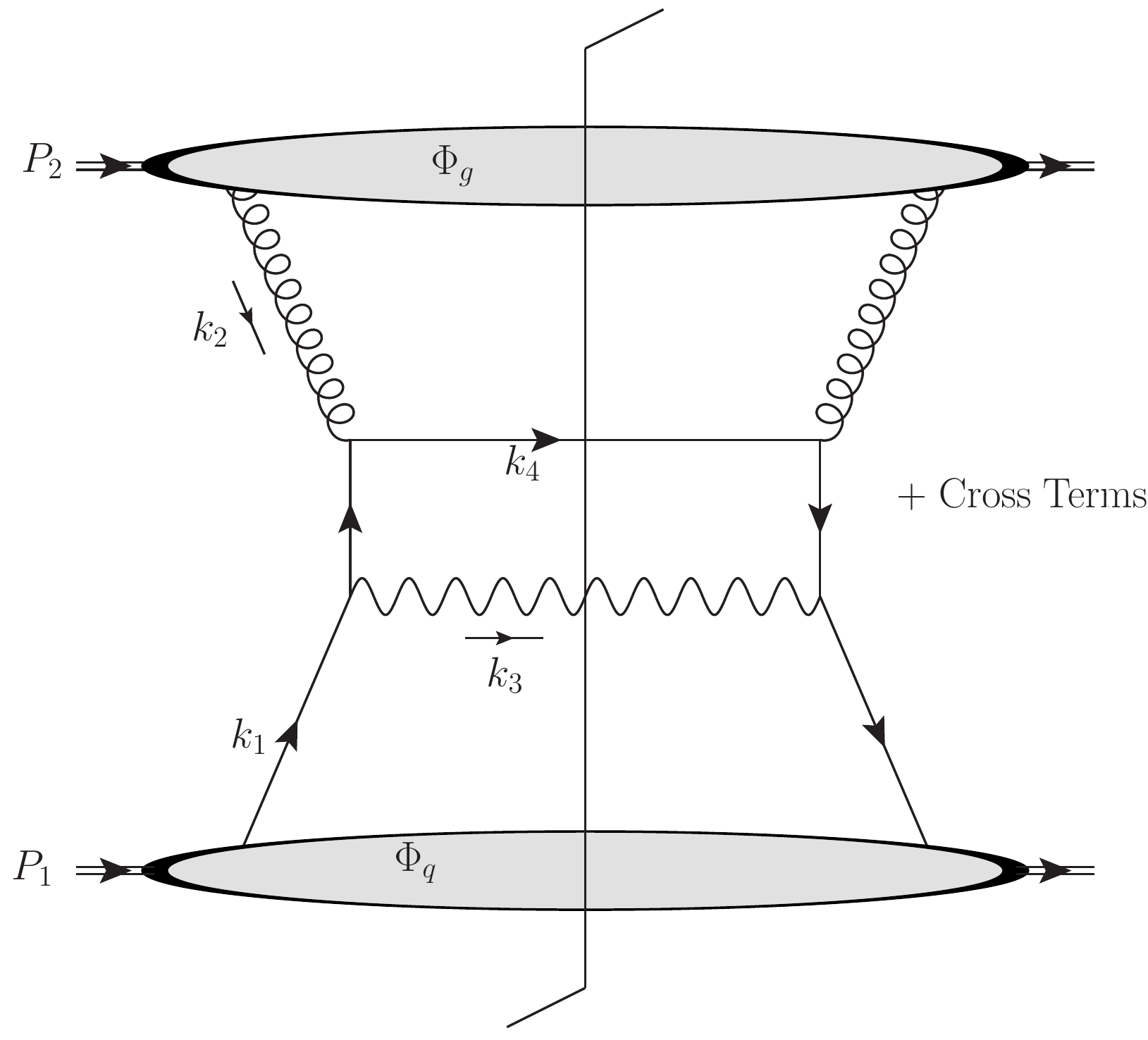}
\caption{A parton-model level picture showing the $2 \to 2$ tree-level subprocesses $q g \to q \gamma$. }
\label{fig:partonmodel}
\end{figure}

The hard subprocess includes both 
$q \bar{q} \to g \gamma$ and $g q  \to \gamma q$ channels, but here
we are most interested in the gluon-initiated subprocess since it is this that (as we will later show) most 
immediately yields a non-trivial extra spin asymmetry from TMD-factorization breaking effects.  
A parton model level depiction of the  $g q \to \gamma q$ hard scattering subprocess is shown 
in Fig.~\ref{fig:partonmodel}; the hard collision is the result of a quark inside
$P_1$ scattering off a gluon inside $P_2$.   For a complete pQCD treatment,
the $q \bar{q} \to g \gamma$ channel must also be accounted for, but an analysis of the  $g q \to \gamma q$ 
subprocess is all that will be needed  to obtain our main result. 

In the treatment of the hard subgraphs, we will often use the kinematical notation:
\begin{align}
\hat{s} \equiv &  (k_1 + k_2)^2 = (k_3 + k_4)^2 = q^2 = Q^2\, , \\
\hat{t} \equiv & (k_4 - k_2)^2 = (k_1 - k_3)^2 = \mathcal{O}(Q^2)\, .
\end{align}  
We will also use a momentum variable defined as
\begin{equation}
h \equiv k_4 - k_2 = k_1 - k_3 \, . \label{eq:hdef}
\end{equation}
The proton mass is $M_p$ and the quark mass is $m_q$.  We are ultimately interested in spin
correlations arising between the final state jet and prompt photon from factorization breaking effects,
so we have labeled the jet/hadron and the photon with a spin vector, $S_4$, and a helicity, $\lambda_3$, in 
Eq.~\eqref{eq:theprocess} and Fig.~\ref{fig:processdiagram}.

A complete kinematical description of the cross section is closely analogous to the usual description of Drell-Yan scattering and 
$e^+ e^-$-annihilation into back-to-back jets.  Our notation will therefore closely follow the standard Drell-Yan conventions.  
(See Ref.~\cite{collins}, chapter 13.2.)  The modifications to the usual notation needed for the special case of Eq.~\eqref{eq:theprocess} 
will be discussed below.

First, we define the relevant coordinate reference frames:

\subsubsection{$P_1$, $P_2$ Center-of-Mass}
\label{sec:f1}
For categorizing the sizes of momentum components of the incoming partons, it will 
be most convenient to work in the center-of-mass of the colliding $P_1$-$P_2$ system, where $P_1$ is moving 
in the extreme light-cone ``$+$'' direction and $P_2$ is in the extreme ``$-$" direction, and the total transverse momentum 
of the $P_1$-$P_2$ system is zero.
We will refer to this as ``frame-1."  Momentum components in frame-1 will be labeled by 
an $f_1$ subscript.
The positive $z$-axis in this frame is along the direction of motion of $P_1$ so that 
the four-momentum components of the incoming hadrons are
\begin{eqnarray}
P_{1,f_1}  = \left(P_{1,f_1}^+,\frac{M_p^2}{2 P_{1,f_1}^+} , {\bf 0}_t \right) \approx   \left( P_{1,f_1}^+,0, {\bf 0}_t \right) \, , \\
P_{2,f_1}  =   \left(\frac{M_p^2}{2 P_{2, f_1}^-} , P_{2, f_1}^-, {\bf 0}_t \right) \approx  \left(0, P_{2, f_1}^-, {\bf 0}_t \right) \, .
\end{eqnarray}
The total four-momentum of the final state jet-photon pair in frame-1 is
\begin{equation}
q_{f_1} = k_{3,f_1} + k_{4,f_1} = (q_{f_1}^+,q_{f_1}^-,{\bf q}_{t, f_1}).
\end{equation}
In frame-1, therefore, small deviations of ${\bf q}_{t, f_1}$ from zero are due to intrinsic transverse momentum in 
the colliding hadrons.  The kinematical region of interest for studies of TMD-factorization and TMD-factorization breaking 
is where ${\bf q}_{t, f_1}$ is of order $\Lambda$.  (For the rest of this paper, $\Lambda$ refers to a general hadronic mass scale like 
$\Lambda_{\rm QCD}$.)

The momentum fractions $x_1$ and $x_2$ are defined as:
\begin{equation}
k_{1, f_1}^+ \equiv x_1 P_1^+; \qquad k_{2,f_1} \equiv x_2 P_2^- \, .
\end{equation}
The incoming partons $k_1$ and $k_2$ are approximately collinear to their parent hadrons, so in frame-1 their components are of size
\begin{equation}
k_{1,f_1} \sim \left(Q,\Lambda^2/Q,{\bf \Lambda}_t \right) \, , \qquad k_{2,f_1} \sim \left(\Lambda^2/Q,Q,{\bf \Lambda}_t \right) \, . \label{eq:k1k2size}
\end{equation}
In frame-1, the components of $k_3$ and $k_4$ are all large:
\begin{equation}
k_{3,f_1} \sim (Q,Q,{\bf Q}_t), \, \qquad k_{4,f_1} \sim (Q,Q,{\bf Q}_t) \, .
\end{equation}

It will also be convenient to define unit four-vectors characterizing the extreme forward and backward directions in frame-1:
\begin{eqnarray}
n_{1,f_1} \equiv (1,0,{\bf 0}_t)_{f_1}, \label{eq:ndefs1} \\  
n_{2,f_1} \equiv (0,1,{\bf 0}_t)_{f_1}\, . \label{eq:ndefs2}
\end{eqnarray}
Most of the analyses of this paper will be performed in frame-1.

At various stages, we will need only the  frame-1 transverse components of a parton's momentum, expressed in four-vector form.
Therefore, we define the four-vector notation:
\begin{equation}
\label{eq:l2perpdef}
v_{ t} \equiv \left(0,0, {\bf v}_{t} \right) \, .
\end{equation}
The transverse components of the metric tensor in frame-1 may be conveniently expressed as
\begin{equation}
g_t^{\mu_1 \mu_2} \equiv g^{\mu_1 \mu_2} - n_{1, \, f_1}^{\mu_1} n_{2, \, f_1}^{\mu_2} - n_{1, \, f_1}^{\mu_2} n_{2, \, f_1}^{\mu_1} \, .
\end{equation}

\subsubsection{$k_3$, $k_4$ Center-of-Mass}
\label{sec:f2}
We define ``frame-2" to be the center-of-mass of the final state jet-photon system, and the corresponding 
momentum space coordinates will be labeled by subscripts $f_2$.  In this frame the orientation of the 
interaction planes is clearest and easiest to visualize.

If the masses of the incoming hadrons (and the mass of the final state jet) are neglected,
then in the center-of-mass of the jet-photon system
\begin{eqnarray}
q_{f_2} & = & (Q,{\bf 0}), \\
k_{3, f_2} & = & (Q/2,  {\bf k}_{f_2} ), \\
k_{4, f_2 } & = &  (Q/2, - {\bf k}_{f_2} ), \\
P_{1,f_2} & = &  | {\bf P}_{1,f_2} | (1, {\bf u}_{1, f_2} ), \\
P_{2,f_2} & = &  | {\bf P}_{2,f_2} | (1, {\bf u}_{2, f_2} ), 
\end{eqnarray}
where here we have reverted to standard four-vector notation $v = (v^0,v^1,v^2,v^3)$.  
The energy of the 
jet-photon pair is $Q$, and ${\bf u}_{1, f_2}$, ${\bf u}_{2, f_2} $ are unit spatial three-vectors in the 
directions of $P_1$ and $P_2$ respectively.
In frame-2, the final state jet and real photon are exactly back-to-back, but the incoming hadrons $P_1$ and $P_2$ are 
slightly away from back-to-back (see Fig.~\ref{fig:processdiagram}).  The region of very small intrinsic transverse momentum 
corresponds to almost exactly back-to-back $P_1$ and $P_2$.  The $z$-axis and $x$-axis can be covariantly fixed by defining 
the following four-vectors:
\begin{equation}
Z^{\mu} \equiv \frac{\left( 0, {\bf u}_{1, f_2}  - {\bf u}_{2, f_2} \right)}{|{\bf u}_{1, f_2}  - {\bf u}_{2, f_2}|}, \quad X^{\mu} \equiv \frac{ \left( 0, {\bf u}_{1, f_2}  +  {\bf u}_{2, f_2} \right)}{|{\bf u}_{1, f_2}  + {\bf u}_{2, f_2}|}.
\end{equation}
The $z$-axis bisects the angle between 
${\bf P}_{1,f_2}$ and $-{\bf P}_{2,f_2}$, and the $x$-axis is orthogonal to the $z$-axis, so that together they form a ``$P_1,P_2$-plane."
Hence, frame-2 is analogous to the Collins-Soper frame~\cite{Collins:1977iv} for Drell-Yan scattering, but with the lepton pair replaced by 
the jet-photon pair.  The other plane, analogous to the lepton plane for Drell-Yan scattering, is formed by the jet-photon pair.  We will call 
it the ``$k_3,k_4$-plane."  An illustration of the process in the coordinate system of frame-2 is shown in Fig.~\ref{fig:processdiagram}.  
The polar angle of the jet axis with respect to the z-axis is $\theta_{f_2}$, and the azimuthal angle with respect to the $P_1, P_2$-plane is $\phi_{f_2}$.

\subsubsection{$k_4$ Boosted in the z-direction}

Finally, it will be useful in some instances to work in the $k_4$, $k_3$ center-of-mass frame with the $z$-axis lying along 
the direction of $k_4$, so that $k_4$ has its plus component boosted to order $\sim Q$ and 
$k_3$ has its minus component boosted to order $\sim Q$.
We will call this ``frame-3" and denote the components of vectors in this frame with an $f_3$.
In light-cone coordinates,
\begin{equation}
k_{4, f_3} \approx \left(k_{4, f_3}^+,0, 0 \right) \, ,
\end{equation} 
 where $k_{4, f_3}^+ \sim Q$.
 We also define a vector 
 \begin{equation}
 \label{eq:n3def}
 n_{3, f_3}  = (0,1,{\bf 0}_t)_{f_3}\, ,
 \end{equation}
 so that $ n_{3} \cdot k_{4} = k_{4, f_3}^+ \sim Q$, and 
  \begin{equation}
 \label{eq:n4def}
 n_{4, f_3}  = (1,0,{\bf 0}_t)_{f_3}\, ,
 \end{equation}
  so that $ n_{4} \cdot k_{3} = k_{3, f_3}^- \sim Q$.
  Frame-3 is simply related to frame-2 by a rotation.  Most importantly for us, the transverse spatial components of $P_1$ and $P_2$ in frame-3 are
  \begin{equation}
 \label{eq:largep2trans}
 |  \left( \bf{P}_{1t} \right)_{f_3} | \approx  |  \left( \bf{P}_{2t} \right)_{f_3} | \approx | {\bf P}_{1,f_2} | \sin \theta_{f_2} = \mathcal{O}(Q) \, ,
  \end{equation}
  for small intrinsic transverse momentum and wide angle scattering (see Fig.~\ref{fig:processdiagram}). 

\subsection{Final State Spin Dependence}
\label{sec:abelian}

In later sections, we will be interested in the dependence on the polarization of the final state $k_4$ in Fig.~\ref{fig:processdiagram}, and 
the representation of spin will follow the notation and conventions in appendix A of Ref.~\cite{collins}.  In particular, the spin vector $S_4$ of 
the final state quark in frame-3, where $k_4$ is exactly in the $z$-direction, is
\begin{align}
S_4^\mu & =  (S_{4,f_3}^0,S_{4,f_3}^x ,S_{4,f_3}^y,S_{4,f_3}^z) \nonumber \\
                  & =    (\lambda_4 k_{4,f_3}^z {\rm sign}(k_{4,f_3}^z), m_q b_{t,f_3}^x,m_q b_{t,f_3}^y,\lambda_4 k_{4,f_3}^0 {\rm sign}(k_{4,f_3}^z) ) \, .
\end{align}
The $\lambda$ and $b_{t,f_3}$ are the components of the Bloch vector.  
The largest component of $S_4$ is of order $Q$.
The spin vector is normalized with a maximum of $-S_4^2 \leq m_q^2$, and the spin sum is
\begin{equation}
\label{eq:spinsum}
\sum_s \, u(k_4,S_4) \bar{u}(k_4,S_4) = \left( \slashed{k}_4 + m_q \right) \left( 1 + \gamma_5 \frac{\slashed{S}_4}{m_q}   \right)\, .
\end{equation}
The polarization of the final state photon will be expressed in the usual way; the polarization four-vectors in frame-3 are
\begin{equation}
\label{eq:photonpol}
\epsilon_{\lambda_3 = 1}^\mu = \left( 0, -\frac{1}{\sqrt{2}}, -\frac{i}{\sqrt{2}} ,0 \right)_{f_3} \, , \qquad \epsilon_{\lambda_3 = -1}^\mu = \left( 0, \frac{1}{\sqrt{2}},-\frac{i}{\sqrt{2}},0 \right)_{f_3} \, ,
\end{equation}
where $\lambda_3 = \pm 1$ refer to opposite helicities.

For a review and a general treatment of spin see, e.g., Ref.~\cite{Leader:2001gr} and for typical conventions see~\cite{Nakamura:2010zzi}.
  
\subsection{Leading Regions and the Classification of Subgraphs}
\label{sec:leadingregions}
In perturbative QCD derivations of factorization in an arbitrary gauge, one must deal with extra soft and collinear gluons that are
allowed in a general leading region analysis, but which do not factorize topologically graph-by-graph.
Where factorization is known to be valid, the derivations show that contributions from these extra gluons either 
cancel in the sum over graphs, or factor into contributions corresponding to separate
Wilson line operators, with each Wilson line operator belonging to a
separately well-defined parton correlation function for each hadron.  
This standard factorization applies to sums over all graphs and relies on applications
 of the QCD Ward identity (gauge invariance) to disentangle the extra gluons that connect different subgraphs with one another. 
A factorization proof must show that, after the application of the gauge invariance arguments and Ward identity cancellations, 
any remaining leading-power factorization breaking terms cancel in the sum over final states for sufficiently inclusive observables.
\begin{figure}
\centering
\includegraphics[scale=.4]{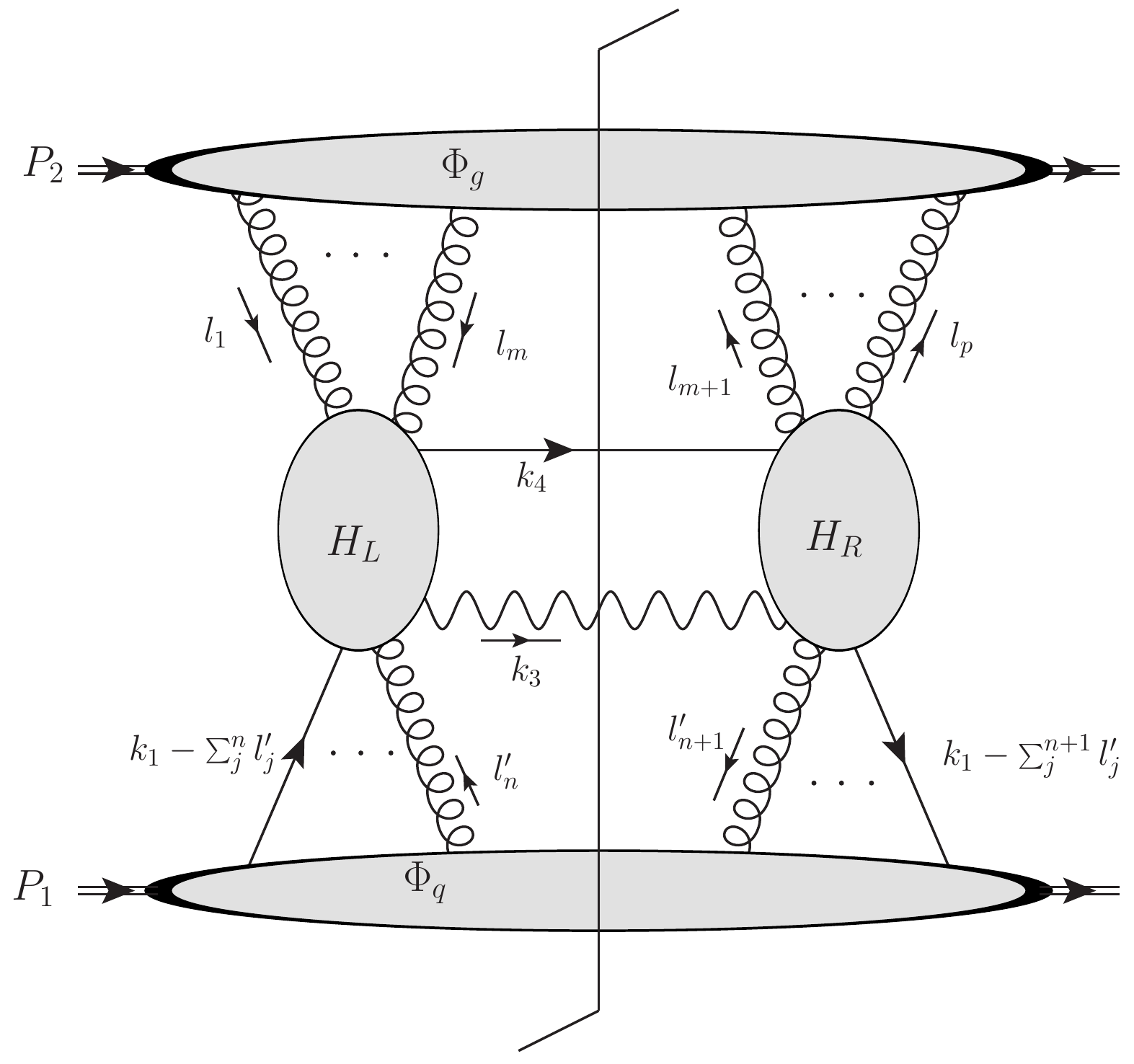}
\caption{The structure of a general leading region. }
\label{fig:leadingregions}
\end{figure}

Standard leading region arguments~\cite{Libby:1978bx} (see also chapter 13 of~\cite{sterman} and chapter 5 of~\cite{collins}), show that graphs with the topology illustrated
in Fig.~\ref{fig:leadingregions} are the dominant ones at leading power for the process in Eq.~\eqref{eq:theprocess}.
A generic Feynman graph may have arbitrarily many 
soft/collinear gluons (labeled ${l_j}$, with $j$ running from $1$ to $m$) entering the hard subgraph $H_L$ from the 
top gluon subgraph, $\Phi_g$.  
A quark enters $H_L$ from the lower quark subgraph, $\Phi_q$, along with arbitrarily many extra 
longitudinally polarized gluons $l_i^\prime$ (with $i$ running from $1$ to $n$).
Analogous statements apply to the right side of the cut in Fig.~\ref{fig:leadingregions}.
We define
\begin{equation}
\label{eq:k2sum}
k_2 \equiv \sum_{j=1}^m l_j\, ,
\end{equation}  
so that the total momentum entering the hard subgraphs, $H_L$ and $H_R$, matches 
the routing of momentum in tree-level graphs like Fig.~\ref{fig:partonmodel}.
There are also important contributions from gluons that directly connect the $\Phi_q$ and $\Phi_g$ subgraphs, but
to keep the diagram simple we have not shown them explicitly in Fig.~\ref{fig:leadingregions} (see, however, Sect.~\ref{sec:spec}). 
The soft and collinear gluons may cross the final state cut, and this is included in the definitions of the $\Phi_q$ and $\Phi_g$ bubbles.

From here forward, the subscripts $f_1$ will be dropped, and it will be assumed to be implicit that we are working in frame-1
except when specified otherwise.  In our notation, Lorentz indices denoted by 
Greek letters are intended to include all four components of a four-momentum, 
while indices denoted with $i$ or $j$ include only the frame-1 transverse spatial components.

\begin{figure}
\centering
\includegraphics[scale=.4]{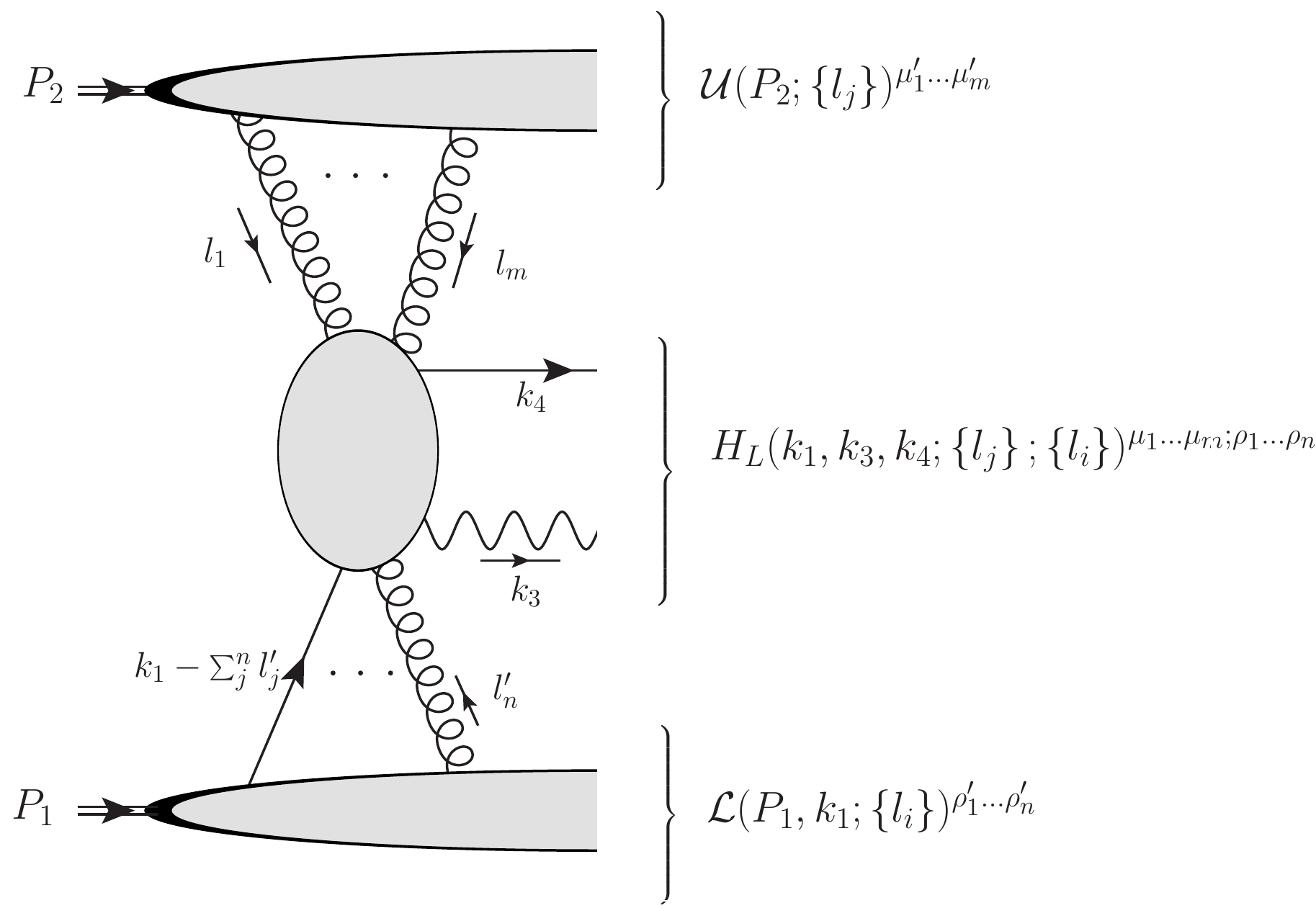}
\caption{A generic contribution to the amplitude with the different pieces of Eq.~\eqref{eq:ampbasic} labeled. }
\label{fig:leadingamplitude}
\end{figure}

The differential cross section for an arbitrary number of extra soft/collinear gluons is expressed as
\begin{multline}
\label{eq:crosssection}
d \sigma =  \mathcal{C} \, Q^2 \, \sum_{X_1} \sum_{X_2} \sum_{m,p,n,q} \int \frac{d^4 k_1}{(2 \pi)^4} \int \frac{d^4 l_1}{(2 \pi)^4} \cdots  \int \frac{d^4 l_m}{(2 \pi)^4} \int \frac{d^4 l_{m+1}}{(2 \pi)^4} \cdots \times \\ \times  \int \frac{d^4 l_p}{(2 \pi)^4}
\int \frac{d^4 l^\prime_1}{(2 \pi)^4} \cdots  \int \frac{d^4 l^\prime_n}{(2 \pi)^4} \int \frac{d^4 l^\prime_{n+1}}{(2 \pi)^4} \cdots \times \\ \times  \int \frac{d^4 l^\prime_q}{(2 \pi)^4}
\; | M |^2 \; \times ({\rm mom. \, conserving}\, \delta{\rm -functions}),
\end{multline}
where $M$ is the amplitude in Fig.~\ref{fig:leadingamplitude}, and the integrals over the momenta 
entering or exiting the hard subgraphs are written explicitly.  The inclusive sums and integral over final states in the upper and lower 
bubbles of Fig.~\ref{fig:leadingregions} are represented by $\sum_{X_1} \sum_{X_2}$.  The momentum conserving 
$\delta$-functions fix $k_1 + k_2  = k_3 + k_4 = q$, with the definition of $k_2$ in Eq.~\eqref{eq:k2sum}.  For maximum generality, we leave
the overall numerical normalization $\mathcal{C}$ unspecified in Eq.~\eqref{eq:crosssection}.  
The essential requirement is only that the cross section is differential in ${\bf q}_t$ --- note that there is 
no integration over ${\bf q}_t$ in Eq.~\eqref{eq:crosssection}.   

Inspection of the tree-level graphs (see, e.g., Sect.~\ref{sec:onegluon}) verifies that the power-law for $M$ is $M \sim Q^0$.  
The phase space integrals of Eq.~\eqref{eq:crosssection} are Lorentz invariant 
and so do not contribute extra powers of $Q$ to the cross section when boosting frames. 
In later sections we will analyze only the relative sizes, in powers of $Q$, of the factors that combine to form $| M |^2$.  
Therefore, the right-hand side of Eq.~\eqref{eq:crosssection} is defined with a normalization factor $Q^2$ so that its 
overall power-law dependence is the same as for the amplitude, $d \sigma \sim Q^0$ (up to logarithmic factors).  
The overall $Q^2$ factor is to compensate for the two powers of $1/Q$ that come 
from the momentum conserving delta functions for $k_1^+$ and $k_2^-$.
We are mainly interested in the dependence of the cross section on powers of $Q$, 
so to keep notation simple we will 
quote the power-behavior of any factor as a power of $Q$ only, with any other factors of order a hadronic 
mass scale, necessary for maintaining correct units, left implicit.

Since the cross section is order $\sim Q^0$, then
when a particular term in the perturbative expansion 
is of a higher power of $Q$ than zero it is called ``superleading."  
In a complete analysis including all graphs, gauge invariance ensures that all such terms cancel exactly against other superleading terms.

In later sections we will find it simplest to work at the amplitude level and to consider 
the contribution to the cross section only in the last step.  
We will organize the analysis of amplitudes according to Fig.~\ref{fig:leadingamplitude}.  A specific
diagrammatic contribution can be broken into upper and lower subgraphs, $ \ublob$ and $\lblob$, and a hard part, $H_L$,  
as follows:
\begin{multline}
\label{eq:ampbasic}
M = \ublob(P_2;\left\{l_j\right\})^{\mu_1^\prime \dots \mu_m^\prime} \, g_{\mu^\prime_1 \mu_1} \cdots g_{\mu_m^\prime \mu_m} \, \times \\ \times
\bar{u}(k_4,S_4) \, H_L(k_1,k_3,k_4;\left\{l_j\right\};\left\{l_i\right\})^{\lambda_3 ; \mu_1 \dots \mu_m;\rho_1 \dots \rho_n} \, \times \\ \times g_{\rho^\prime_1 \rho_1} \cdots g_{\rho_n^\prime \rho_n} \, 
\lblob(P_1,k_1;\left\{l_i\right\})^{\rho_1^\prime \dots \rho_n^\prime}.
\end{multline}  
The Lorentz indices for gluons that connect $\ublob$ and $\lblob$ to the hard subgraph $H_L$ are shown explicitly.  In our notation, the final state quark 
wave function $\bar{u}(k_4,S_4)$ is written separately rather than being included in $H_L$ because this will be convenient for power counting arguments in
later sections.
$H_L$ and $\lblob$ also carry Dirac indices, though to maintain manageable notation, we do
not show these explicitly.  
The subgraphs $\ublob$,  $\lblob$ and $H_L$ are defined to include the sum of all subgraphs corresponding 
to a given set of $m$ external gluons for $\ublob$, $n$ external gluons for $\lblob$, and $m+n$ external gluons for $H_L$.  However, this over counts the number of
graphs by a factor of $m!  \times n!$.  
To compensate for this, therefore, we define the $\ublob$ subgraph to also include a 
factor of $1 / m!$, and the $\lblob$ subgraph to include a factor of $1 / n!$.

All of our analysis will be done using Feynman gauge where the analytic properties of Feynman diagrams reflect relativistic causality and the main issues concerning   
TMD-factorization and TMD-factorization breaking are clearest.  Lorentz indices will by denoted by $\mu,\nu,\rho,\sigma$, transverse components will be 
denoted by $ij$, and color indices will be denoted by $\alpha, \beta, \delta, \kappa$.

Actual derivations of factorization, where they are valid, involve a number of important points that will not be touched on in this paper, both 
because they are not directly relevant to the specific issues under consideration and because generalized TMD-factorization is anyway known 
to be violated for processes like Eq.~\eqref{eq:theprocess}.  
Those other issues include, for example, rapidity 
divergences and other topological Wilson line issues that must 
be fully addressed in situations where TMD-factorization is valid 
(see, for example, Refs.~\cite{Collins:1999dz,Belitsky:2002sm,Collins:2003fm,Hautmann:2007uw,Collins:2008ht,Cherednikov:2007tw,Cherednikov:2012zi}), 
and their relationship to 
the evolution of separate, well-defined TMD PDFs~\cite{collins}.  Furthermore, in the integrations over
intrinsic transverse momentum in Eq.~\eqref{eq:crosssection}, propagators in the hard part may go on-shell whenever any of the transverse momenta become 
too large.  Therefore, in this paper it will be assumed that all such transverse momentum have upper cutoffs to prevent this, since the main concern is 
with transverse momenta of the order of hadronic mass scales or smaller.  
 
\section{Superleading Terms, Gauge Invariance, and the Glauber Region}
\label{sec:superleading}

\subsection{Dominant Subdiagrams}
\label{sec:dominant}

Recall that, in frame-1, $P_1$ is highly boosted in the plus direction while $P_2$ is highly boosted in the minus direction.
The graphs that individually contribute superleading terms to the 
cross section --- i.e., with a power-law behavior higher than that of the cross section itself --- are those in which all the components of $\ublob$ are in the minus direction:
\begin{equation}
\label{eq:superleading}
\ublob(P_2;\left\{l_j\right\})^{\mu_1^\prime \dots \mu_m^\prime} \to \ublob(P_2;\left\{l_j\right\})^{-,-,-, \dots, -}.
\end{equation}
The situation here is the same as was addressed in Ref.~\cite{Collins:2008sg} in 
connection with ordinary collinear factorization and collinear gluon PDFs.
Gauge invariance ensures that, in the sum over all diagrams, any superleading 
contributions to a physical cross section cancel.  The way this is demonstrated
relies on the same basic Ward identity arguments that are also needed to 
identify gauge link contributions and verify normal factorization theorems in processes where they apply~\cite{Collins:1981uw,Labastida:1984gy}.

The observation from Ref.~\cite{Collins:2008sg} that is most relevant to this paper is that, 
although the superleading powers of $Q$ cancel after the sum over all graphs, 
contributions from subgraphs like Eq.~\eqref{eq:superleading} generally do not.  (See, e.g., the third term in braces in Eq.~(83) of Ref.~\cite{Collins:2008sg}.) 
To understand this, recall that the minus component of polarization of the virtual $k_2$-gluon in hadron-2 is not 
exactly the same as its component along the direction of $k_2$ because $k_2$ also 
has order $\sim {\bf \Lambda}_t$ transverse components.  
So a minus polarization is not exactly the same as a polarization along the direction of $k_2$, making 
the Ward identity cancellations more delicate.

In spite of the non-vanishing leading-power contribution that arises from $\ublob$ subgraphs like Eq.~\eqref{eq:superleading}, 
the cross section for the \emph{collinear} case addressed in Ref.~\cite{Collins:2008sg} does factorize, as expected, into the well-known and well-defined collinear
gauge-invariant correlation functions.
But even in that case the uncanceled leading-power terms from subdiagrams like Eq.~\eqref{eq:superleading} remain as contributions from the 
gauge link, and are therefore critical for maintaining consistency with gauge invariance at leading power.

In later sections, we will encounter other complications from subgraphs like 
Eq.~\eqref{eq:superleading} in the treatment of generalized TMD-factorization breaking scenarios like Eq.~\eqref{eq:theprocess}.  

\subsection{Grammer-Yennie Method}
\label{sec:grammeryennie}

Factorization in covariant gauges is very conveniently addressed with the Grammar-Yennie approach~\cite{Grammer:1973db}, which we 
briefly review here.  
In this method, each extra gluon's propagator numerator is split into two terms that we call the ``G-term" and the ``K-term." 
For a gluon with momentum $l$ attaching $\ublob$ to the rest of the graph in Fig.~\ref{fig:leadingamplitude}, we write the decomposition as
\begin{equation}
\label{eq:gramyen}
g^{\mu^\prime \mu} = K(l)^{\mu^\prime \mu} + G(l)^{\mu^\prime \mu} \, .
\end{equation}
The $G$ and $K$ terms are defined as
\begin{eqnarray}
K(l)^{\mu^\prime \mu} & \equiv & \frac{n_1^{\mu^\prime} l^\mu}{n_1 \cdot l}  \label{eq:Kdef} \, \\
G(l)^{\mu^\prime \mu} & \equiv & g^{\mu^\prime \mu} -  \frac{n_1^{\mu^\prime} l^\mu}{n_1 \cdot l} \label{eq:Gdef}  \, .
\end{eqnarray}
Note the explicit dependence of $G(l)$ and $K(l)$ on gluon four-momentum.
The decomposition breaks the diagram into terms whose properties have a 
clear interpretation in terms of Ward identities and gauge invariance.  
The motivation for the separation is that it systematizes the identification of   
dominant and subdominant contributions to a subgraph, while automatically (and exactly) implementing 
any Ward identity cancellations that come from contracting the gluon momenta $l^\mu_{\{j\}}$ with 
the rest of the graph in Fig.~\ref{fig:leadingamplitude}.  Notice that in Eqs.~(\ref{eq:Kdef},~\ref{eq:Gdef}) it is the exact $l^\mu$ that appears in the numerators rather 
than just $l^-$.
After expressing the graphs in terms of $G$ and $K$ terms, it becomes clear how to apply valid approximations while at the same time maintaining 
gauge invariance.

Equation~\eqref{eq:ampbasic} has been written with  the metric tensors from the 
propagator numerators of extra gluons displayed explicitly.  This is in preparation for later sections where we will 
implement the Grammer-Yennie decomposition on each extra gluon and examine the 
power law behavior of all $K$ and $G$ combinations in each Feynman graph.  

If $l \sim \left(\Lambda^2/Q, Q, {\bf \Lambda}_t \right)$ in Eq.~(\ref{eq:Kdef},\ref{eq:Gdef}) so that it is collinear, then power counting with the 
$G$ and $K$ terms is straightforward.  
For example, it is the $g^{+ -}$ component of its propagator numerator in Eq.~\eqref{eq:ampbasic} that gives the dominant power, and  
the $K^{+ -}$ term in Eq.~\eqref{eq:gramyen} reproduces this exactly.   As long as $l \sim \left(\Lambda^2/Q, Q, {\bf \Lambda}_t \right)$, the other components 
of $K^{\mu^\prime \mu}$ and $G^{\mu^\prime \mu}$ are suppressed relative to the $K^{+ -}$ components.  However, the simple power counting fails to apply directly if $l$ is 
integrated into different momentum regions such as the Glauber region. 

\subsection{The Glauber Region}
\label{sec:glauber}
Many issues surrounding factorization and TMD factorization breaking are closely connected to the treatment of the Glauber region 
in the integrations over extra gluons in Fig.~\ref{fig:leadingamplitude}, so we briefly review the basic issue here.  

Consider, for example, a virtual gluon $l^\mu$ attaching $\ublob$ and $H_L$ in Fig.~\ref{fig:leadingamplitude}.  
The transverse momentum may be of any size, though for addressing TMD-factorization 
we are especially interested in the contribution from the region where ${\bf l}_t \sim {\bf \Lambda}_t$.   The integral over the $l^+$ component is then 
trapped by spectator poles at sizes of order $\sim \Lambda^2 /Q$.  
To recover factorization, $l$ must either be strictly soft (all components of order $\Lambda$) or strictly collinear.
However, in the integration over $l^-$ there are generally leading regions where $|l^+ l^-| << |l_t^2|$.  This is the Glauber region, where
the Ward identities arguments 
that apply to ordinary soft or collinear gluons fail.  
To justify the steps of a factorization derivation, it must either be shown that it is possible to deform all contour integrals 
away from any Glauber regions so that they can be treated as properly soft or properly collinear, 
or the cross section must be sufficiently inclusive that any Glauber 
pole contributions cancel in the inclusive sum over final states. 
The potential for Glauber gluons to spoil factorization derivations has been noted long ago (e.g., Ref.~\cite{Bodwin:1981fv}).

\section{Constraints from TMD-Factorization}
\label{sec:partonmodel}

To ensure the clarity of later results, we must begin with a precise and detailed statement
of what is meant when we speak of the usual TMD-factorization-based assumption as applied to Eq.~\eqref{eq:theprocess}.  The purpose of this 
section is to make a clear statement of what, for this paper, we will call the``maximally general" criteria necessary for a version 
of TMD-factorization to be said to hold.  We will also summarize the constraints imposed by these assumptions and briefly discuss what 
it would mean for them to be violated.

\subsection{TMD-Factorization With Maximally General Criteria}
\label{sec:genfact}

Our initial statement of the TMD-factorization hypothesis must be very general
since the goal of later sections is to  demonstrate that TMD-factorization 
fails for Eq.~\eqref{eq:theprocess} even in a very loose sense. 
As reviewed in Sect.~\ref{sec:leadingregions}, a full derivation of TMD-factorization must account for the extra soft/collinear gluon connections 
of the type shown in Fig.~\ref{fig:leadingregions}.  Thus, our statement of the maximally generalized TMD-factorization hypothesis is 
that, for $N$ such extra gluons, in the sum over all graphs at a given order, the differential cross section 
in Eq.~\eqref{eq:crosssection} must be expressible in the form 
\begin{multline}
\label{eq:vgenfact}
d \sigma =  {\bf \mathbb{M}} \, \sum_{ij} \, \int \frac{d^{2} {\bf k}_{1t}}{(2 \pi)^2}\, \underbrace{\Phi_{g/P_2} \left( P_2;\hat{k}_2(x_2,{\bf q}_t - {\bf k}_{1t}) \right)^{ij}}_\text{$\sim Q^{0} $}  
\tr{ \begin{array}{c}  \! \\  \! \end{array} \right. \left. \underbrace{\bar{u}(k_4,S_4)}_\text{$\sim Q^{1/2} $}  
\; \underbrace{\hat{H}^{i, \, \lambda_3}_L \left( \hat{k}_1(x_1,{\bf k}_{1t}),\hat{k}_2(x_2,{\bf q}_t - {\bf k}_{1t}),k_3,k_4 \right)}_\text{$\sim 1/Q $}  \right. \times \; \\ \left. \times 
\underbrace{\Phi_{q/P_1} \left( P_1;\hat{k}_1(x_1,{\bf k}_{1t}) \right)}_\text{$\sim Q $}  \; \underbrace{\hat{H}^{j, \, \lambda_3}_{R} \left( \hat{k}_1(x_1,{\bf k}_1),\hat{k}_2(x_2,{\bf q}_t - {\bf k}_{1t}),k_3,k_4 \right)}_\text{$\sim 1/Q $}
 \; \underbrace{u(k_4,S_4)}_\text{$\sim Q^{1/2} $} \right. \left. \begin{array}{c}  \! \\  \! \end{array} } \\ 
+ \mathcal{O}((1/Q)^a) \, ,
\end{multline}
with $a > 0$ so that error terms are power suppressed.  

We will now discuss the meaning of each factor in Eq.~\eqref{eq:vgenfact}.  
The basic structure is reminiscent of a TMD-factorization parton model, and it is a basic 
starting point for many phenomenological approaches.
At this stage, the effective TMD PDFs, $\Phi_{g/P_2}$ and $\Phi_{q/P_1}$, need not necessarily correspond to 
gauge invariant operator definitions, within our maximally general criteria, 
and they may be highly process dependent.  
They may also carry any number of non-spacetime indices such as color (not shown explicitly).
The polarization projections on Lorentz and Dirac indices for partons entering the hard 
subgraphs, $\hat{H}^{i, \, \lambda_3}_L$ and $\hat{H}^{j, \, \lambda_3}_{R}$, are left as yet unspecified. Dirac indices will not be written explicitly.
The ${\bf \mathbb{M}}$ in front of the integration is intended to represent a possible collection of matrices in non-spacetime indices such as color, and it includes 
any overall numerical factors.  The power law of Eq.~\eqref{eq:vgenfact} 
matches that of Eq.~\eqref{eq:crosssection} for tree-level graphs
like Fig.~\ref{fig:partonmodel}.  With the horizontal braces beneath the equation, we have indicated the powers of $Q$ for each factor, and we will use 
this notation throughout the paper to aid the reader in following the counting of powers of $Q$.\footnote{The way powers of large momentum $\sim Q$ are
partitioned between the hard factors and the effective TMD PDFs differs from some conventions --- ours is chosen to 
simplify the power counting of later sections.}  
The way that the factors in Eq.~\eqref{eq:vgenfact} arise in a tree-level, parton model treatment will be reviewed in Sect.~\ref{sec:onegluon}.
   
The ordinary, bare-minimum kinematical parton model approximations needed to write down a factorized equation have been applied to initial state parton momenta 
in Eq.~\eqref{eq:vgenfact}.
That is, the minus component of $k_1$ is ignored outside its parent subgraph, $\Phi_{q/P_1}$, and is integrated over within 
the definition of $\Phi_{q/P_1}$ (whatever that exact definition might be).  Likewise, the plus component of $k_2$ is 
ignored outside $\Phi_{g/P_2}$ and is integrated over in the definition of
$\Phi_{g/P_2}$.
The use of approximate parton momenta outside $\Phi_{q/P_1}$ and $\Phi_{g/P_2}$ is symbolized by the hats on $\hat{k}_1$, $\hat{k}_2$ and $\hat{H}_{L/R}$.  
Thus, the factorization applies after a mapping from exact to approximate $k_1$, 
\begin{equation}
k_1 \to \hat{k}_1(x_1,{\bf k}_{1t}) = \left(\hat{k}_1^+ \left( k_1^+, {\bf k}_{1t} \right), \hat{k}_1^-\left( k_1^+, {\bf k}_{1t} \right) ,{\bf \hat{k}}_{1t} \left( k_1^+, {\bf k}_{1t} \right) \right) \, ,  \label{eq:hat1} \\
\end{equation}
is implemented 
outside $\Phi_{q/P_1}$, and a mapping from  exact to approximate $k_2$,
\begin{eqnarray}
k_2 \to \hat{k}_2(x_2,{\bf q}_t - {\bf k}_{t1})  = \left(\hat{k}_2^+\left( k_2^-, {\bf k}_{2t} \right), \hat{k}_2^-\left( k_2^-, {\bf k}_{2t} \right), {\bf \hat{k}}_{2t} \left( k_2^-, {\bf k}_{2t} \right) \right) \, , \label{eq:hat2}
\end{eqnarray}
is implemented 
outside $\Phi_{g/P_2}$. 
The components of the approximate hatted momentum variables, $\hat{k}_1^\mu$ and $\hat{k}_2^\mu$, are written as functions only of $k_1^+,{\bf k}_{1t}$ and $k_2^-,{\bf k}_{2t}$ respectively, 
as demanded by the minimal kinematical approximations.  
The only requirement needed in order for maximally generalized TMD factorization to be said to be valid is that the mapping from exact 
to approximate momenta be independent of $k_1^-$ and $k_2^+$, and that the approximate momenta obey 
the basic partonic four momentum conservation law, $\hat{k}_1 + \hat{k}_2 = k_1 + k_2 = q$.  
The precise definitions of the $\hat{k}_1$ and $\hat{k}_2$ mappings may also depend on physical external hadron momenta, though this is not shown explicitly.
The hard subgraphs in Eq.~\eqref{eq:vgenfact} have been replaced by their approximate versions, which use the hatted momentum variables:
\begin{equation}
H_{L/R}(k_1,k_2,k_3,k_4) \to \hat{H}_{L/R}\left( \hat{k}_1(x_1,{\bf k}_{1t}),\hat{k}_2(x_2,{\bf q}_t - {\bf k}_{1t}),k_3,k_4 \right) \, . \label{eq:hardsub}
\end{equation}

In a complete derivation of factorization, a basic step in the derivation would be to precisely formulate the 
exact-to-approximate mapping in Eqs.~(\ref{eq:hat1},~\ref{eq:hat2}) such that a useful 
version of factorization is maintained to arbitrary order in small $\alpha_s$ in the hard part.  
For our purposes, simply dropping the $k_1^-$ and and $k_2^+$ components outside 
their respective parent subgraphs would be a sufficient prescription, 
but we have left the exact transformation laws unspecified in Eqs.~(\ref{eq:hat1},~\ref{eq:hat2})  
to ensure that the later demonstration of TMD-factorization breaking is completely general.
Our only requirement for the mapping is that $k_1^-$ and $k_2^+$ are neglected outside $\Phi_{q / P_1}^{(1g)}$ and  $\Phi_{g / P_2}^{(1g)}$.

To summarize, we say that the maximally general TMD-factorization hypothesis is respected if, for $N$ extra soft/collinear gluons in Fig.~\ref{fig:leadingregions},
\begin{quote}
\begin{enumerate}[{\bf 1.)}] 
\item $\Phi_{q/P_1} \left( P_1;\hat{k}_1(x_1,{\bf k}_{1t}) \right)$ depends only on the momenta $P_1$ and $\hat{k}_1(x_1,{\bf k}_{1t})$ and has 
the power-law behavior in frame-1 of an ordinary quark TMD PDF.  It may be any matrix in non-spacetime indices, such as color, that contract with other factors in the 
factorization formula or with ${\bf \mathbb{M}}$.  The Dirac polarization components of $\Phi_{q/P_1}$ may depend only on the target hadron momentum $P_1$ and a target quark momentum $k_1$.
\item $\Phi_{g/P_2} \left( P_2;\hat{k}_2(x_2, {\bf k}_{2t}) \right)^{ij}$ depends only on the momenta $P_2$ and $\hat{k}_2(x_2,{\bf k}_{2t}= {\bf q}_t - {\bf k}_1)$ and 
has the power-law behavior in frame-1 of an ordinary gluon TMD PDF.  It may be any matrix in non-spacetime indices that contract with the other factors in the 
factorization formula or with ${\bf \mathbb{M}}$.  The frame-1 transverse Lorentz components, $ij$, may depend only on the
total transverse momentum, ${\bf k}_{2t}$, of the incident gluon from hadron-2. 
\item   $\hat{H}_{L/R}\left( \hat{k}_1(x_1,{\bf k}_{1t}),\hat{k}_2(x_2,{\bf k}_{2t}),k_3,k_4 \right)^{i j}$ depends only on the momenta $ \hat{k}_1(x_1,{\bf k}_{1t})$, 
$\hat{k}_2(x_2,{\bf k}_{2t})$, $k_3$, and $k_4$, and has the power-law behavior of the tree-level hard subgraph.  It may also be an arbitrary matrix in non-spacetime indices that contract with other factors in the 
factorization formula or with ${\bf \mathbb{M}}$.
\item   The only exception to the first two criteria is that $\Phi_{q/P_1} \left( P_1;\hat{k}_1(x_1,{\bf k}_{1t}) \right)$ may include dependence 
on an auxiliary unit vector approximately equal to $n_2$, to account for possible gauge link operators, 
and $\Phi_{g/P_2} \left( P_2;\hat{k}_2(x_2, {\bf k}_{2t}) \right)^{ij}$ may include dependence on an auxiliary unit vector approximately equal to $n_1$. 
\end{enumerate}
\end{quote}
To be consistent with the maximally generally TMD-factorization 
assumption, the effective TMD PDFs and the hard parts in Eq.~\eqref{eq:vgenfact} may have 
complicated process dependence and may break gauge invariance as long as they satisfy these general conditions.

Finally, note that the gluon TMD PDF in Eq.~\eqref{eq:vgenfact} only includes a sum over the transverse indices, $i$ and $j$, for the gluon polarization, and not a sum over 
longitudinal indices.  Within generalized TMD-factorization approaches, it is usually assumed from the outset that the basic polarization structure of the gluon subgraph is 
like that of an ordinary gluon number density, with only transverse components for a single on-shell gluon moving in the large minus direction contributing. 
Therefore, we include in the definition of maximally generalized TMD-factorization the assumption that effective gluon TMD PDFs always have two \emph{transverse} Lorentz indices and depend on 
only one small transverse gluon momentum.

Notice that the criteria enumerated above are less constraining
than in what was referred to as ``generalized TMD-factorization" in Ref.~\cite{Rogers:2010dm}.  There it was required at least that separate gauge invariant 
(albeit process dependent) TMD PDFs be identifiable for generalized TMD-factorization to be said to hold.  In this article, we require only a general separation 
into different transverse momentum dependent blocks in Eq.~\eqref{eq:vgenfact} with the properties listed above.

\subsection{Standard Classification Scheme}
\label{sec:azimuthal}

In this subsection, we review the conventional steps for extracting specific
azimuthal angular or spin dependence in a TMD-factorization formalism based on the set of 
very general criteria laid out in the last subsection.
The types of possible angular and spin dependence are attributed to special TMD PDFs that are defined with non-trivial polarization projections.
The standard classification strategy, therefore, is to first
enumerate the leading-power projections of Dirac structures in the effective unpolarized quark factor, $\Phi_{q/P_1}$, and the leading projections on 
transverse Lorentz components in the effective unpolarized gluon TMD PDF $\Phi_{g/P_2}$ in Eq.~\eqref{eq:vgenfact}. 
The result is a set of functions that are then identified with the various possible TMD PDFs.

For the leading power unpolarized gluon TMD PDF, the only dependence on transverse 
momentum allowed by condition 2.) of Sect.~\ref{sec:genfact} is from 
the total gluon transverse momentum ${\bf k}_{2t}$, and the only polarization 
dependence is from the transverse components $i$ and $j$.  Therefore, the leading power unpolarized 
gluon TMD PDF may be decomposed into the sum of a polarization-independent term and a polarization-dependent term:
\begin{multline}
\label{eq:gluondecomp}
 \left[ \Phi_{g/P_2} \left( P_2;\hat{k}_2(x_2,{\bf q}_t - {\bf k}_{1t}) \right) \right]^{i j}  = \\ 
 -g_t^{ij} \; \underbrace{\Phi_{g/P_2}^{ {\rm U}} \left( P_2;\hat{k}_2(x_2,{\bf q}_t - {\bf k}_{1t}) \right)}_\text{$\sim Q^{0}$}  + \left(\frac{k_{2t}^i k_{2t}^j}{M_{P_2}^2} 
 + \frac{g_t^{ij} k_{2t}^2}{2 M_{P_2}^2}  \right) \; \underbrace{\Phi_{g/P_2}^{{\rm BM}} \left( P_2;\hat{k}_2(x_2,{\bf q}_t - {\bf k}_{1t}) \right)}_\text{$\sim Q^{0} $}  \\
= {\rm P}_{{\rm U}}^{ij} \; \Phi_{g/P_2}^{{\rm U}} \left( P_2;\hat{k}_2(x_2,{\bf q}_t - {\bf k}_{1t}) \right) + {\rm P}_{{\rm BM}}({\bf k_{2t}})^{i j} \; \Phi_{g/P_2}^{{\rm BM}} \left( P_2;\hat{k}_2(x_2,{\bf q}_t - {\bf k}_{1t}) \right)   \, .
\end{multline}
Here, $\Phi_{g/P_2}^{{\rm U}}$ is the gluon TMD PDF for an unpolarized gluon and $\Phi_{g/P_2}^{{\rm BM}} $ is 
a gluon Boer-Mulders~\cite{Boer:1997nt,Mulders:2000sh} function.
In the last line, we have defined projections onto transverse indices:
\begin{align}
{\rm P}_{{\rm U}}^{ij} \equiv &  -g_t^{ij} \, , \\
{\rm P}_{{\rm BM}}({\bf k}_{2t})^{ij} \equiv & \left(\frac{k_{2t}^i k_{2t}^j}{M_{P_2}^2} 
 + \frac{g_t^{ij} k_{2t}^2}{2 M_{P_2}^2}  \right) \, .
\end{align}
An auxiliary gauge-link vector $n_1$ adds no other structure since it has no transverse components.

The unpolarized quark TMD PDF, $\Phi_{q/P_1}$, is decomposed into a complete basis of Dirac matrices, 
\begin{equation}
\label{eq:clifford}
\Phi_{q/P_1} \left( P_1;\hat{k}_1(x_1,{\bf k}_{1t}) \right) = S + \gamma_5 P + \gamma_\mu V^\mu + \gamma_\mu \gamma_5 A^\mu + \frac{1}{2} \sigma_{\mu \nu} T^{\mu \nu} \, \sim Q \, ,
\end{equation}
and we keep only the leading power components.
Since $\Phi_{q/P_1} \left( P_1;\hat{k}_1(x_1,{\bf k}_{1t}) \right)$ depends only on $P_1$ and $\hat{k}_1(x_1,{\bf k}_{1t})$, 
which in frame-1 have large (order $Q$) components in the plus direction,   
then the leading terms in Eq.~\eqref{eq:clifford} are those with a $\mu = +$ contravariant index.  
The TMD PDFs are therefore obtained from Dirac traces that project those large components.
For example, the largest component of the vector term, $\gamma_\mu V^\mu$ , is 
\begin{equation}
\label{eq:largeV}
\gamma^- V^+ \sim \, Q \, .
\end{equation}
The ordinary azimuthally symmetric and unpolarized TMD PDF is therefore identified with the projection
\begin{equation}
\label{eq:vplus}
\Phi^{\left[ \gamma^+ \right]}_{q/P_1}\left( P_1;\hat{k}_1(x_1,{\bf k}_1) \right) \equiv V^+ 
= \frac{1}{4} \tr{\gamma^+ \Phi_{q/P_1}\left( P_1;\hat{k}_1(x_1,{\bf k}_1) \right) } \sim Q \, . 
\end{equation}
In this expression, we have adopted the notation of Ref.~\cite{Mulders:1995dh} by including 
a $\left[ \gamma^+ \right]$ to indicate the specific Dirac projection.

The only dependence on external momenta is through $P_1$ and $k_1$, so the largest Dirac components 
of $\Phi^{\left[ \gamma^+ \right]}_{q/P_1}$ are proportional 
to $\slashed{P} \sim \slashed{k}_1 \sim Q$.
The trace in Eq.~\eqref{eq:vplus} then takes the general form,
\begin{equation}
\label{eq:largetracesunpol}
\Phi^{\left[ \gamma^+ \right]}_{q/P_1}\left( P_1;\hat{k}_1(x_1,{\bf k}_1) \right) \propto  \frac{1}{4} \tr{\gamma^+ \slashed{P}_1} \sim Q \, .
\end{equation}
An auxiliary gauge-link vector $\slashed{n}_2$ introduces no further leading-power projections because it only
leads to additional factors of $\gamma^+$ in Eq.~\eqref{eq:largetracesunpol}.

For a projection with a generic Dirac structure $\Gamma_q$, the corresponding quark TMD PDF is
\begin{equation}
\label{eq:genproj}
\Phi^{ \left[ \Gamma_q \right]}_{q/P_1}\left( P_1;\hat{k}_1(x_1,{\bf k}_1) \right) 
= \frac{1}{4} \tr{ \Gamma_q \, \Phi_{q/P_1}\left( P_1;\hat{k}_1(x_1,{\bf k}_1) \right) } \sim Q \, .
\end{equation}
The decomposition in Eq.~\eqref{eq:clifford} spans the set of possible independent Dirac structures for a quark TMD PDF, and each $\Gamma_q$
is associated with a characteristic type of angular or spin dependence in the cross section.
In the TMD-factorization-based classification scheme, 
the types of leading-power quark TMD PDFs are 
categorized according to the possible Dirac projections $\Gamma_q$ that contribute at leading power.  
For example, the projection,
\begin{equation}
\Gamma_q \to \gamma^+   \label{eq:gammaplus}
\end{equation} 
reproduces the ordinary azimuthally symmetric and unpolarized 
TMD PDF in Eq.~\eqref{eq:vplus}.
Another well-known example is the quark Boer-Mulders~\cite{Boer:1997nt} function, obtained from the projection,
\begin{equation}
\Gamma_q \to \gamma^+ \gamma^j \gamma_5  \, . \label{eq:bm}
\end{equation} 
If hadrons 1 and 2 are also polarized then there is a large collection of additional leading-power TMD PDF structures involving hadron spins.
These have been thoroughly classified at least to leading twist~\cite{Mulders:1995dh,Tangerman:1994eh}.
 
So the maximally general statement of the TMD-factorization conjecture of 
Sect.~\ref{sec:genfact}, when specialized to a particular set of projections, $\Gamma_g$ and $\Gamma_q$, is
\begin{multline}
\label{eq:genproj2}
d \sigma^{\left[\Gamma_g,\Gamma_q \right]} =  
 {\bf \mathbb{M}} \, \sum_{i j} \int \frac{d^{2} {\bf k}_{1t}}{(2 \pi)^2}\,  \underbrace{\Phi^{\left[ \Gamma_g \right]}_{g/P_2} \left( P_2;\hat{k}_2(x_2,{\bf q}_t - {\bf k}_{1t}) \right)}_\text{$\sim Q^0 $} \, 
  \underbrace{\Phi^{\left[ \Gamma_q \right]}_{q/P_1}\left( P_1;\hat{k}_1(x_1,{\bf k}_{1t}) \right)}_\text{$\sim Q $} \,  
 \\ 
 \times {\rm P}_{\Gamma_g}({\bf q}_t - {\bf k}_{1t})^{ij} \; \tr{ \begin{array}{c}  \! \\  \! \end{array} \right. \left. \underbrace{\bar{u}(k_4,S_4)}_\text{$\sim Q^{1/2} $} 
 \; \underbrace{H^{i, \, \lambda_3}_L \left( \hat{k}_1(x_1,{\bf k}_{1t}),\hat{k}_2(x_2,{\bf q}_t - {\bf k}_{1t}),k_3,k_4 \right)}_\text{$\sim 1/Q $} \;  \times \right. \\ \left. \times \; {\rm P}_{\Gamma_q}  \;  
 \underbrace{H^{j, \, \lambda_3}_{R} \left( \hat{k}_{1}(x_1,{\bf k}_{1t}),\hat{k}_2(x_2,{\bf q}_t - {\bf k}_{1t}),k_3,k_4 \right)}_\text{$\sim 1/Q $}  \;  \underbrace{u(k_4,S_4)}_\text{$\sim Q^{1/2} $} \right. \left. \begin{array}{c}  \! \\  \! \end{array} } \, .
\end{multline}
This is obtained by using Eqs.~(\ref{eq:gluondecomp},\ref{eq:clifford},\ref{eq:genproj}) in Eq.~\eqref{eq:vgenfact}.
The ${\rm P}_{\Gamma_q}$ on the last line of Eq.~\eqref{eq:genproj2} is the relevant Dirac matrix 
from Eq.~\eqref{eq:clifford} corresponding to projection $\Gamma_q$.  (In the unpolarized azimuthally symmetric case, ${\rm P}_{\Gamma_q} \to \gamma^-$.)
The $\Gamma_g = \{ {\rm U}, {\rm BM} \}$ labels which gluon polarization projection is taken from Eq.~\eqref{eq:gluondecomp}.
The superscript $\left[ \Gamma_q, \Gamma_g \right]$ on $d \sigma$ is to indicate that Eq.~\eqref{eq:genproj2} is the contribution to the cross section
from that particular combination of $\Gamma_q$ and $\Gamma_g$ target quark and gluon polarization projections.  

Throughout this section, we have used the language of a generalized parton model by
referring to $\Phi^{\left[ \Gamma_q \right]}_{q/P_1}$ and $\Phi^{ \left[ \Gamma_g \right]}_{g/P_2}$ as though they were ordinary TMD PDFs.  
However, in general pQCD
diagrams like Fig.~\ref{fig:leadingregions}, the $\Phi$'s are merely labels for the target subgraphs corresponding to hadrons 1 and 2, each of which may be 
linked to the rest of the process via arbitrarily many soft and collinear gluon exchanges.  Strictly speaking, TMD PDFs acquire a well-defined 
meaning in pQCD only after a factorization derivation has shown
how those extra soft and collinear gluons separate into independently defined factors in the sum over all graphs.

Because of its very general form, there is a strong temptation
to apply the factorized classification scheme of this subsection very broadly, even in
processes like Eq.~\eqref{eq:theprocess} where
generalized TMD-factorization formally fails~\cite{Rogers:2010dm}.
But even the maximally general version of the TMD-factorization conjecture from this section
imposes significant constraints on the  
possible general behavior of the cross section at leading power.
For example, the power counting logic used to identify the leading TMD functions in Eq.~\eqref{eq:gluondecomp} and  Eq.~\eqref{eq:clifford} 
relies crucially on the assumptions that, a.) $[\Phi_{g/P_2}]^{ij}$ depends at leading power only on 
transverse Lorentz indices and the momenta $P_2$, $k_2^-$ and ${\bf k}_{2t}$ and, b.)
$\Phi_{q/P_1}$ depends only on momenta $P_1$, $k_1^+$ and ${\bf k}_{1t}$.
When generalized TMD-factorization is broken, we must in principle account for 
the possibility that dependence on extra external momenta 
leaks into the $\Phi_{g/P_2}$ and $\Phi_{q/P_1}$ from other subgraphs via the extra soft and collinear gluon exchanges in Fig.~\ref{fig:leadingregions}.
However, in confronting this possibility we are no longer justified in assuming from the outset that, for example, the $\mu = +$ components in Eq.~\eqref{eq:clifford} are the only dominant ones.
Any Dirac structure projected from Eq.~\eqref{eq:clifford}
must be viewed as a potential candidate leading-power effect.  Analogously, the decomposition of $[\Phi_{g/P_2}]^{ij}$ may 
include terms beyond those of Eq.~\eqref{eq:gluondecomp}, involving polarizations induced by internal intrinsic transverse momenta other than ${\bf k}_{2t}$, 
and polarization configurations like Eq.~\eqref{eq:superleading} must also be accounted for.  We will discuss the implications of this in more detail in the next subsection.

For the sake of clarity, our discussion has so far focused on the TMD PDFs, though the same observations apply to a treatment  
of final state hadronization, i.e. to the treatment of a quark fragmentation or jet function.
The use of TMD-factorization, in the very general form  
summarized in this section, has become standard for classifying leading-power TMD effects.
It is a valid method, of course, in processes where a form of TMD-factorization holds  or is expected to hold (such as in SIDIS or Drell-Yan). 
For TMD-factorization breaking processes, however, we will find that the structure in Eq.~\eqref{eq:genproj2} turns out to be overly restrictive.

\subsection{Parity, TMD PDFs and TMD-Factorization Breaking}
\label{eq:symmetries}

A TMD-factorization conjecture imposes even stronger constraints on the qualitative structure of 
cross section when considered 
in combination with discrete symmetries of QCD.  Discrete symmetry arguments
have long been useful for constraining the properties of parton correlation functions in pQCD.  It is now known, however, that 
such arguments are more subtle in processes that involve TMD-factorization than in similar collinear cases, due to the non-trivial 
role of gauge links in TMD-factorization.
Early on, for example, $TP$ invariance was applied in Ref.~\cite{Collins:1992kk} 
to argue that the Sivers function should vanish in hard processes.  That derivation, however, neglected effects from gauge links in the definitions of 
TMD PDFs.  As the role of gauge links in the derivation of TMD-factorization became better understood, it was realized that 
$TP$ invariance implies not that the Sivers function vanishes, but rather that it reverses sign between the Drell-Yan process and SIDIS~\cite{Brodsky:2002rv,Brodsky:2002cx,Collins:2002kn}, thus 
providing a non-trivial prediction for a specific type of non-universality.

Such observations provide motivation to also examine how arguments based on discrete symmetries are affected when TMD-factorization breaks down altogether.
As an example, let us return to the quark TMD PDF's Dirac decomposition in Eq.~\eqref{eq:clifford}.  In addition to Eq.~\eqref{eq:gammaplus}, 
naive power counting would imply another leading-power Dirac structure formed by the axial vector term
\begin{equation}
\gamma_5  \gamma^-  A^+\, \label{eq:npvproj}  \; \sim Q\, .
\end{equation}
So one should also include in the categorization of leading projections, 
\begin{equation}
\label{eq:longproj}
\Gamma_q \to \gamma_5 \gamma^+ \, .
\end{equation}
Therefore, by analogy with Eq.~\eqref{eq:vplus}, a TMD function defined as
\begin{equation}
\label{eq:long}
\Phi^{\left[ \gamma_5 \gamma^+ \right]}_{q/P_1}\left( P_1;\hat{k}_1(x_1,{\bf k}_1) \right) 
\equiv \frac{1}{4} \tr{\gamma_5 \gamma^+ \Phi_{q/P_1}\left( P_1;\hat{k}_1(x_1,{\bf k}_1) \right) } \; \sim Q \, 
\end{equation}
should be included within the classification of leading-power quark TMD PDFs.
The interpretation of this TMD PDF in a generalized parton model/TMD-factorization approach
would be that it describes a distribution of quarks with an intrinsic helicity polarization
inside an unpolarized hadron.  
Such TMD PDFs are forbidden by parity invariance for an unpolarized parent hadron
for any TMD PDF definitions that satisfy 
the basic requirements of the generalized TMD-factorization hypothesis enumerated in Sect.~\ref{sec:genfact}. 
In Feynman graph calculations, this appears in the inability to construct a pseudo-scalar using only the four-momenta from hadron-1. 
At least four different four-momentum vectors are needed to give a non-vanishing trace.
Specifically, an evaluation of Eq.~\eqref{eq:long} always leads to a Dirac trace of the form,
\begin{equation}
\label{eq:long2}
\Phi^{\left[ \gamma_5 \gamma^+ \right]}_{q/P_1}\left( P_1;\hat{k}_1(x_1,{\bf k}_1) \right)  \propto \frac{1}{4} 
\tr{ \gamma_5 \gamma^+ \slashed{\hat{k}}_{1t} \slashed{P}_1 } = 0\, .
\end{equation}
There are not enough four momentum vectors involved to give a non-vanishing trace.
A gauge link simply introduces more $\sim n_2$ vectors, which only project extra factors of $\gamma^+$ in the Dirac trace:
\begin{equation}
\label{eq:long3}
\Phi^{\left[ \gamma_5 \gamma^+ \right]}_{q/P_1}\left( P_1;\hat{k}_1(x_1,{\bf k}_1) \right)  \propto \frac{1}{4} 
\tr{ \gamma_5 \gamma^+ \slashed{\hat{k}}_{1t} \slashed{P}_1 \slashed{n}_2 } = \frac{1}{4} \tr{ \gamma_5 \gamma^+ \slashed{\hat{k}}_{1t} \slashed{P}_1 \gamma^+ } = 0\, .
\end{equation}
So even including gauge links does not allow for projections of this type within a TMD-factorization conjecture. 
TMD PDFs like Eq.~\eqref{eq:long2}, and any effects that would follow from them,   
are constrained to vanish at leading power within the minimal TMD-factorization criteria of Sect.~\ref{sec:genfact} and~\ref{sec:azimuthal}.  
For this reason, they might be referred to as ``naively parity violating."   

To further investigate the consequences of discrete symmetries in a generalized TMD-factorization 
framework, we will focus in later sections on the final state parton polarizations --- the helicity $\lambda_3$ of the prompt photon and the 
the transverse spin $S_{4, \perp}$ of the outgoing quark in Fig.~\ref{fig:processdiagram}.  If the initial state partons have no helicity then 
the hard subprocess
\begin{equation}
\label{eq:hardproc}
q + g \to q^{(S_{4, \perp})} + \gamma^{(\lambda_3)} 
\end{equation}
violates the helicity conservation of massless QCD.   
For the $S_{4, \perp}-\lambda_3$ correlation to be possible, 
a helicity must be carried by the $q$ or $g$ inside its parent unpolarized hadron before the hard collision.
In a generalized TMD factorization framework, this would imply a type of naively parity violating TMD PDF.

After showing that the criteria of Sect.~\ref{sec:genfact} fail to hold in a more detailed 
treatment of the extra soft/collinear gluons of Fig.~\ref{fig:leadingregions},
we will ultimately demonstrate in Sect.~\ref{sec:extra} that final state $S_{4, \perp}$-$\lambda_3$ correlations
are actually leading power in spite of the apparent helicity non-conservation in the naive $2 \to 2$ partonic description in Eq.~\eqref{eq:hardproc}. 

If one abandons the maximally general TMD-factorization criteria of Sect.~\ref{sec:genfact}, and allows 
the effective TMD PDFs to acquire dependence on additional extra external momenta, then
any Dirac structure projected from Eq.~\eqref{eq:clifford} is a potential candidate 
for a leading-power quark TMD effect, rather than only those with a ``$+$'' Lorentz component.

For example, the scalar term, $S$, in Eq.~\eqref{eq:clifford} is at most of order a hadronic mass within 
the TMD-factorization framework of Sect.~\ref{sec:genfact}, and so would be counted as subleading.
But if, because of the extra gluon connections in Fig.~\ref{fig:leadingregions}, $\Phi_{q/P_1}\left( P_1;\hat{k}_1(x_1,{\bf k}_1) \right)$ were allowed to depend on other external 
momenta,
\begin{equation}
\Phi_{q/P_1}\left( P_1;\hat{k}_1(x_1,{\bf k}_1) \right) \longrightarrow \Phi_{q/P_1}\left( P_1;\hat{k}_1(x_1,{\bf k}_1),q,P_2,k_2\dots \right),
\end{equation}
then $S$ could acquire dependence on large scales like $P_1 \cdot P_2$ and $q \cdot k_{1t}$.  Then the usual power 
counting arguments for $\Phi_{q/P_1}$ become invalid.

The pseudo-scalar term, $\gamma_5 P$, in Eq.~\eqref{eq:clifford} is another example that is both subleading and forbidden by parity invariance
in the TMD-factorization framework.  It will be the main example that we will use in 
Sect.~\ref{sec:extra} to demonstrate a leading-power TMD-factorization breaking final state spin correlation.
The projection is simply
\begin{equation}
\label{eq:longprojj}
\Gamma_q \to \gamma_5
\end{equation}
and the substitution in Eq.~\eqref{eq:genproj2} is 
\begin{equation}
\label{eq:longprojjj}
{\rm P}_{\Gamma_q} \stackrel{\Gamma_q \to \gamma_5}{=} \gamma_5  \, .
\end{equation}
If one works within the TMD-factorization conjecture by following the classification scheme of Sect.~\ref{sec:azimuthal}, then one would 
define a TMD PDF corresponding to Eq.~\eqref{eq:genproj} using the projection in Eq.~\eqref{eq:longprojj}:
\begin{equation}
\Phi^{\left[ \gamma_5 \right]}_{q/P_1}\left( P_1;\hat{k}_1(x_1,{\bf k}_1) \right) 
\equiv \frac{1}{4} \tr{\gamma_5 \Phi_{q/P_1}\left( P_1;\hat{k}_1(x_1,{\bf k}_1) \right) } \propto m_q \tr{\gamma_5} = 0 \, .
\end{equation}
So, the possibility of any spin correlation effects arising from such projections would be discounted in the 
standard TMD-factorization classification scheme.

We will return again to the projection in Eq.~\eqref{eq:longprojj} in Sect.~\ref{sec:extra} where we will confront TMD-factorization 
breaking effects.

\subsection{Maximally General TMD-factorization at the Amplitude Level}
\label{sec:maxgen}

The conditions enumerated in Sect.~\ref{sec:genfact} are sufficiently weak that it becomes straightforward to separate, at 
the amplitude level, contributions that are definitely consistent with the maximally 
general TMD-factorization criteria from those which may violate 
TMD-factorization if left uncanceled.  This will be useful for later sections where much of the analysis is performed at 
the amplitude level.

Assume that for a fixed number $m$ and $n$ of extra gluons in Eq.~\eqref{eq:ampbasic} one is able to sum a set of Feynman diagrams to obtain a 
contribution to the amplitude of the form
\begin{equation}
\label{eq:vgfactamp}
M = \mathcal{N} \; \sum_i \underbrace{\ublob_{eff}(P_2;k_2,\left\{l_2,\dots,l_m\right\})^i}_\text{$\sim Q^0$} \; \underbrace{\bar{u}(k_4,S_4)}_\text{$\sim Q^{1/2}$} \; 
\underbrace{H_L(k_1,k_3,k_4,k_2)^{i, \, \lambda_3}}_\text{$\sim 1/Q$} \; \underbrace{\lblob_{eff}(P_1;k_1,\left\{l_1^\prime, \dots, l_n^\prime\right\})}_\text{$\sim Q^{1/2}$} \, . 
\end{equation}
Here we have changed variables from $l_1$ to $k_2 \equiv \sum_j^m l_j$.  The factor $\mathcal{N}$ can represent any matrix of non-spacetime indices such as color.   The hard subgraph $H_L$ depends, 
as in Sect.~\ref{sec:genfact},  only on the momenta $k_1$, $k_2$, $k_3$ 
and $k_4$.  Assume further that the effective $\ublob_{eff}$ and $\lblob_{eff}$ factors take the forms,
\begin{equation}
\label{eq:ueffposs}
\ublob_{eff}(P_2;k_2,\left\{l_2,\dots,l_m\right\})^i = \mathcal{F}_{\ublob}(P_2;k_2,\left\{l_2,\dots,l_m\right\})^{\mu_1^\prime \cdots \mu_m^\prime} \; \mathcal{P}_{\ublob}({\bf k}_{2t})^i_{\mu_1^\prime \cdots \mu_m^\prime} \, ,
\end{equation}
and 
\begin{equation}
\label{eq:leffposs}
\lblob_{eff}(P_2;k_2,\left\{l_1^\prime,\dots,l_n^\prime\right\}) = \mathcal{F}_{\lblob}(P_1;k_1,\left\{l_1^\prime,\dots,l_n^\prime \right\})^{\rho_p^\prime \cdots \rho_n^\prime} \; \mathcal{P}_{\lblob}(P_1,{\bf k}_{1t})_{\rho_p^\prime \cdots \rho_n^\prime} \, .
\end{equation}
By substituting Eq.~\eqref{eq:vgfactamp} into Eq,~\eqref{eq:crosssection} and applying the minimal 
kinematical approximations of Eqs.~(\ref{eq:hat1}-\ref{eq:hardsub}), one immediately recovers the structure in Eq.~\eqref{eq:vgenfact}.

Note that $\mathcal{F}_{\ublob}$ and $\mathcal{F}_{\lblob}$ do not share any $l$-momentum arguments.
Since there is no direct dependence on $\left\{l_2, \dots \, l_m \right\}$ outside $\ublob_{eff}$, the part of Eq.~\eqref{eq:ueffposs} labeled by $\mathcal{F}_{\ublob}$, which
does not depend on transverse polarization components (labeled ``$i$"), may have dependence on 
any of these extra gluon momenta and still maintain consistency with the 
maximally general TMD-factorization criteria of Sect.~\ref{sec:genfact}.  In the squared amplitude, these momenta are simply integrated over 
inside $\Phi_{g/P_2}$.   However, the polarization-dependent part, labeled by 
$\mathcal{P}_{\ublob}({\bf k}_{2t})^i$, couples to the transverse ``$i$" component of the hard part in Eq.~\eqref{eq:vgfactamp}, 
so it is allowed to depend \emph{only} on the total transverse momentum, ${\bf k}_{2t}$, coming out of the gluon subgraph.
Without such a requirement, the criteria from Sect.~\ref{sec:genfact} can be violated, and the decomposition of the effective gluon TMD PDF into polarization dependent functions in terms 
of ${\bf k}_{2t}$, as in Eq.~\eqref{eq:gluondecomp}, would in general be incomplete.  The integrations over ${\bf l}_{2t} \cdots $ could not be performed independently 
of the contractions in the hard part.

Similarly, there is no direct dependence on $\{l_1^\prime, \dots \, l_n^\prime \}$ outside $\lblob_{eff}$, so the non-polarization dependent part of Eq.~\eqref{eq:leffposs}, labeled by $\mathcal{F}_{\lblob}$, may depend on 
any of these extra gluon momenta.  In the squared amplitude, they will be integrated over inside the definition of $\Phi_{q/P_1}$.  However, the projection onto Dirac components, 
labeled by $\mathcal{P}_{\lblob}(P_1,{\bf k}_{1t})$, can only depend on $P_1$ and the total momentum $k_1$ of the quark coming out of hadron-1.  

A potential violation of the criteria of Sect.~\ref{sec:genfact} would arise, for example, from a contribution 
in which one is forced to write Eq.~\eqref{eq:ueffposs} in the form,
\begin{equation}
\label{eq:ueffpossviol}
\ublob_{eff}(P_2;k_2,\left\{l_2,\dots,l_m\right\})^i = \mathcal{F}_{\ublob}(P_2;k_2,\left\{l_2,\dots,l_m\right\})^{\mu_1^\prime \cdots \mu_m^\prime} \; \mathcal{P}_{\ublob}({\bf l}_{2t},{\bf k}_{2t},\dots)^i_{\mu_1^\prime \cdots \mu_m^\prime} \, ,
\end{equation} 
i.e., where the polarization projection $\mathcal{P}_{\ublob}$ depends separately on the transverse components of individual
extra gluon transverse momentum, rather than only on their sum.  Recall that it is only ${\bf k}_{2t}$ that enters the hard scattering.
In Eq.~\eqref{eq:ueffpossviol} the transverse gluon momentum ${\bf k}_{2t}$ that enters the hard subgraph may differ from the 
transverse gluon momentum ${\bf l}_{2t}$ that induces a polarization dependence.
If such terms fail to cancel, then they contribute to a breakdown of TMD-factorization, even under the maximally general criteria of Sect.~\ref{sec:genfact}.  

\section{Analysis of One Gluon}
\label{sec:onegluon}
Before addressing the case of multiple gluon interactions, we will use this section to give a detailed demonstration of 
how Ward identity arguments apply to the simplest case of just one gluon attaching hadron-2 to the 
rest of the graph (Fig.~\ref{fig:onegluon}).
In the absence of extra gluons, there are few enough complications that the problems with 
TMD-factorization will not appear  --- extra soft and collinear gluons of the type normally associated with gauge links 
are needed for the violation of TMD-factorization to be apparent.   Therefore, the results of this section will be found to 
be consistent with the conditions for maximally general TMD-factorization from Sect.~\ref{sec:partonmodel}.  
Nevertheless, the steps will help establish the basic framework needed for dealing more 
generally with factorization issues in Sect.~\ref{sec:twogluons}.  For simplicity, in this section we will restrict 
consideration to the case where the final states in Fig.~\ref{fig:processdiagram} are totally unpolarized.

\subsection{Separation into Subgraphs}
\label{sec:onegluesep}
\begin{figure*}
\centering
  \begin{tabular}{c@{\hspace*{25mm}}c}
    \includegraphics[scale=0.2]{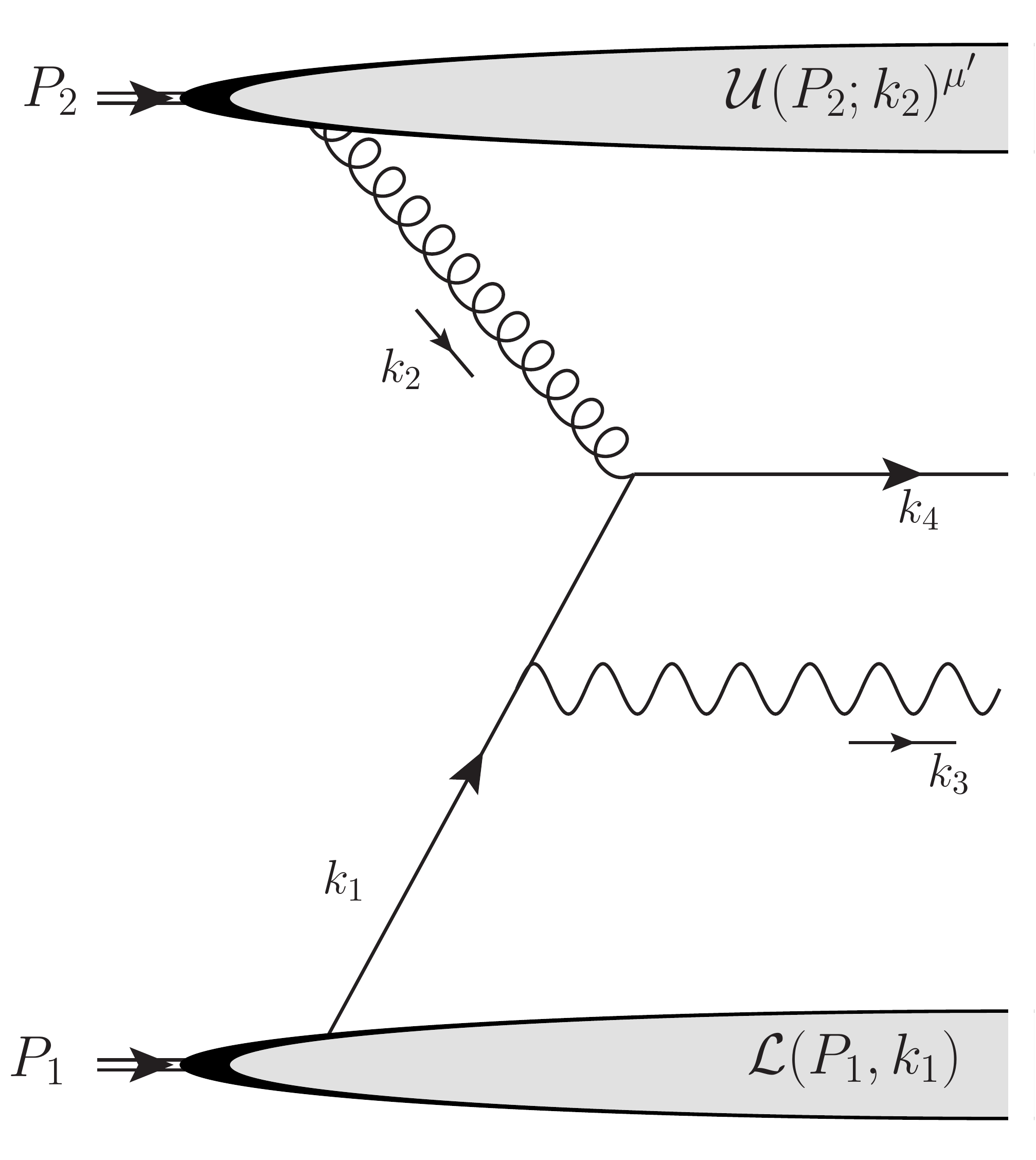}
    &
    \includegraphics[scale=0.2]{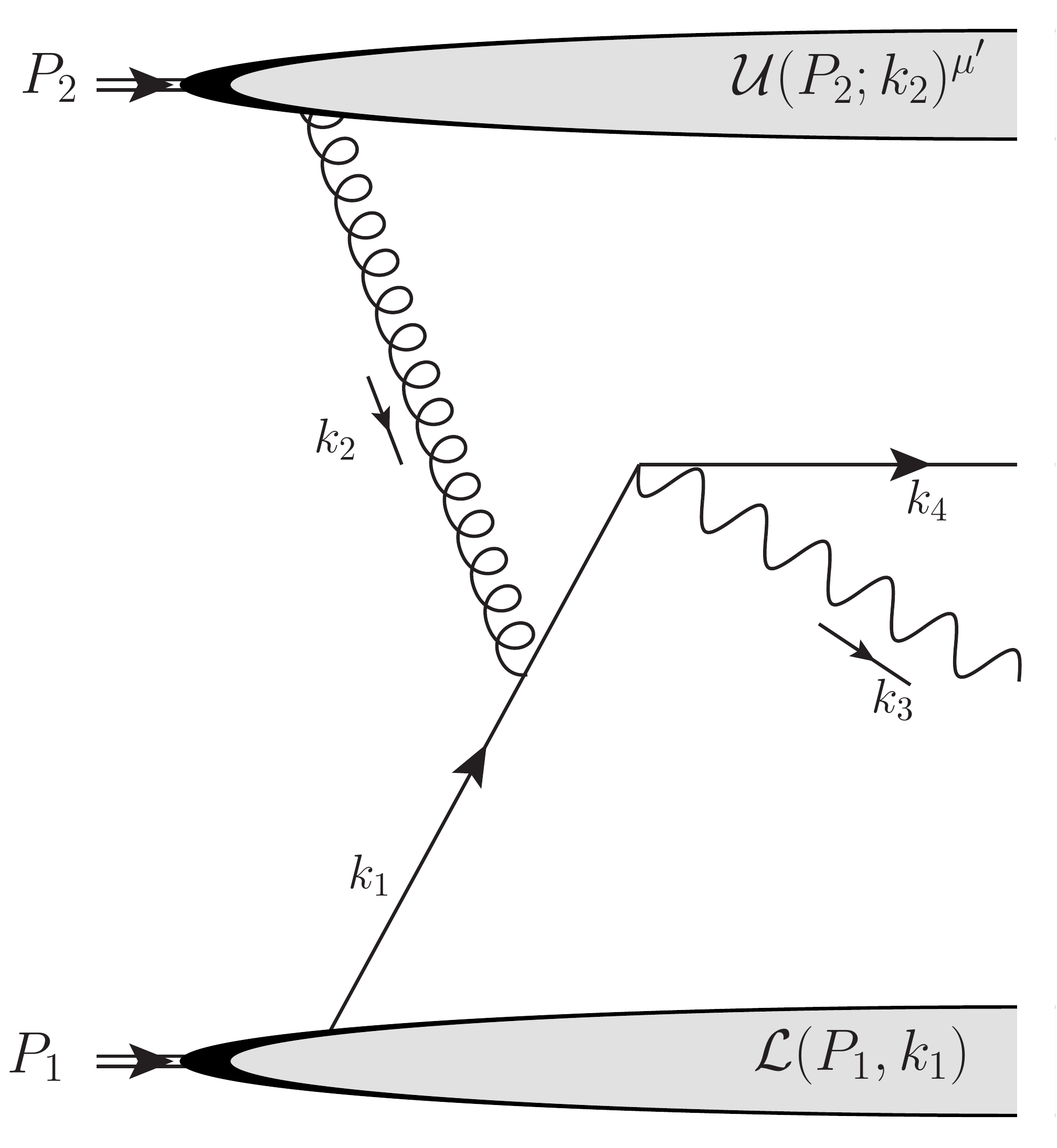}
    \\
    (a) & (b)
    \\[5mm]
   \end{tabular}
\caption{The single gluon parton-model-level contributions to Fig.~\ref{fig:leadingamplitude}.}
\label{fig:onegluon}
\end{figure*}
We begin by applying the analysis of Sect.~\ref{sec:superleading} to the special case of Fig.~\ref{fig:onegluon}.
The separation into subgraphs according to Eq.~\eqref{eq:ampbasic} is
\begin{equation}
\label{eq:m1gluon}
M_{(1g)} = \ublob(P_2;k_2)^{\kappa ; \mu^\prime} \, g_{\mu^\prime \mu} \,  \bar{u}(k_4) \, H_L(k_1,k_2,k_3,k_4)^{\kappa ; \mu \sigma} \,  \lblob(P_1;k_1) \, .
\end{equation}
The Lorentz index for the gluon is $\mu$, that for the photon is $\sigma$, and $\kappa$ is the color-octet index for the gluon.  Repeated color indices are summed over.
We will drop the explicit $\sigma$ index on $M_{(1g)}$.

In frame-1, the powers of $Q$ for the components of $\ublob^{\mu^\prime}$ are
\begin{align}
\label{eq:1gluonublob}
\ublob^- \sim Q, \qquad \ublob^i \sim Q^0, \qquad \ublob^+ \sim 1/Q \,.
\end{align}
As usual, superscripts ``$i,j$" denote transverse Lorentz components in frame-1.
The final state momenta, $k_3$ and $k_4$, are at wide angles so all components of $H_L^{\mu}$ are of comparable size:
\begin{equation}
\label{eq:1gluonh}
H_L^- \sim H_L^+ \sim H_L^i \sim 1/ Q \, .
\end{equation}
For this section, the Lorentz index on $H_L$ and $\ublob$ always 
refers to the gluon unless otherwise specified.

The final state quark wave function behaves as $\bar{u}(k_4) \sim Q^{1/2}$.
The dominant contribution from $\lblob$ comes from leading-power projections onto the Dirac components of 
$\Phi_{q/P_1}$, as in Eq.~\eqref{eq:vplus}, which are at most of order $Q$.
So the dominant component of $\lblob$ also has a power-law behavior $\lblob \sim Q^{1/2}$.

Since we are mainly interested the contraction of $\ublob$ with the rest of the graph, let us introduce the notation,
\begin{equation}
\label{eq:ronegluon}
R^{\kappa ; \mu \sigma} \equiv  \underbrace{\bar{u}(k_4)}_\text{$\sim Q^{1/2} $} \, \underbrace{H_L(k_1,k_2,k_3,k_4)^{\kappa ; \mu \sigma}}_\text{$\sim 1/Q $} \, \underbrace{\lblob(P_1;k_1)}_\text{$\sim Q^{1/2} $} \, ,
\end{equation}
so that
\begin{equation}
\label{eq:1gluonR}
M_{(1g)} = \ublob(P_2;k_2)^{\kappa ; \mu^\prime} g_{\mu^\prime \mu} R^{\kappa ; \mu \sigma} \,. 
\end{equation}
Then all the components of $R^{\kappa ; \mu \sigma}$ have the power-law behavior, 
\begin{equation}
\label{eq:r1gluon}
R^{\kappa ; - \sigma} \sim R^{\kappa ; + \sigma} \sim R^{\kappa ; i \sigma} \sim Q^0 \, ,
\end{equation}
and the largest power-law in the full amplitude from the tree-level graphs of Fig.~\ref{fig:onegluon} is from the contraction
\begin{equation}
\label{eq:largest}
M_{(1g)}^{\rm max} \sim \ublob^{\kappa ; -} R^{\kappa ; + \sigma}  \sim Q  \, .
\end{equation}
This is a contribution from the $K$-gluon term in the Grammer-Yennie decomposition.  It is a
superleading contribution of the type discussed in Sect.~\ref{sec:superleading}.

We next implement the Grammer-Yennie separation in Eqs.~(\ref{eq:gramyen}-\ref{eq:Gdef}) in Eq.~\eqref{eq:1gluonR} by writing
\begin{equation}
M_{(1g)} = M_{(1g)}^G + M_{(1g)}^K
\end{equation}
with
\begin{align}
M_{(1g)}^G &= \ublob(P_2;k_2)^{\kappa ; \mu^\prime}  G(k_2)_{\mu^\prime \mu} R^{\kappa ; \mu \sigma} \qquad \text{G-term} \, , \label{eq:m1gG} \\
M_{(1g)}^K  &= \ublob(P_2;k_2)^{\kappa ; \mu^\prime} K(k_2)_{\mu^\prime \mu}  R^{\kappa ; \mu \sigma} \qquad \text{K-term} \, ,    \label{eq:m1gK}
\end{align} 
and,
\begin{eqnarray}
K(k_2)^{\mu^\prime \mu} & \equiv & \frac{n_1^{\mu^\prime} k_2^\mu}{n_1 \cdot k_2}  \label{eq:Kdefk2} \\
G(k_2)^{\mu^\prime \mu} & \equiv & g^{\mu^\prime \mu} -  \frac{n_1^{\mu^\prime} k_2^\mu}{n_1 \cdot k_2} \label{eq:Gdefk2}  \, .
\end{eqnarray}
The $K$-term gives superleading as well as subleading contributions in the amplitude.  
The $G$-term, however, has been deliberately constructed so that its
``$+-$" component is removed.  The highest power-law contributed by the $G$-term when contracted with $\ublob$ is therefore
\begin{equation}
\underbrace{ \ublob^{\kappa ; i}}_\text{$\sim Q^{0} $} \,  \underbrace{R^{\kappa ; i \sigma}}_\text{$\sim Q^{0} $} \;  \sim \; \underbrace{\ublob^{\kappa ; -}}_\text{$\sim Q $} \, 
\underbrace{R^{\kappa ; i \sigma} (k_{2 t}^i/k_2^-)}_\text{$\sim 1/Q$} \;  \sim \; Q^0 \, . \label{eq:leadingonegluon}
\end{equation}
These are the leading-power $G$-term contributions to the amplitude. 
The remaining $G$-term contributions, in the 
contraction of $\ublob$ with $R$, are power suppressed relative to the leading terms:
\begin{equation}
\label{eq:1gluonsupp}
\ublob^{\kappa ; +} R^{\kappa ; - \sigma}  \sim 1 / Q \, ,
\end{equation}
in accordance with the power-laws in Eqs.~\eqref{eq:1gluonublob} and~\eqref{eq:r1gluon}.
The explicit expression for $R^{\kappa ; \mu \sigma}$, from the two graphs in Fig.~\ref{fig:onegluon}, is 
\begin{equation}
R^{\kappa ; \mu \sigma} = 
i g_s e_q \bar{u}(k_4) \left\{ \frac{\gamma^\mu (\slashed{k}_1 - \slashed{k}_3 + m_q) \gamma^\sigma t^\kappa}{(k_1 - k_3)^2 - m_q^2 + i0}  +
\frac{\gamma^\sigma (\slashed{k}_1 + \slashed{k}_2 + m_q) \gamma^\mu t^\kappa}{(k_1 + k_2)^2 - m_q^2 + i0} 
 \right\} \lblob(P_1;k_1) \, . \label{eq:Rtwoterms}
\end{equation}

\subsection{Cancellation of Superleading Terms}
\label{sec:supcan}
The superleading $K$-term in Eq.~\eqref{eq:1gluonR} can be seen to vanish in the sum over graphs in the full amplitude by first noting that Eq.~\eqref{eq:m1gK} can be written as
\begin{equation}
 \ublob(P_2;k_2)^{\kappa ; \mu^\prime} K(k_2)_{\mu^\prime \mu}  R^{\kappa ; \mu \sigma}
 =  \ublob(P_2;k_2)^{\kappa ; \mu^\prime} \frac{n_{1 \; \mu^\prime} k_{2 \; \mu}}{n_1 \cdot k_2}  R^{\kappa ; \mu \sigma} 
 =  \underbrace{\frac{\ublob(P_2;k_2)^{\kappa ; -}}{k_2^-}}_\text{$\sim Q^0$} \times \underbrace{k_{2 \mu} R^{\kappa ; \mu \sigma}}_\text{$\sim Q $} \; \sim Q \, , \label{eq:oneKdecomp}
\end{equation}
so that attention may be focused solely on $k_{2 \mu} R^{\kappa ; \mu \sigma}$.  (The underbraces in Eq.~\eqref{eq:oneKdecomp} only denote the maximum powers of $Q$ \emph{graph-by-graph}.
Since they are superleading, the terms of order $Q$ have to cancel in the sum over all graphs.)
In order to argue that all the $K$-gluon contributions in Eq.~\eqref{eq:oneKdecomp} are power suppressed, it must be shown that both the order $\sim Q$ (superleading) 
and the order $\sim Q^0$ (leading) contributions to the $k_{2 \mu} R^{\kappa ; \mu \sigma}$ subgraph cancel in the sum over graphs
when contracted with $\ublob(P_2;k_2)^{\kappa ; \mu^\prime}$.
Including both terms from Eq.~(\ref{eq:Rtwoterms}), $k_{2 \mu} R^{\kappa ; \mu \sigma}$ becomes
\begin{align}
\label{eq:separation}
k_{2 \mu} & R^{\kappa ; \mu \sigma} =  
i g_s e_q \bar{u}(k_1 + k_2 - k_3) & &   \nonumber \\ & \times  
\left\{ \frac{\slashed{k}_2 (\slashed{k}_1 - \slashed{k}_3 + m_q) \gamma^\sigma t^\kappa}{(k_1 - k_3)^2 - m_q^2 + i0}  \right. +  && \; \text{Term 1} \nonumber \\ \left.  \right. &  + \left.
\frac{\gamma^\sigma (\slashed{k}_1 + \slashed{k}_2 + m_q) \slashed{k}_2 t^\kappa }{(k_1 + k_2)^2 - m_q^2 + i0} 
 \right\} \times  && \; \text{Term 2} & \nonumber \\ &   \times \lblob(P_1;k_1) \, .
\end{align}
From here forward, the steps for dealing with Fig.~\ref{fig:onegluon} are similar to standard arguments for a Ward identity cancellation.
For Term 1 in braces in Eq.~\eqref{eq:separation}, we apply the Feynman identity replacement,
\begin{equation}
\label{eq:onegluondecomp1}
\slashed{k}_2 = (\slashed{k}_1 + \slashed{k}_2 - \slashed{k}_3 - m_q ) - (\slashed{k}_1 - \slashed{k}_3 - m_q) \, ,
\end{equation}
while in Term 2 we use
\begin{equation}
\label{eq:onegluondecomp2}
\slashed{k}_2 = (\slashed{k}_2 + \slashed{k}_1  - m_q ) - (\slashed{k}_1 - m_q) \, .
\end{equation}
When the $(\slashed{k}_1 + \slashed{k}_2 - \slashed{k}_3 - m_q )$ in Eq.~\eqref{eq:onegluondecomp1} acts to the 
left on $\bar{u}(k_1 + k_2 - k_3)$, in Term 1 of Eq.~\eqref{eq:separation}, the resulting contribution vanishes exactly by the action of the 
Dirac equation on the outgoing quark wavefunction.  

When the $-(\slashed{k}_1 - m_q)$ in Eq.~\eqref{eq:onegluondecomp2} acts to the right on $\lblob(P_1;k_1)$, in Term 2 of Eq.~\eqref{eq:onegluondecomp1}, the resulting contribution is suppressed by two powers 
of $Q$.  (It would be exactly zero if the target quark were taken to be exactly on-shell.)
To see this, let us rewrite Term 2 from Eq.~\eqref{eq:separation},  but with the $k_1$-propagator inside $\lblob(P_1;k_1)$ explicitly displayed:
\begin{multline}
i g_s e_q \underbrace{\bar{u}(k_1 + k_2 - k_3)}_\text{$\sim Q^{1/2}$}  \, \underbrace{\frac{\gamma^\sigma (\slashed{k}_1 + \slashed{k}_2 + m_q) \slashed{k}_2 t^\kappa}{(k_1 + k_2)^2 - m_q^2 + i0}}_\text{$\sim Q^{0}$} 
\; \underbrace{\lblob(P_1;k_1)}_\text{$\sim Q^{1/2}$} 
= \\
i g_s e_q \underbrace{\bar{u}(k_1 + k_2 - k_3)}_\text{$\sim Q^{1/2}$} \, \underbrace{\frac{\gamma^\sigma (\slashed{k}_1 + \slashed{k}_2 + m_q) t^\kappa }{(k_1 + k_2)^2 - m_q^2 + i0}}_\text{$\sim 1/Q$} 
\; \underbrace{\left( \slashed{k}_2 \frac{(\slashed{k}_1 + m_q)}{k_1^2 - m_q^2 + i0} \right)}_\text{$\sim Q^2$} \; \underbrace{\cdots \;\;\;\;\;}_\text{$\sim Q^{-1/2}$}  \; \; 
\sim Q \; . \label{eq:k2sub}
\end{multline}
In the last line, the ``$\cdots $" symbolizes the other factors that make up $\lblob(P_1;k_1)$, and these are order $\sim Q^{-1/2}$ because of the overall $\lblob(P_1;k_1) \sim Q^{1/2}$ power-law.  
Note that in the second line of Eq.~\eqref{eq:k2sub}, the $\slashed{k}_2$ has been moved outside the hard propagator and is included in the parentheses with the $k_1$-propagator.
 
When the Feynman identity substitution is made for $\slashed{k}_2$, the $-(\slashed{k}_1 - m_q)$ from the second term of Eq.~\eqref{eq:onegluondecomp2} gives
\begin{equation}
i g_s e_q \underbrace{\bar{u}(k_1 + k_2 - k_3)}_\text{$\sim Q^{1/2}$} \, \underbrace{\frac{\gamma^\sigma (\slashed{k}_1 + \slashed{k}_2 + m_q) t^\kappa }{(k_1 + k_2)^2 - m_q^2 + i0}}_\text{$\sim 1/Q$} 
\; \times \underbrace{\left( [ - (\slashed{k}_1 - m_q) ] \frac{(\slashed{k}_1 + m_q)}{k_1^2 - m_q^2 + i0} \right) }_\text{$\sim Q^{0}$} \; \underbrace{\cdots \;\;\;\;\;}_\text{$\sim Q^{-1/2}$}
\; \sim 1 / Q \; . \label{eq:k2subb}
\end{equation}
So the factor in parentheses looses two powers of $Q$ relative to Eq.~\eqref{eq:k2sub}, and the contribution to $k_{2 \mu} R^{\kappa ; \mu \sigma}$
from the second term in Eq.~\eqref{eq:onegluondecomp2} has a power law $\sim 1/Q$.  
That it is suppressed by \emph{two} powers of $Q$ instead of only one is crucial because we will ultimately
multiply it with the superleading $\ublob^-$ in Eq.~\eqref{eq:oneKdecomp}.
We will need this result again in Sect.~\ref{sec:twogluons} when we take into account effects from an additional soft/collinear gluon radiated from hadron-2.

The remaining $K$-term contributions are from the second term in Eq.~\eqref{eq:onegluondecomp1} and the first term in Eq.~\eqref{eq:onegluondecomp2} in the Feynman identity substitutions.  These
exactly cancel against each other when we use quark propagator denominator cancellations like
\begin{equation}
\label{eq:simplecancel}
\frac{ (\slashed{k}_1 + \slashed{k}_2 + m_q) }{(k_1 + k_2)^2 - m_q^2 + i0} \times (\slashed{k}_1 + \slashed{k}_2  - m_q ) = 1 \, .
\end{equation}
The exact cancellation is not put in danger by the $i0$ because $(k_1 + k_2)^2$ is constrained by kinematics to be always of order $Q^2$.

Thus, there is no leading-power $K$-gluon contribution in the one-gluon case, and the superleading contributions cancel as expected.
The only remaining contribution is from the $G$-gluon in Eq.~\eqref{eq:m1gG}.  
From Eq.~\eqref{eq:leadingonegluon}, it immediately follows that the leading-power $G$-gluon contributions to $M_{(1g)}$ involve only the transverse $\mu$-components of $R^{\kappa ; \mu \sigma}$.
The plus component is already entirely accounted for above in the treatment of the $K$-term and is, by construction, exactly removed in the $G$-term.  
Because of the suppression in Eq.~\eqref{eq:1gluonsupp}, the minus $\mu$-component of $R^{\kappa ; \mu \sigma}$ gives a contribution suppressed by a power of $Q$ 
relative to the leading power.  Therefore, the Lorentz components in the $G$-term contraction may be restricted to the frame-1 transverse $\mu$-components.  
That is, we may replace
\begin{equation}
G(k_2)^{\mu^\prime \mu} \to G(k_2)^{\mu^\prime i} = g^{\mu^\prime i} -  \frac{n_{1}^{\mu^\prime} k_{2t}^{i}}{n_{1} \cdot k_2} \, , \label{eq:Greplace}
\end{equation}
to leading power, which is in agreement with the normal gluon PDF vertex with no gauge link contribution (see, e.g., Eq.~(45) of Ref.~\cite{Collins:2008sg} and Fig.~3.4 of Ref.~\cite{Collins:1981uw}).
The lowest-order uncanceled contribution to the amplitude is therefore consistent with typical expectations for the parton model gluon density:
\begin{align}
M_{(1g)}
& =i g_s e_q \sum_i \ublob(P_2;k_2)_{\mu^\prime}^{\kappa}  \left( g^{\mu^\prime i} -  \frac{n_{1}^{\mu^\prime} k_{2t}^{i}}{n_{1} \cdot k_2} \right) \;
 \bar{u}(k_4) \left\{ \frac{\gamma^i (\slashed{k}_1 - \slashed{k}_3 + m_q) \gamma^\sigma t^\kappa}{(k_1 - k_3)^2 - m_q^2 + i0}  +
\frac{\gamma^\sigma (\slashed{k}_1 + \slashed{k}_2 + m_q) \gamma^i t^\kappa}{(k_1 + k_2)^2 - m_q^2 + i0} 
 \right\} \lblob(P_1;k_1) \nonumber \\ \nonumber \\
& = i g_s e_q \sum_i \underbrace{\ublob(P_2;k_2)_{\mu^\prime}^{\kappa}   \; G(k_2)^{\mu^\prime i}}_\text{$\sim Q^{0}$} \;
 \underbrace{\bar{u}(k_4)}_\text{$\sim Q^{1/2}$} \; \underbrace{H_{\rm LO}(k_1,k_3,k_4,k_2)^{\kappa ; i \sigma}}_\text{$\sim 1/Q$}  \; \underbrace{\lblob(P_1;k_1)}_\text{$\sim Q^{1/2}$}  \; \sim Q^0 \, , \label{eq:1gluonampresult}
\end{align}
where we have used Eqs.~(\ref{eq:Greplace},~\ref{eq:Rtwoterms}) in Eq.~\eqref{eq:m1gG}.
In the second line, we have defined $H_{\rm LO}(k_1,k_3,k_4,k_2)^{\kappa ; \mu \sigma}$ as the sum of the lowest order contributions to the hard parts in Fig.~\ref{fig:onegluon}:
\begin{equation}
\label{eq:treehardpart}
H_{\rm LO}(k_1,k_3,k_4,k_2)^{\kappa ; \mu \sigma} \equiv \frac{\gamma^\mu (\slashed{k}_1 - \slashed{k}_3 + m_q) \gamma^\sigma t^\kappa}{(k_1 - k_3)^2 - m_q^2 + i0}  +
\frac{\gamma^\sigma (\slashed{k}_1 + \slashed{k}_2 + m_q) \gamma^\mu t^\kappa}{(k_1 + k_2)^2 - m_q^2 + i0} \, .
\end{equation}
Checking Eqs.~(\ref{eq:1gluonampresult},\ref{eq:treehardpart}) with Sect.~\ref{sec:maxgen} confirms that
the amplitude is now in a form consistent with the TMD-factorization conjecture, Eqs.~(\ref{eq:vgfactamp}-\ref{eq:leffposs}), 
in the maximally general form.  Tallying the powers of $Q$ indicated with braces underneath Eq.~\eqref{eq:1gluonampresult} verifies that the combined powers give 
a leading contribution to the amplitude.

\subsection{Tree Level TMD-Factorization}
\label{sec:tmdfactorization}

Now nothing obstructs the immediate
recovery of the normal TMD-factorization formula characteristic of parton model expectations.  
Using Eq.~\eqref{eq:1gluonampresult} in Eq.~\eqref{eq:crosssection} and averaging over initial hadron spins gives 
\begin{widetext}
\begin{align}
d \sigma & = \mathcal{C} Q^2 \, \sum_{s} \sum_{X_1} \sum_{X_2} \int \frac{d^4 k_1}{(2 \pi)^4} \frac{d^4 k_2}{(2 \pi)^4} \, | M_{(1g)} |^2  \, (2 \pi)^4 \delta^{(4)} (k_1 + k_2 - k_3 - k_4) \nonumber  \\ & =
\mathcal{C} Q^2 e_q^2 g_s^2 \sum_s \sum_{X_1} \sum_{X_2} \sum_{ij} \int \frac{d^4 k_1}{(2 \pi)^4} \frac{d^4 k_2}{(2 \pi)^4} \, \nonumber \\ & 
\qquad \qquad \qquad \qquad \qquad \qquad \times  \underbrace{\ublob_{\mu^\prime}^\kappa \ublob^{\dagger \, \kappa^\prime}_{ \nu^\prime} 
G^{\mu^\prime i} G^{\nu^\prime j}
\tr{\bar{u}(k_4) \; H^{\kappa ; i  \sigma}_{\rm LO} \;  \lblob \bar{\lblob}  \; H^{\dagger \; \kappa^\prime ; j}_{{\rm LO}, \, \sigma} \; u(k_4)}}_\text{$\sim Q^0$} \, (2 \pi)^4 \delta^{(4)} (k_1 + k_2 - k_3 - k_4)  \, . \label{eq:1gluoncross}
\end{align}
The sums over $X_1$ and $X_2$ represent the sums and integrals over the final states in the $\lblob$ and $\ublob$ subgraphs.  The spin sums 
for the incoming hadrons are included in these sums as well.  In Eq.~\eqref{eq:1gluoncross}, the 
explicit momentum arguments of the various factors have been dropped for convenience.  The outgoing quark has spin label $s$. 
The minus sign from the spin sum on the prompt photon has been absorbed into the definition of $\mathcal{C}$.

To put Eq.~\eqref{eq:1gluoncross} into a form closer to Eq.~\eqref{eq:genproj2}, we now apply the minimal partonic 
approximations of Eqs.~(\ref{eq:hat1}-\ref{eq:hardsub}). 
Specifically,  we neglect the smallest (order $\sim \Lambda^2 / Q$) momenta $k_1^-$ and $k_2^+$ everywhere except within their own 
respective parent subgraphs, $\lblob(P_1;k_1)$ and $\ublob(P_2;k_2)$.   Then the right side of Eq.~\eqref{eq:1gluoncross} becomes
\begin{align}
\mathcal{C} Q^2 e_q^2 g_s^2 \sum_{s} \sum_{X_1} \sum_{X_2} \sum_{ij}  & \int \frac{d^2 {\bf k}_{1t}}{(2 \pi)^2} \, \left( \int \frac{d k_2^+}{2 \pi} \ublob_{\mu^\prime}^{ \kappa} \ublob^{\dagger \; \kappa^\prime}_{\nu^\prime} 
G^{\mu^\prime i} G^{\nu^\prime j} \right) 
\left( \int \frac{d k_1^-}{2 \pi} \tr{\bar{u}(k_4) \; \hat{H}^{\kappa ; i \sigma}_{\rm LO} \;  \lblob \bar{\lblob}  \; \hat{H}^{\dagger \; \kappa^\prime ; j}_{{\rm LO}, \, \sigma} \; u(k_4)}  \right) \nonumber \\ 
& = \mathcal{C} Q^2 e_q^2 g_s^2 \sum_{s} \sum_{ij} \int \frac{d^2 {\bf k}_{1t}}{(2 \pi)^2} \left(  \sum_{X_2} \int \frac{d k_2^+}{2 \pi} \ublob_{\mu^\prime}^{\kappa} \ublob^{\dagger \; \kappa^\prime}_{\nu^\prime} 
\left(  g^{\mu^\prime i} -  \frac{n_{1}^{\mu^\prime} k_{2t}^{i}}{n_{1} \cdot k_2}   \right)  \left( g^{\nu^\prime j} -  \frac{n_{1}^{\nu^\prime} k_{2t}^{j}}{n_{1} \cdot k_2}  \right) \right)  \nonumber \\
& \qquad \times \left( \sum_{X_1} \int \frac{d k_1^-}{2 \pi} \tr{ \bar{u}(k_4) \; \hat{H}^{\kappa ; i \sigma}_{\rm LO} \;  \lblob \bar{\lblob}  \; \hat{H}^{\dagger \; \kappa^\prime ; j}_{{\rm LO}, \, \sigma} \; u(k_4)}  \right) \nonumber \\&
 = \mathcal{C} e_q^2 g_s^2 \sum_{s} \sum_{ij} \int \frac{d^2 {\bf k}_{1t}}{(2 \pi)^2} \, \Phi_{g / P_2}^{(1g)}\left( P_2;\hat{k}_2(x_2,{\bf q}_t - {\bf k}_{1t}) \right)^{\kappa \kappa^\prime;ij} \times  \nonumber \\ & \qquad \times 
 {\rm Tr} \left[ \bar{u}(k_4) \; \hat{H}^{\kappa ; i \sigma}_{\rm LO}\left( \hat{k}_1(x_1,{\bf k}_{1t}),\hat{k}_2(x_2,{\bf q}_t - {\bf k}_{1t}),k_3,k_4 \right) \;  \Phi_{q / P_1}^{(1g)} \left( P_1;\hat{k}_1(x_1,{\bf k}_{1t}) \right)  \right. \times \nonumber \\ 
 \; & \qquad \times \left. \hat{H}^{\dagger \; \kappa^\prime ; j}_{{\rm LO}, \, \sigma}\left( \hat{k}_1(x_1,{\bf k}_{1t}),\hat{k}_2(x_2,{\bf q}_t - {\bf k}_{1t}),k_3,k_4 \right) \; u(k_4) \right]  \, . \label{eq:basicdecom}
 \end{align}
In lines 1-3 we have used the integrals over ${\bf k}_{2t}$, $k_1^+$, and $k_2^-$ to evaluate the $\delta$-functions in Eq.~\eqref{eq:1gluoncross}, and the 
positions of the integrals over $k_1^-$ and $k_2^+$ 
have been arranged into factors corresponding to the respective subgraphs for hadron-1 and hadron-2, with the factors that correspond to effective TMD PDFs separated by parentheses.  
The use of the kinematical approximations in  Eqs.~(\ref{eq:hat1}-\ref{eq:hardsub}) is symbolized by the ``hats'' on $H_{\rm LO}$.
After the second equality in Eq.~\eqref{eq:basicdecom}, we have identified effective tree-level TMD PDFs:
\begin{align}
\Phi_{q / P_1}^{(1g)}\left( P_1;\hat{k}_{1}(x_1,{\bf k}_{1t}) \right)  & \equiv   Q  \sum_{X_1} \int  \frac{d k_1^-}{2 \pi} \lblob \bar{\lblob}  \; \sim Q \, , \label{eq:1gluondef} \\
 \Phi_{g / P_2}^{(1g)}\left( P_2;\hat{k}_{2}(x_2,{\bf k}_{2t}) \right)^{\kappa \kappa^\prime;i j} &  \equiv  Q \sum_{X_2} \int \frac{d k_2^+}{2 \pi} \ublob_{\mu^\prime}^\kappa \ublob^{\dagger \; \kappa^\prime}_{\nu^\prime} 
\left(  g^{\mu^\prime i} -  \frac{n_{1}^{\mu^\prime} k_{2t}^{i}}{n_{1} \cdot k_2}   \right) \left( g^{\nu^\prime j} -  \frac{n_{1}^{\nu^\prime} k_{2t}^{j}}{n_{1} \cdot k_2}  \right) \; \sim Q^0 \, . \label{eq:2gluondef}
\end{align}
Also, after the last equality of Eq.~\eqref{eq:basicdecom} we have restored the explicit momentum arguments to make clear the connection with 
the statement of TMD-factorization in Eq.~\eqref{eq:vgenfact}.  
Equations~\eqref{eq:2gluondef} are consistent with the basic structure of TMD PDF operator definitions
that are typical in a generalized TMD parton model, and they are consistent with criteria of Sect.~\ref{sec:genfact}.  
Hence, in Eq.~\eqref{eq:basicdecom} we have recovered a result consistent with TMD-factorization under the maximally general criteria.
In these graphs there is not yet sensitivity 
to gauge links type effects; for that we will need to consider the extra soft gluons of Fig.~\ref{fig:leadingregions}.

With the tree level cross section now in the form of Eq.~\eqref{eq:vgenfact}, one may proceed with the classification of 
quark and gluon polarization dependence following the normal steps in Sect.~\ref{sec:azimuthal}.
The contribution to the cross section that corresponds to a generic  quark polarization projection, $\Gamma_q$, 
and gluon polarization, $\Gamma_g$, is written as in Eq.~\eqref{eq:genproj2}:
\begin{align}
d & \sigma^{\left[ \Gamma_q, \Gamma_g \right] } =  \mathcal{C} e_q^2 g_s^2
\sum_{s} \sum_{i j} \int \frac{d^{2} {\bf k}_{1t}}{(2 \pi)^2}\, \underbrace{\Phi^{(1g) \left[ \Gamma_g \right]}_{g/P_2}\left( P_2;\hat{k}_2(x_2,{\bf k}_{2t}) \right)^{\kappa \kappa^\prime ; i j}}_\text{$\sim Q^0$}
\; \underbrace{\Phi^{(1g) \left[ \Gamma_q \right]}_{q/P_1}\left( P_1;\hat{k}_1(x_1,{\bf k}_{1t}) \right)}_\text{$\sim Q$} \times \nonumber  \\ &  \times  
{\rm Tr} [ \underbrace{\bar{u}(k_4)}_\text{$\sim Q^{1/2}$} \; \underbrace{\hat{H}^{\kappa ; i \sigma}_{\rm LO}\left( \hat{k}_1(x_1,{\bf k}_{1t}),\hat{k}_2(x_2,{\bf q}_t - {\bf k}_{1t}),k_3,k_4 \right)}_\text{$\sim 1/Q$} \;  
{\rm P}_{\Gamma_q} \; 
\underbrace{\hat{H}^{\dagger \; \kappa^\prime ; j}_{{\rm LO}, \, \sigma}\left( \hat{k}_1(x_1,{\bf k}_{1t}),\hat{k}_2(x_2,{\bf q}_t - {\bf k}_{1t}),k_3,k_4 \right)}_\text{$\sim 1/Q$} \; \underbrace{u(k_4)}_\text{$\sim Q^{1/2}$} ] 
\nonumber \\ 
& \sim Q^0 \, . \label{eq:basicdecom2}
 \end{align}
For example, the standard azimuthally symmetric contribution is, from Eqs.~(\ref{eq:gluondecomp},\ref{eq:gammaplus}):
\begin{multline}
\label{eq:basicdecom3}
d \sigma^{\left[ \gamma^+, {\rm U} \right]} = \mathcal{C} e_q^2 g_s^2 \sum_{s} \sum_{j} \int \frac{d^{2} {\bf k}_{1t}}{(2 \pi)^2}\, \Phi^{(1g), {\rm U}}_{g/P_2}\left( P_2;\hat{k}_2(x_2,{\bf k}_{2t}) \right)^{\kappa \kappa^\prime}
\Phi^{(1g)\left[ \gamma^+ \right]}_{q/P_1}\left( P_1;\hat{k}_1(x_1,{\bf k}_{1t}) \right) \times \\ \times \tr{\bar{u}(k_4) 
\hat{H}^{\kappa ; i \sigma}_{\rm LO} \left( \hat{k}_1(x_1,{\bf k}_{1t}),\hat{k}_2(x_2,{\bf q}_t - {\bf k}_{1t}),k_3,k_4 \right) \;  
\gamma^- \; \hat{H}^{\dagger \; \kappa^\prime ; j}_{{\rm LO}, \, \sigma}\left( \hat{k}_1(x_1,{\bf k}_{1t}),\hat{k}_2(x_2,{\bf q}_t - {\bf k}_{1t}),k_3,k_4 \right) \; u(k_4)}  \, .
 \end{multline}
As a final step, one may average over quark and gluon color to reproduce the standard expressions for the quark and gluon TMD PDFs with a color averaged hard part, 
though this is not necessary to satisfy the criteria of Sect.~\ref{sec:partonmodel}.

To summarize, in this section we have verified that the generalized parton model picture of TMD-factorization arises
in a detailed tree-level,  single-gluon treatment, and we have shown that it is consistent with the maximally general statement of 
TMD-factorization from Sect.~\ref{sec:partonmodel}.  Tallying the powers of $Q$ in the underbraces of Eq.~\eqref{eq:basicdecom2} verifies
that it is a leading-power contribution.  

Later we will follow similar steps when we include one extra gluon from hadron-2, but there they will fail to 
reproduce a factorized form, even under the loose criteria of Sect.~\ref{sec:genfact}.

\section{Spectator-spectator Interactions}
\label{sec:spec}
\begin{figure}
\centering
\includegraphics[scale=.3]{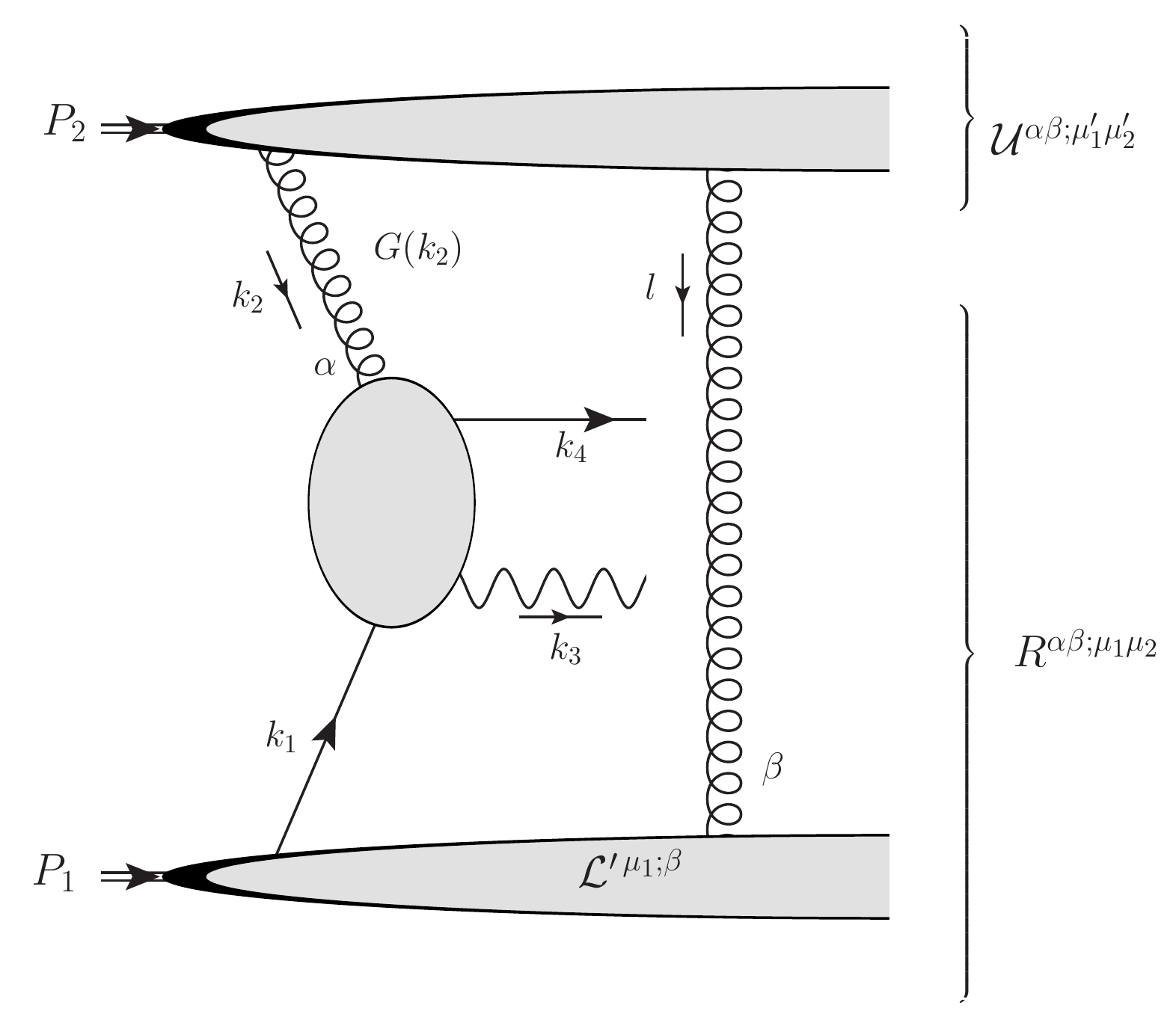}
\caption{The amplitude in Eq.~\eqref{eq:m2gluonspecspec} with a single spectator-spectator interaction $l$.}
\label{fig:specspec}
\end{figure}
\begin{figure*}
\centering
  \begin{tabular}{c@{\hspace*{25mm}}c}
    \includegraphics[scale=0.3]{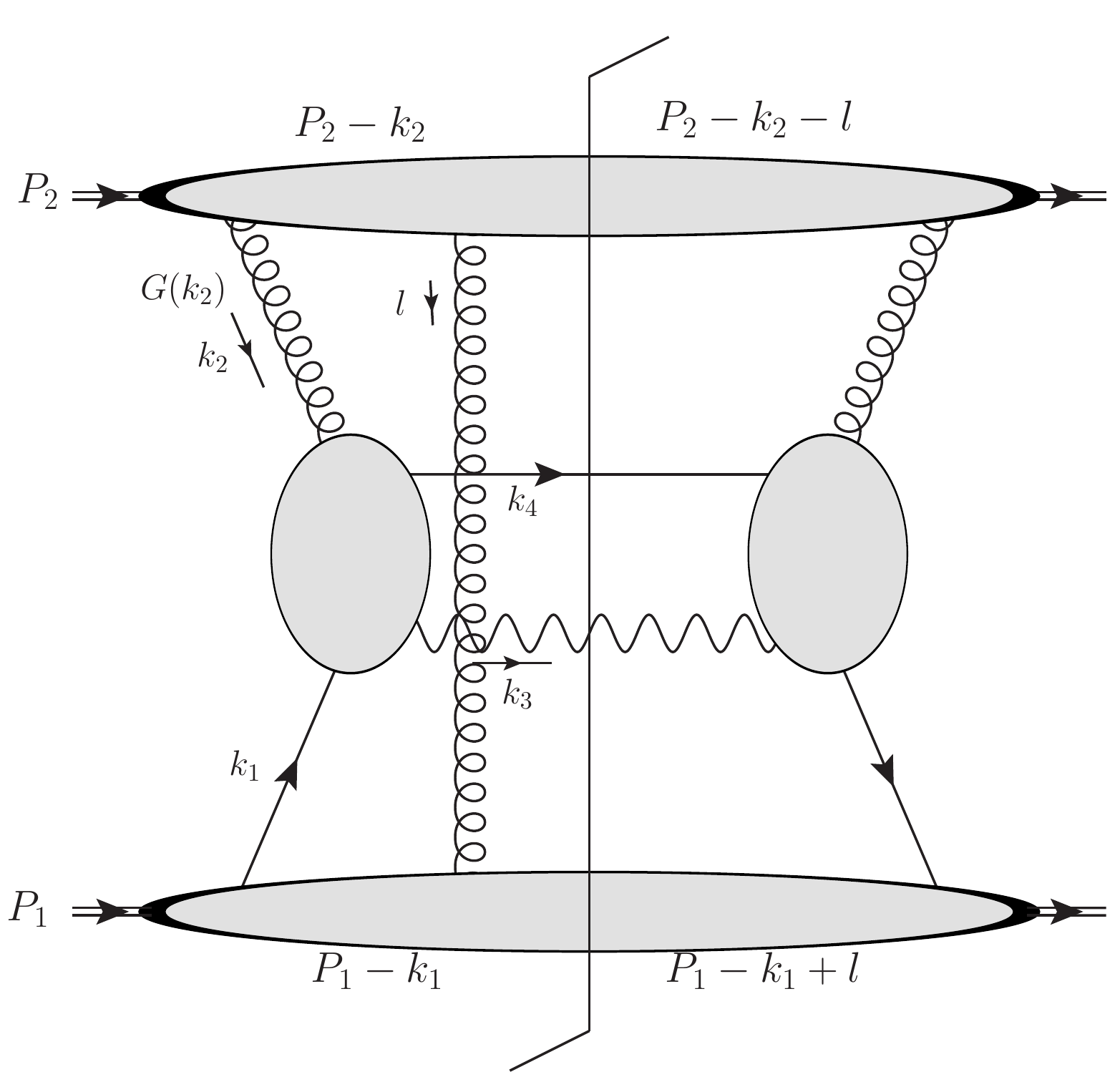}
    &
    \includegraphics[scale=0.3]{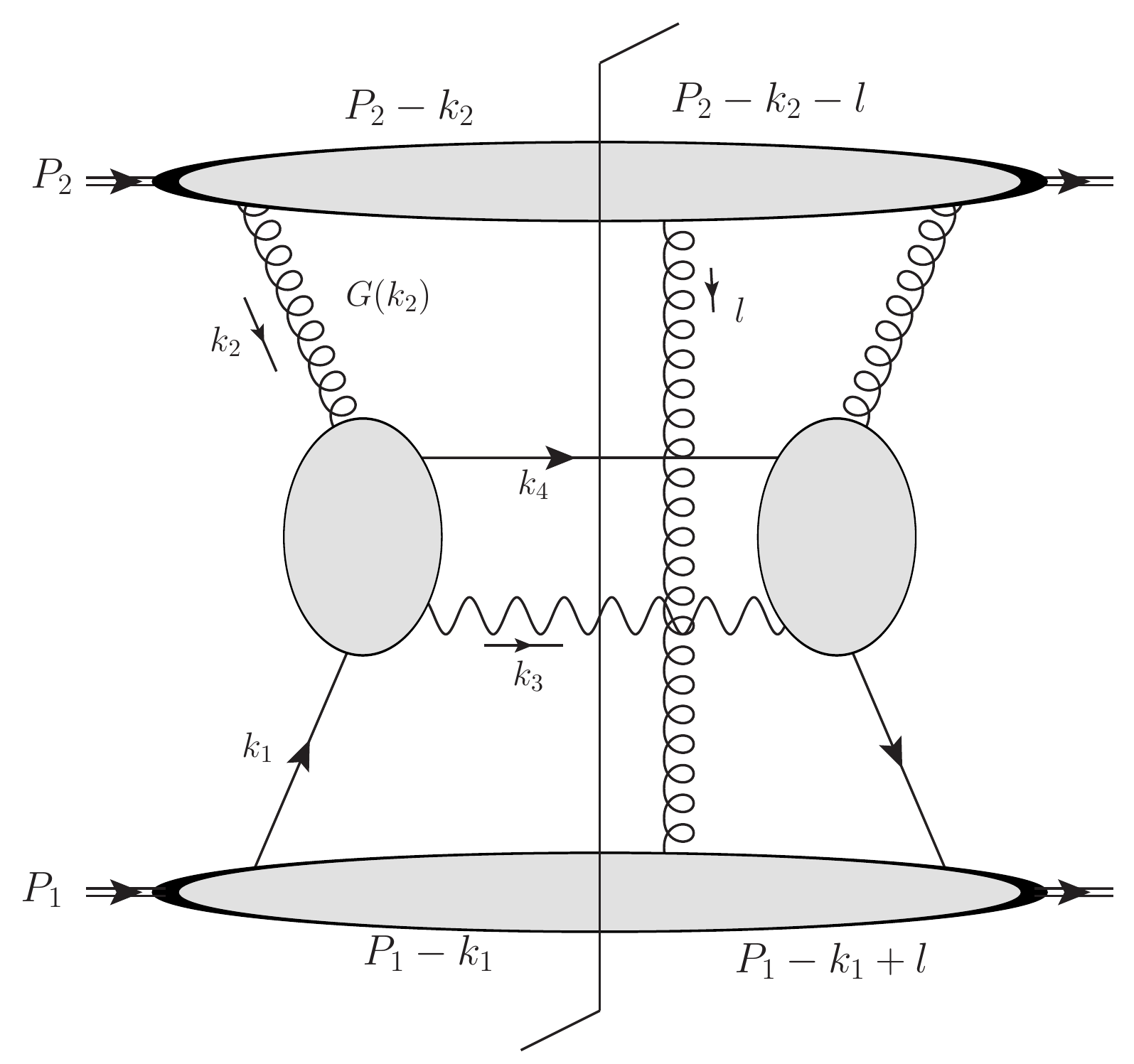}
    \\
    (a) & (b)
    \\[5mm]
   \end{tabular}
\caption{Spectator-spectator interactions that cancel in the inclusive cross section.  The active gluon is a $G(k_2)$ gluon which yields only 
leading (non superleading) contributions.  (As in Fig.~\ref{fig:specspec}, we restrict consideration here to spectator attachments for $l$ in the upper and lower bubbles.)}
\label{fig:specspec2}
\end{figure*}

The next step is to examine the contribution from one extra soft gluon in Fig.~\ref{fig:leadingregions}.  If the generalized criteria 
for TMD-factorization are to be respected, then the sum over all such graphs must allow contributions from the 
extra gluon to be factored into separate contributions in the upper and lower subgraphs.
The single extra soft gluon contributions include both spectator attachments as well as attachments to 
active partons.   In this section we will deal separately with the spectator-spectator type graphs, arguing that any leading or superleading contributions cancel.  
Later we will deal with the active parton attachments and find that they lead to TMD-factorization breaking.

The relevant amplitude for the spectator-spectator case is shown in Fig.~\ref{fig:specspec}, and we extend the notation of Sect.~\ref{sec:onegluesep} by writing it as
\begin{equation}
\label{eq:m2gluonspecspec}
M_{(2g)}^{\rm spec-spec} =  
 \ublob(P_2;k_2,l)^{\alpha \beta ; \mu_1^\prime \mu_2^\prime} \, g_{\mu_1^\prime \mu_1} \, g_{\mu_2^\prime \mu_2} R^{\alpha \beta ; \mu_1 \mu_2}(k_1,k_2,k_3,k_4,l,P_1) \, ,
\end{equation}
where
\begin{equation}
\label{eq:m2gluonspecspec2}
R^{\alpha \beta ; \mu_1 \mu_2}(k_1,k_2,k_3,k_4,l,P_1) \equiv  \bar{u}(k_4) \, H_{\rm LO}(k_1,k_2,k_3,k_4)^{\alpha ; \mu_1 \sigma} \, \lblob^\prime(P_1,k_1,l)^{\beta ; \mu_2} \, 
\end{equation}
includes both the bottom bubble $\lblob^\prime$ and the hard subgraph $H_{\rm LO}$ from Eq.~\eqref{eq:treehardpart}.  (See the labeling in Fig.~\ref{fig:specspec}.)  
The partons are nearly on-shell, $k_2^2 \sim k_1^2 \sim l^2 \lesssim \Lambda^2$, but now 
the four-momentum may be shared between 
$k_2$ and $l$.  As usual, we are interested in the region $k_{1,t} \lesssim \Lambda$ and $k_{2,t} \lesssim \Lambda$.
In Fig.~\ref{fig:specspec}, we restrict consideration to graphs with spectator attachments for $l$
inside the upper and lower bubbles.

The superleading contributions come from the contraction of $R^{\alpha \beta ; + +}$ with $ \ublob(P_2;k_2,l)^{\alpha \beta ; - -}$.  The region $k_2^- \ll Q$
is forbidden because it pushes the quark propagator $k_1$, and propagators with momentum $P_1 - k_1$ inside $\lblob^\prime$, far off shell.
Thus, in the $k_2^- \ll Q$ region, there are at least two extra powers of suppression so that the overall contribution is suppressed by at 
least one power of $Q$ relative to the leading power, even in the contraction of $R^{\alpha \beta ; + +}(k_1,k_2,k_3,k_4,l,P_1)$ with $ \ublob(P_2;k_2,l)^{\alpha \beta ; - -}$.
With $k_2^- \sim Q$, both the $l^+$ and the $l^-$ components of the $l$-gluon are trapped at 
order $\Lambda^2/Q$ by initial and final state poles in $\ublob(P_2;k_2,l)^{\alpha \beta ; \mu_1^\prime \mu_2^\prime} $ 
and in $\lblob^\prime(P_1,k_1,l)^{\mu_2; \beta}$.  That is, $| l^+ l^-| \ll |{\bf l}_T|^2 \lesssim \Lambda^2$.  (This is the Glauber region 
discussed in Sect.~\ref{sec:glauber}.)  But then 
the kinematics of the hard subprocess becomes identical to the single gluon case of Sect.~\ref{sec:onegluon}, with the only 
difference being the slightly shifted momenta $P_2 - k_2 - l$ and $P_1 - k_1 + l$ of the inclusive final state remnants.  Therefore, 
we may once again decompose the $g_{\mu_1^\prime \mu_1}$ in Eq.~\eqref{eq:m2gluonspecspec} into 
$G(k_2)$ and $K(k_2)$ gluons as in Eqs.~(\ref{eq:Kdefk2},\ref{eq:Gdefk2}) to find that the $K(k_2)$ terms cancel to leading power in 
an argument identical to that of the previous section.  Since $k_2^- \sim Q$, the remaining $G(k_2)$ term leaves only leading, not superleading, contributions.

At this stage, the cancellation of the remaining spectator-spectator Glauber interactions in the inclusive sum over the final states of $\lblob$ and $\ublob$ in 
the inclusive cross section follows steps familiar from other hadron-hadron processes like the Drell-Yan process. 
If a single extra spectator-spectator gluon is exchanged, the graphs that contribute at leading power to the 
cross section are as shown in  Fig.~\ref{fig:specspec2}, where the $k_2$ gluon is labeled by $G(k_2)$ to emphasize that these 
graphs are leading and not superleading, according to the argument above, and involve only $G$-gluon attachments in the hard subgraphs.
The extra $l$ momentum is routed entirely through the hadron-1 and hadron-2 bubbles, which are summed inclusively in the cross section. 
The cancellation is between the two final state cuts shown in Fig.~\ref{fig:specspec2} with different final state momenta inside $\ublob$ and $\lblob^\prime$.   \\ \\

\section{Analysis of Two Gluons}
\label{sec:twogluons}
\begin{figure*}
  \centering
  \begin{tabular}{c@{\hspace*{12mm}}c}
    \includegraphics[scale=0.4]{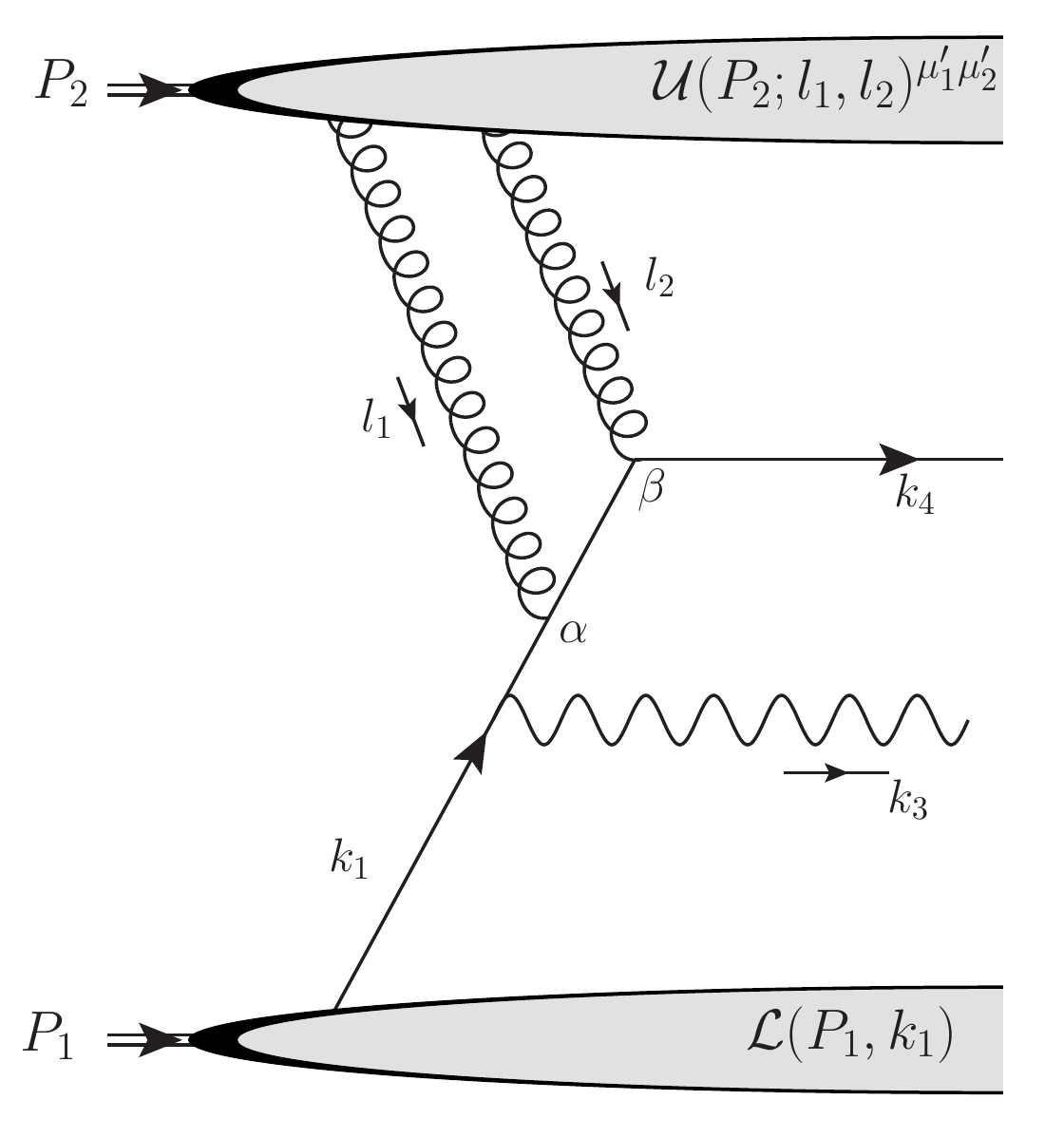}
    &
    \includegraphics[scale=0.4]{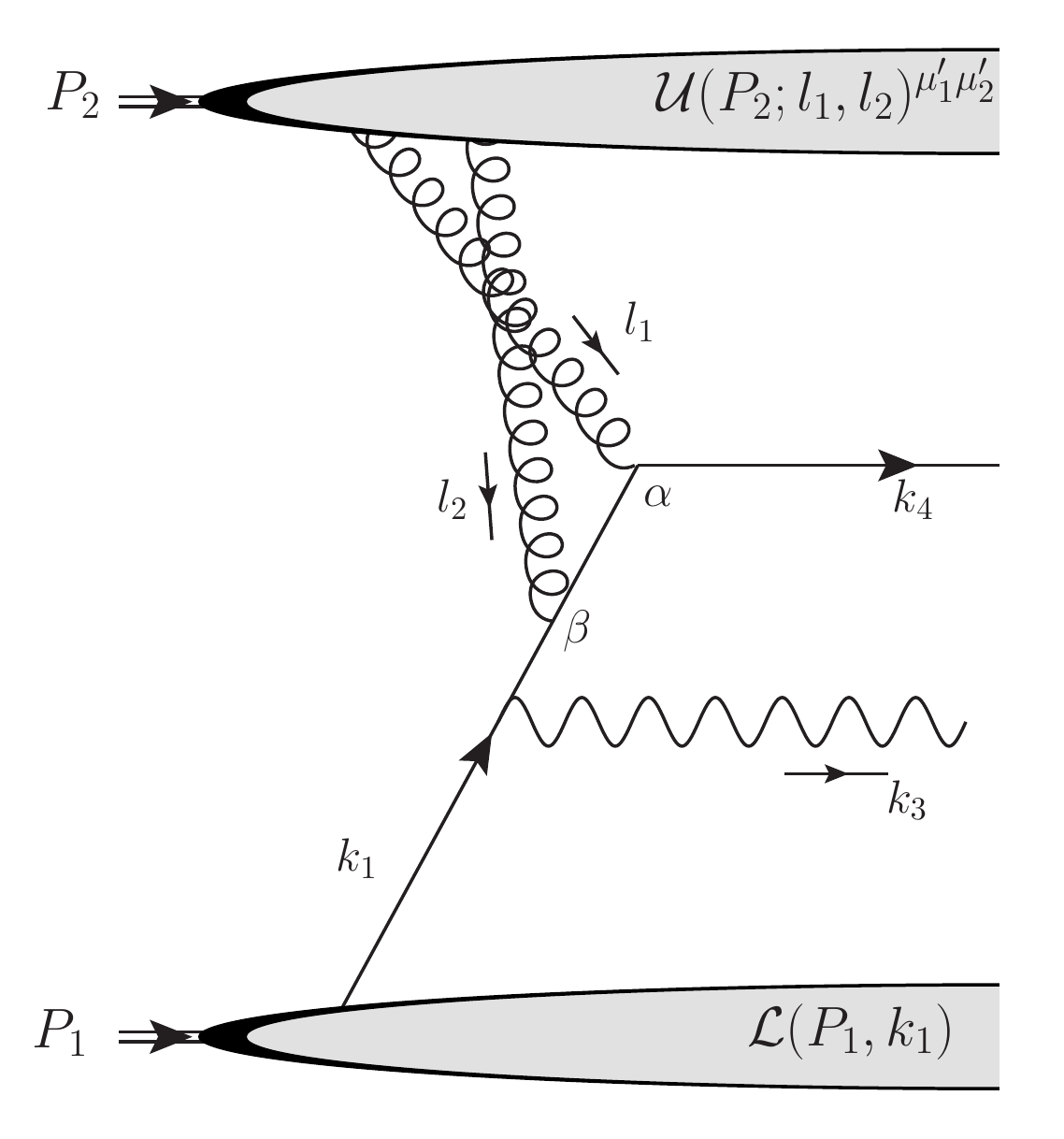}
\\
(a) & (b)
 \\[7mm]
    \includegraphics[scale=0.4]{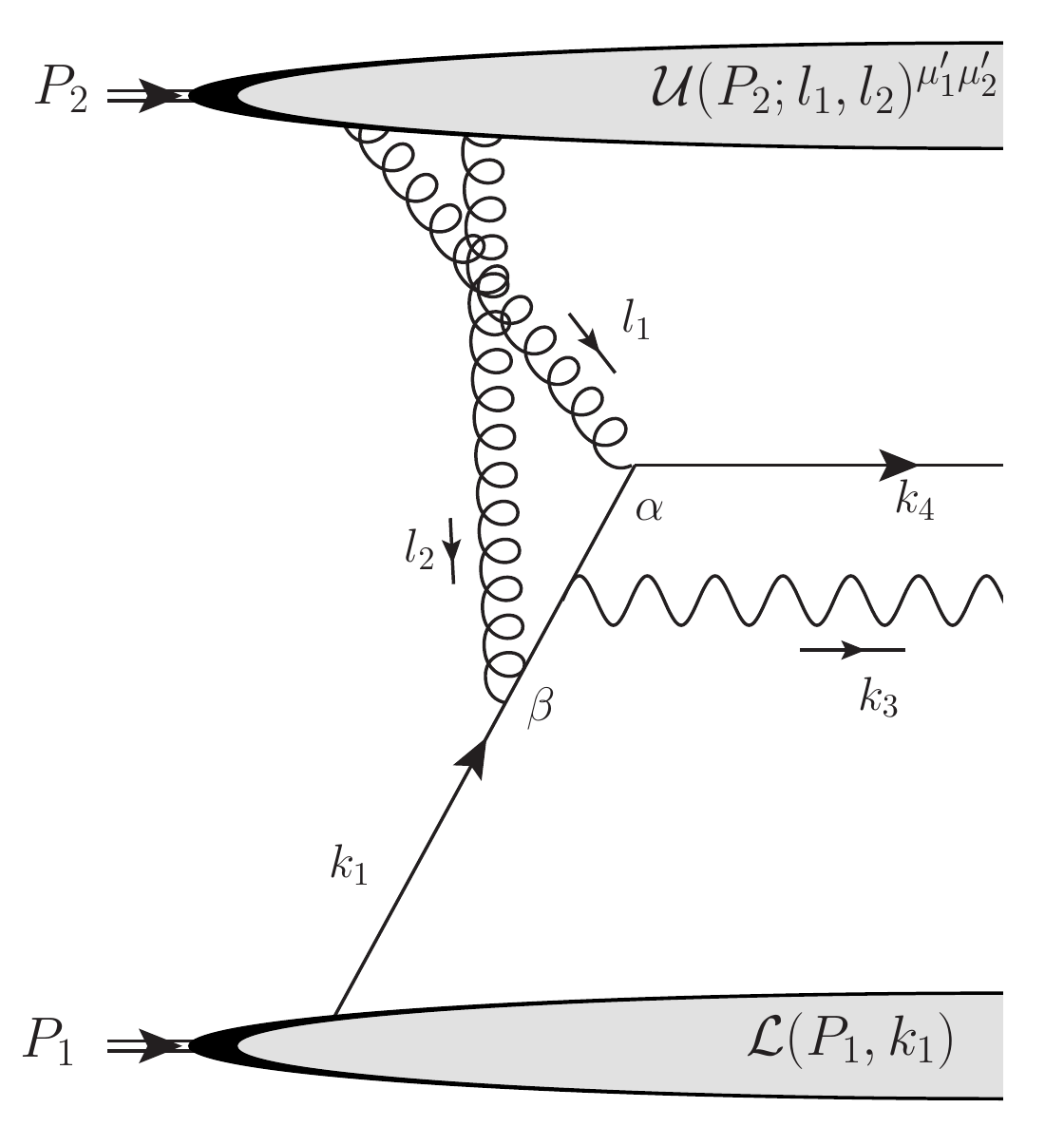}
    &
    \includegraphics[scale=0.4]{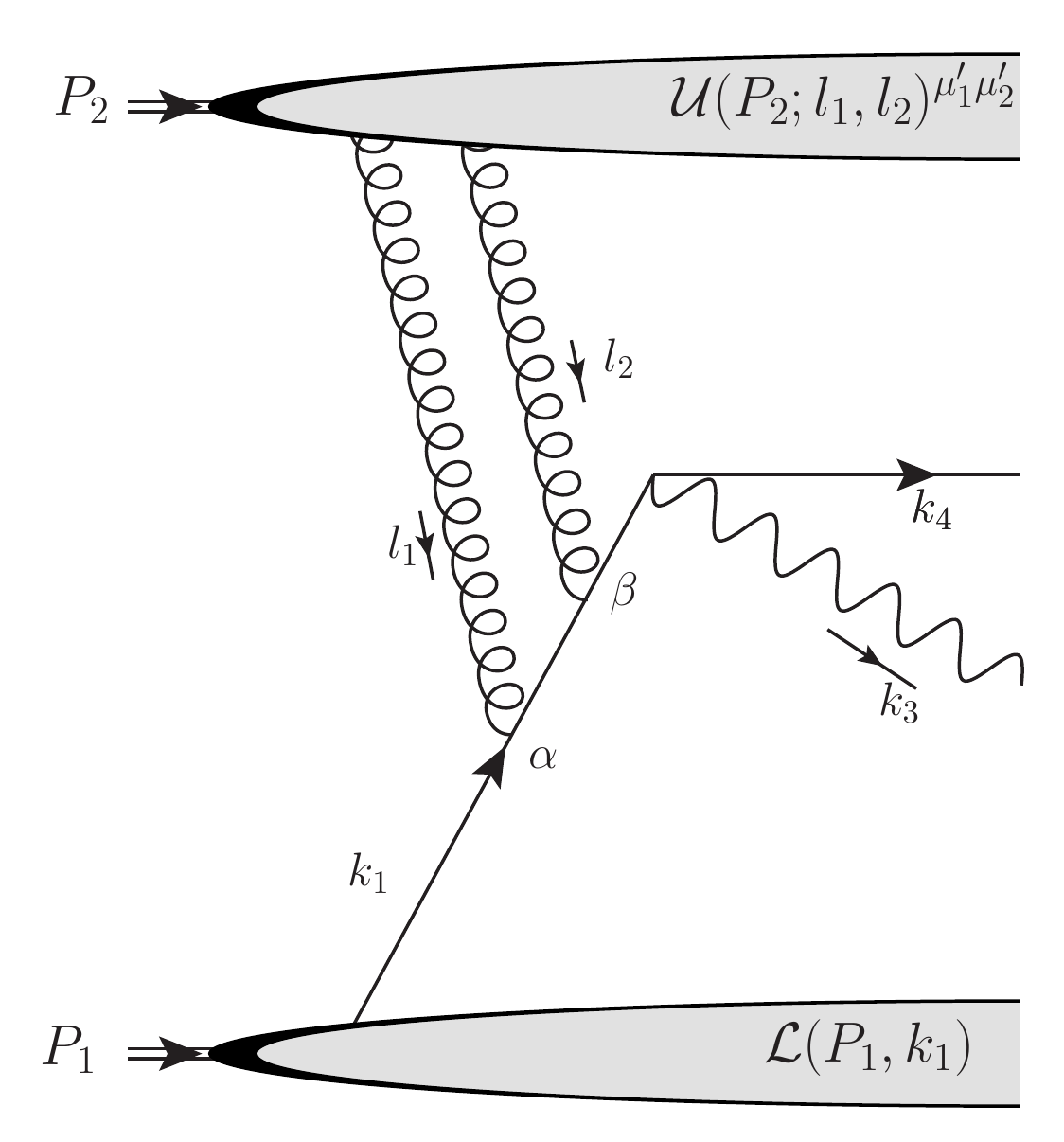}
\\
(c) & (d)
\\[7mm]
    \includegraphics[scale=0.4]{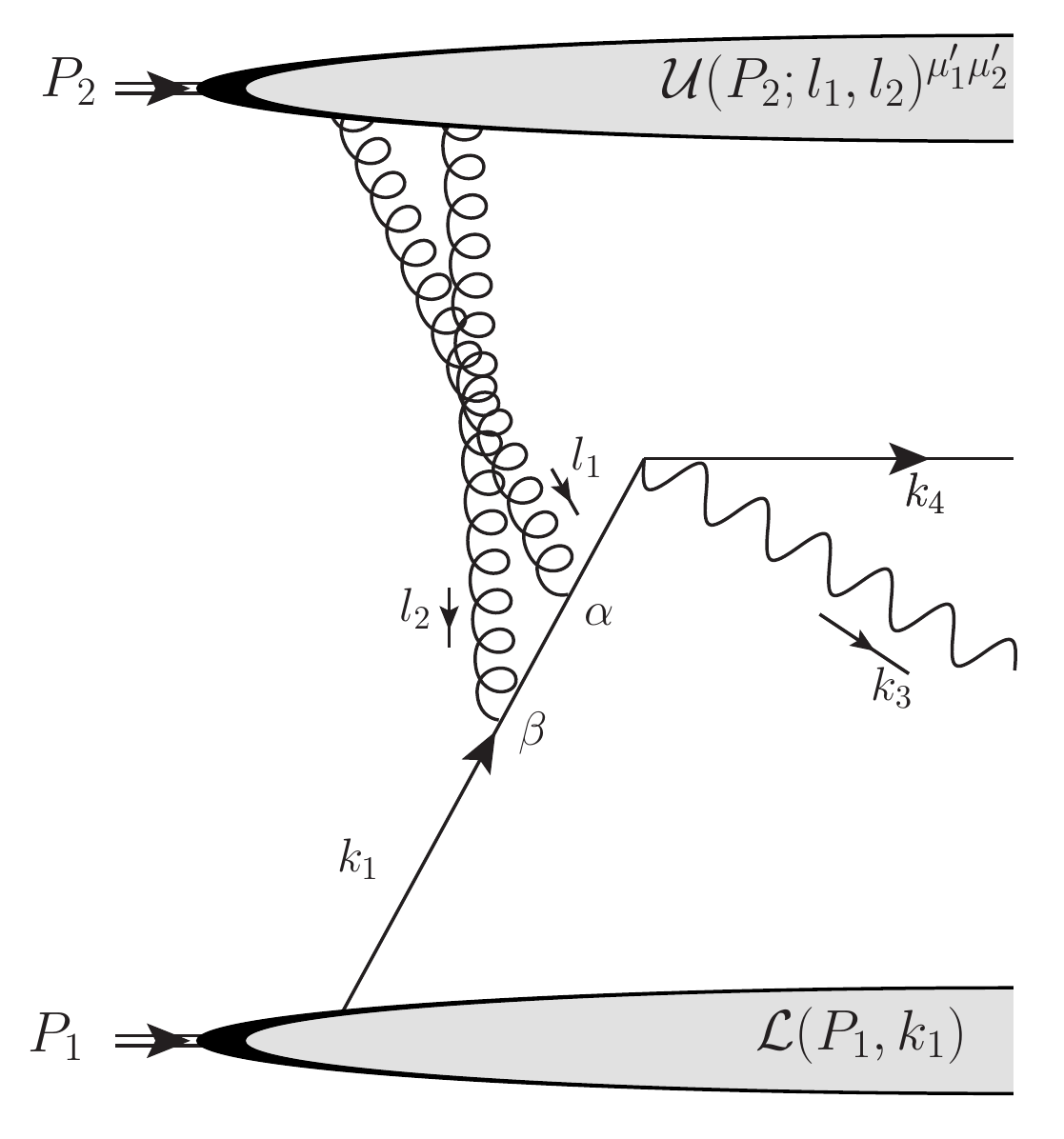}
    &
    \includegraphics[scale=0.4]{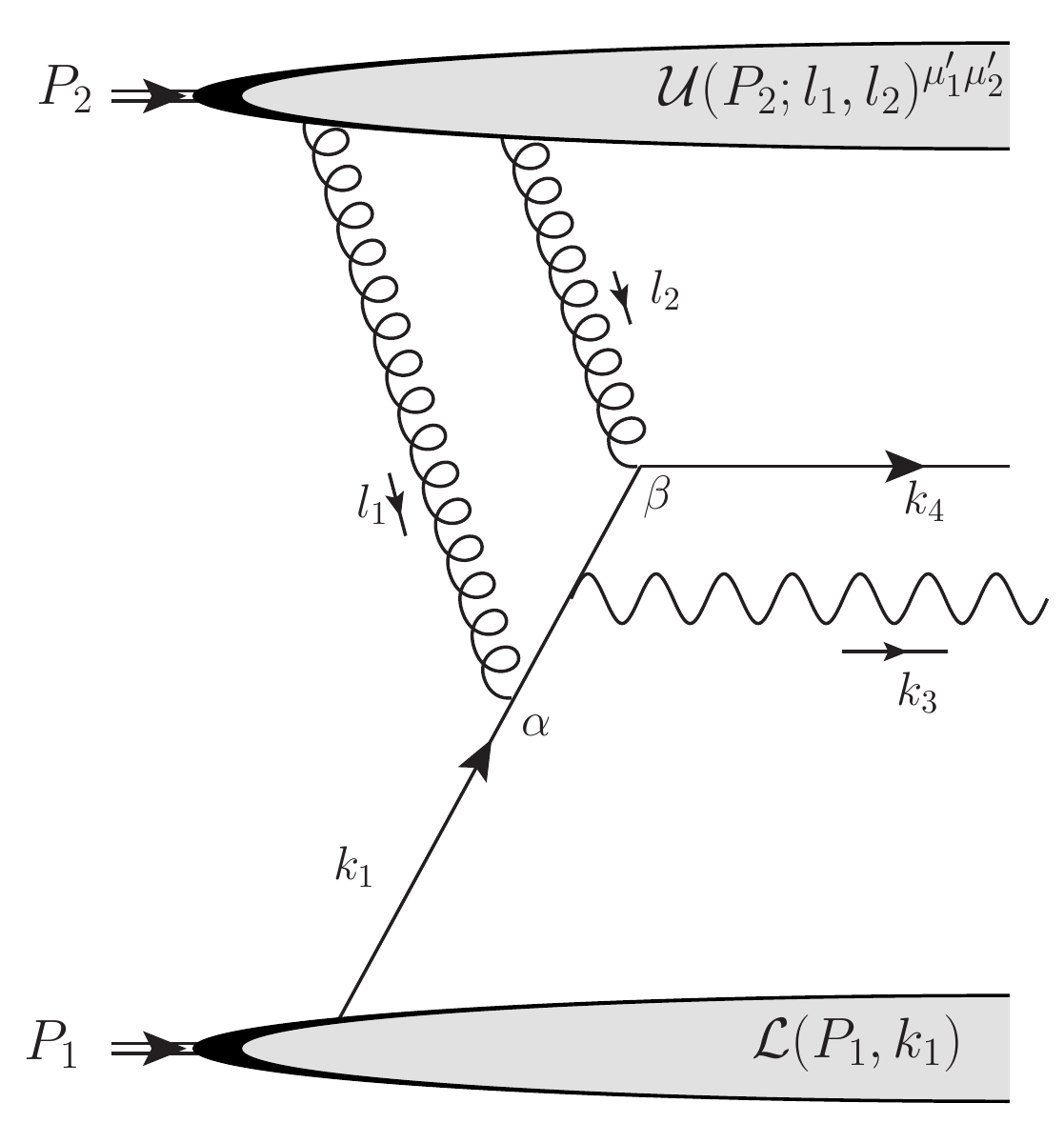}
\\
(e) & (f) 
\\[5mm]
  \end{tabular}
  \caption{Graphs with an extra gluon radiated from hadron-2.}
  \label{fig:twogluons}
\end{figure*}

The objective for the next two sections is to show, for the process of Sect.~\ref{sec:theprocess}, that there is a leading-power violation 
of the loose conditions for TMD-factorization from Sect.~\ref{sec:partonmodel}.
The demonstration will be for the case of a single extra soft gluon radiated from hadron-2 and attaching to active partons, 
with the relevant Feynman graphs at the amplitude level shown in Fig.~\ref{fig:twogluons}.

Graphs where one gluon couples to the other before the hard collision were already included in the 
single gluon case of Sect.~\ref{sec:onegluon}.
The possibility of TMD-factorization violation coming from spectator-spectator type interactions was 
addressed in Sect.~\ref{sec:spec}.  Spectator attachments in hadron-1 attaching to active partons 
are sensitive to non-perturbative structure inside $\lblob^\prime$.  
Therefore, we now focus only on gluons attaching the hadron-2 subgraph to active partons, as shown in the graphs in Fig.~\ref{fig:twogluons}.  
There are only gluon-fermion attachments in the hard part so ghosts do not contribute at this order.  

The steps in this section are nearly the same as those of Ref.~\cite{Collins:2008sg}, but are tailored to exhibit
the role of intrinsic transverse momentum.  

\subsection{Power Counting}
 \label{sec:twopowercounting}
For the two-gluon case, the separation into blocks according to Eq.~\eqref{eq:ampbasic} becomes
\begin{align}
\label{eq:m2gluon}
M_{(2g)} =  &  \; \ublob(P_2;l_1,l_2)^{\alpha \beta ; \mu_1^\prime \mu_2^\prime} \, g_{\mu_1^\prime \mu_1} \, 
g_{\mu_2^\prime \mu_2} \bar{u}(k_4) \, H_L(k_1,k_2,k_3,k_4;l_1,l_2)^{\alpha \beta ; \mu_1 \mu_2} \lblob(P_1;k_1) \nonumber \\ = & \;
 \ublob(P_2;l_1,l_2)^{\alpha \beta ; \mu_1^\prime \mu_2^\prime} \, g_{\mu_1^\prime \mu_1} \, g_{\mu_2^\prime \mu_2} R^{\alpha \beta ; \mu_1 \mu_2}(k_1,k_2,k_3,k_4;l_1,l_2,P_1) \, .
\end{align}
On the second line we have defined an $R^{\alpha \beta ; \mu_1 \mu_2}$ for the two $\ublob$-gluon case, analogous to Eq.~\eqref{eq:ronegluon} for the one $\ublob$-gluon case. 
With an extra soft/collinear gluon now entering the $H_L(k_1,k_3,k_4;l_1,l_2)^{\alpha \beta ; \mu_1 \mu_2}$ subgraph, there is another fermion propagator in the unfactorized hard subprocess, so 
each Lorentz component of $H_L(k_1,k_3,k_4;l_1,l_2)^{\alpha \beta ; \mu_1 \mu_2}$ now acquires an extra power of $1/Q$ relative to the one-gluon case in Eq.~\eqref{eq:1gluonh}.
The lower quark subgraph, $\lblob(P_1;k_1)$, is exactly the same as in Sect.~\ref{sec:onegluon}, 
so all components of $R^{\alpha \beta ; \mu_1 \mu_2}$ are now of size $\sim 1/Q$ (as usual, up to factors of order $\Lambda$ and logarithms from renormalization).
By contrast, the one-gluon case in Eq.~\eqref{eq:r1gluon} was order $\sim Q^0$.

The two-gluon subgraph $\ublob^{\alpha \beta ; \mu_1^\prime \mu_2^\prime}$ now has two Lorentz indices, and in frame-1 its highest 
power-law behavior is from the $\ublob^{- -} \sim Q^2$ components.  
The powers for the individual Lorentz components are 
\begin{align}
 \ublob^{--}  &\sim Q^2 \, , \label{eq:twolargest} \\
 \ublob^{-j} \sim  \ublob^{j-}  &\sim Q \, , \label{eq:twoleading} \\
\ublob^{ij}  \sim  \ublob^{+-}  \sim \ublob^{-+}  &\sim Q^0 \, , \label{eq:twosubleading} \\
\ublob^{j+}  \sim \ublob^{+j} &\sim 1/Q \, , \\
\ublob^{++} &\sim 1/Q^{2} \,.
\end{align}
The components in Eq.~\eqref{eq:twolargest} contribute superleading $\sim Q$ terms graph-by-graph in the amplitude; the $\ublob^{\alpha \beta ; - -} \sim Q^2$ 
components of the $\ublob^{\alpha \beta ; \mu_1^\prime \mu_2^\prime}$ subgraph multiply $R^{\alpha \beta ; \mu_1 \mu_2} \sim 1/Q$ components in Eq.~\eqref{eq:m2gluon}.
Recall that, while overall superleading contributions must cancel in the final analysis, superleading \emph{subdiagrams} can 
contribute uncanceled leading terms to the amplitude when they multiply subleading factors in the rest of the graph.    

Next we implement the Grammer-Yennie decomposition on Eq.~\eqref{eq:m2gluon}, writing 
\begin{eqnarray}
g_{\mu_1^\prime \mu_1} & = & K(l_1)_{\mu_1^\prime \mu_1} + G(l_1)_{\mu_1^\prime \mu_1} \, , \label{eq:gramyentwoa} \\
g_{\mu_2^\prime \mu_2} & = & K(l_2)_{\mu_2^\prime \mu_2} + G(l_2)_{\mu_2^\prime \mu_2} \, , \label{eq:gramyentwob}
\end{eqnarray}
with the $K$'s and $G$'s defined as in Sect.~\ref{sec:grammeryennie}, but now with the four-momenta that correspond to each gluon:
\begin{eqnarray}
G(l_1)_{\mu_1^\prime \mu_1} & = & g_{\mu_1^\prime \mu_1} -  \frac{n_{1 \, , \mu_1^\prime} l_{1 \, , \mu_1}}{n_1 \cdot l_1} \, ,\label{eq:g1def}  \\
G(l_2)_{\mu_2^\prime \mu_2} & = & g_{\mu_2^\prime \mu_2} -  \frac{n_{1 \, , \mu_2^\prime} l_{2 \, , \mu_2}}{n_1 \cdot l_2} \, , \label{eq:g2} \\
K(l_1)_{\mu_1^\prime \mu_1} & = & \frac{n_{1\, , \mu_1^\prime} l_{1 \, , \mu_1}}{n_1 \cdot l_1} \, , \\
K(l_2)_{\mu_2^\prime \mu_2} & = & \frac{n_{1\, , \mu_2^\prime} l_{2 \, , \mu_2}}{n_1 \cdot l_2} \, . \label{eq:k2}
\end{eqnarray}
Using Eqs.~(\ref{eq:gramyentwoa},~\ref{eq:gramyentwob}) in Eq.~\eqref{eq:m2gluon} separates $M_{(2g)}$ into the following set of $G$/$K$ terms:
\begin{multline}
\label{eq:gydecomptwo}
M_{(2g)} = \underbrace{\ublob(P_2;l_1,l_2)^{\alpha \beta ; \mu_1^\prime \mu_2^\prime}}_\text{{\small $\sim Q^2$}} \, \left(  G(l_1)_{\mu_1^\prime \mu_1}  G(l_2)_{\mu_2^\prime \mu_2} + 
G(l_1)_{\mu_1^\prime \mu_1} K(l_2)_{\mu_2^\prime \mu_2} + \right. \\ \left.  K(l_1)_{\mu_1^\prime \mu_1} G(l_2)_{\mu_2^\prime \mu_2} 
+ K(l_1)_{\mu_1^\prime \mu_1} K(l_2)_{\mu_2^\prime \mu_2}  \right)  \,  \underbrace{R^{\alpha \beta ; \mu_1 \mu_2}}_\text{{\small $\sim 1/Q$}} 
\; \sim Q \, .
\end{multline}
The power-laws indicated with each underbrace denote maximum powers term-by-term.  Thus, the largest contribution to the amplitude from any individual 
graph is order $\sim Q$, which is superleading.  As in Sect.~\ref{sec:onegluon}, we must verify the cancelation of such superleading terms in the sum of all graphs 
by analyzing each of the $G$/$K$ combinations.

\subsection{Basic Expression}
\label{sec:basic}
We begin by directly writing down the expression for $R^{\alpha \beta ; \mu_1 \mu_2}$ from the sum of diagrams in Fig.~\eqref{fig:twogluons}.
\begin{align}
\label{eq:rexpression}
R^{\alpha \beta ; \mu_1 \mu_2} =  
i e_q g_s^2 \, \bar{u}(k_4,S_4) \left\{
\frac{\gamma^{\mu_2} ( \slashed{k}_4 - \slashed{l}_2 + m_q) \gamma^{\mu_1} (\slashed{k}_4 - \slashed{k}_2  + m_q) \gamma^{\sigma} t^\beta t^\alpha}{\left[ (k_4 - l_2)^2 - m_q^2 + i0 \right] \left[ (k_4 - k_2)^2 - m_q^2 + i0 \right]} \right. + & && \text{Graph a.} \nonumber \\ + \left.
\frac{\gamma^{\mu_1} ( \slashed{k}_4 - \slashed{l}_1 + m_q) \gamma^{\mu_2} (\slashed{k}_4 - \slashed{k}_2 + m_q) \gamma^{\sigma} t^\alpha t^\beta }{\left[ (k_4 - l_1)^2 - m_q^2 + i0 \right] \left[ (k_4 - k_2 )^2 - m_q^2 + i0 \right]} \right. + & && \text{Graph b.} \nonumber \\  + \left. 
\frac{\gamma^{\mu_1} ( \slashed{k}_4 - \slashed{l}_1 + m_q) \gamma^{\sigma} (\slashed{k}_1 + \slashed{l}_2 + m_q) \gamma^{\mu_2} t^\alpha t^\beta }{\left[ (k_4 - l_1)^2 - m_q^2 + i0 \right] \left[ (k_1 + l_2 )^2 - m_q^2 + i0 \right]} \right. + & && \text{Graph c.} \nonumber\\  + \left. 
\frac{\gamma^\sigma ( \slashed{k}_4 + \slashed{k}_3 + m_q) \gamma^{\mu_2} (\slashed{k}_1 + \slashed{l}_1  + m_q) \gamma^{\mu_1} t^\beta t^\alpha}{\left[ (k_4 + k_3)^2 - m_q^2 + i0 \right] \left[ (k_1 + l_1)^2 - m_q^2 + i0 \right]} \right. + & && \text{Graph d.} \nonumber \\  + \left. 
\frac{\gamma^\sigma ( \slashed{k}_4 + \slashed{k}_3 + m_q) \gamma^{\mu_1} (\slashed{k}_1 + \slashed{l}_2  + m_q) \gamma^{\mu_2} t^\alpha t^\beta }{\left[ (k_4 + k_3)^2 - m_q^2 + i0 \right] \left[ (k_1 + l_2)^2 - m_q^2 + i0 \right]} \right. +  & && \text{Graph e.} \nonumber \\ + \left.
\frac{\gamma^{\mu_2} ( \slashed{k}_4 - \slashed{l}_2 + m_q) \gamma^{\sigma} (\slashed{k}_1 + \slashed{l}_1 + m_q) \gamma^{\mu_1} t^\beta t^\alpha }{\left[ (k_4 - l_2)^2 - m_q^2 + i0 \right] \left[ (k_1 + l_1)^2 - m_q^2 + i0 \right]} \right\} \times  & && \text{Graph f.} \nonumber \\ \times
\lblob(P_1;k_1) \; .
\end{align}
For ease of reference, we have labeled each term in braces with its corresponding diagram from Fig.~\ref{fig:twogluons}.
Note that we have restored the $S_4$ argument of the outgoing quark wavefunction.  

\subsection{Glauber regions}
\label{sec:l2glaub}

The Glauber regions correspond to intermediate states in the hard subgraph going on shell,
so $l_2^-$ and $l_1^-$ must be deformed far into the complex plane, as reviewed in Sect.~\ref{sec:glauber}, to ensure that $\left| l_2^- \right|$ and $\left| l_1^- \right|$
may treated as order $\sim Q$.

In the two-gluon case of Fig.~\ref{fig:twogluons}, only one gluon at a time can be Glauber.  If both gluons are Glauber, then $k_2^- << k_{2 t}$, which is forbidden by 
the kinematical restriction $k_2^- \sim Q$.

As an example, assume the case where $l_1^-$ is outside the Glauber and consider the integration on $l_2^-$.
For the final state $l_2$-interactions in Figs.~\ref{fig:twogluons}(a,f), the deformations away from the $l_2^-$-Glauber poles are downward, 
into the negative half of the complex plane, while for the initial state $l_2^-$-interactions in Figs.~\ref{fig:twogluons}(c,e), the deformations are upward.
Therefore, in Figs.~\ref{fig:twogluons}(a,f) 
we apply the downward deformation on $l_2^-$ until $\left| l_2^- \right| \sim Q$, and we 
may replace $l_2^-$ by $(l_2^- - i0)$ in Eqs.~(\ref{eq:g2},~\ref{eq:k2}):
\begin{eqnarray}
\left[ G(l_2)_{\mu_2^\prime \mu_2} \right]_{\rm F.S.} & = & g_{\mu_2^\prime \mu_2} -  \frac{n_{1 \, , \mu_2^\prime} l_{2 \, , \mu_2}}{l_2^- - i0} \, , \label{eq:g23} \\
\left[ K(l_2)_{\mu_2^\prime \mu_2} \right]_{\rm F.S.} & = & \frac{n_{1\, , \mu_2^\prime} l_{2 \, , \mu_2}}{l_2^- - i0} \, . \label{eq:k23}
\end{eqnarray}
The subscripts ``${\rm F.S.}$" are to symbolize that 
these are the $G(l_2)$ and $K(l_2)$ expressions we will use in graphs with final state interactions and downwardly deformed $l_2$ 
contours.

For the initial state $l_2$-interactions in Figs.~\ref{fig:twogluons}(c,e), we implement the upward deformations on $l_2^-$ until $|l_2^-| \sim Q$.  In that case we use
\begin{eqnarray}
\left[ G(l_2)_{\mu_2^\prime \mu_2} \right]_{\rm I.S.} & = & g_{\mu_2^\prime \mu_2} -  \frac{n_{1 \, , \mu_2^\prime} l_{2 \, , \mu_2}}{l_2^- + i0} \, , \label{eq:g22} \\
\left[ K(l_2)_{\mu_2^\prime \mu_2} \right]_{\rm I.S.} & = & \frac{n_{1\, , \mu_2^\prime} l_{2 \, , \mu_2}}{l_2^- + i0} \, . \label{eq:k22}
\end{eqnarray}
The subscripts ``${\rm I.S.}$" are to symbolize that 
these are the $G(l_2)$ and $K(l_2)$ expressions we will use in graphs with initial state interactions and upwardly deformed $l_2$ 
contours.

The requirement is that the signs on the $i0$'s near the $l_2^- = 0$ poles must be such that they do not obstruct the required 
deformations on $l_2^-$ away from the Glauber region.
Expressions exactly analogous to Eqs.~(\ref{eq:g23},~\ref{eq:k22}) apply for deformations of $l_1$ out of the Glauber region.

Note that the directions of the deformations are different depending on whether the extra gluon attachments 
are in the initial or final state.  So arguments using Glauber deformations need to be applied separately to each graph.

\subsection{$G(l_1)G(l_2)$ Terms}
Recall from the one-gluon case in Eq.~\eqref{eq:leadingonegluon} that the $G^{\mu^\prime \mu}$ factors projected only components of $\ublob$ that were power suppressed 
relative to the largest components.  In this subsection, we show that this generalizes to the two-gluon case, so the $G(l_1) G(l_2)$ terms are doubly suppressed relative to 
the largest terms and contribute only subleading terms to $M_{(2g)}$.  

If we apply Eqs.~(\ref{eq:g23},~\ref{eq:g22}), with the deformed contours, to Eq.~\eqref{eq:rexpression} and recall 
Eqs.~(\ref{eq:twolargest}--\ref{eq:twosubleading}), then we may identify the leading power-law for each contribution from the 
$G(l_1)  G(l_2)$ projection:
\begin{align}
\left[  \vphantom{\ublob(P_1;l_1,l_2)^{\alpha \beta ; \mu_1^\prime \mu_2^\prime} } \right.
\ublob(P_1;l_1,l_2)^{\alpha \beta ; \mu_1^\prime \mu_2^\prime} \; G(l_1)_{\mu_1^\prime \mu_1}  & G(l_2)_{\mu_2^\prime \mu_2} \;  R(k_1,k_2,k_3,k_4;l_1,l_2,P_1)^{\alpha \beta ; \mu_1 \mu_2} 
\left. \vphantom{\ublob(P_1;l_1,l_2)^{\alpha \beta ; \mu_1^\prime \mu_2^\prime} } \right]_{\rm I.S. / F.S} \nonumber \\
\; \nonumber \\
& \sim  \underbrace{\ublob(P_1;l_1,l_2)^{\alpha \beta ; ij}}_\text{{\small $\sim Q^0$}} \; \underbrace{R(k_1,k_2,k_3,k_4;l_1,l_2,P_1)^{\alpha \beta ; i j}}_\text{{\small $\sim 1/Q$}} \nonumber \\
\; \nonumber \\
& \sim  \underbrace{\ublob(P_1;l_1,l_2)^{\alpha \beta ; -j}}_\text{{\small $\sim Q$}} \; \underbrace{(l_1^i/(l_1^- \pm i0)) R(k_1,k_2,k_3,k_4;l_1,l_2,P_1)^{\alpha \beta ; i j}}_\text{{\small $\sim 1/Q^2$}} \nonumber \\
\; \nonumber \\
& \sim  \underbrace{\ublob(P_1;l_1,l_2)^{\alpha \beta ; i-}}_\text{{\small $\sim Q$}} \; \underbrace{(l_2^j/(l_2^-\pm i0)) R(k_1,k_2,k_3,k_4;l_1,l_2,P_1)^{\alpha \beta ; i j}}_\text{{\small $\sim 1/Q^2$}} \nonumber \\
\; \nonumber \\
& \sim  \underbrace{\ublob(P_1;l_1,l_2)^{\alpha \beta ; --}}_\text{{\small $\sim Q^2$}} \; \underbrace{(l_1^i/(l_1^- \pm i0)) (l_2^j/(l_2^-\pm i0)) R(k_1,k_2,k_3,k_4;l_1,l_2,P_1)^{\alpha \beta ; i j}}_\text{{\small $\sim 1/Q^3$}} & \nonumber \\
\, \nonumber \\
& \sim  1/Q   \, . &  \label{eq:GG}
\end{align}
Here we have written the largest powers, of order $\sim 1/Q$, in the contraction with the $G(l_1)G(l_2)$ term.
The ``${\rm I.S. / F.S}$" subscript is included 
to symbolize that we apply the $l$-contour deformations upward or downward appropriately in each term
in accordance with whether the gluon attaches to the initial or final quark.  

The leading powers in $M_{(2g)}$ are of order $\sim Q^0$ while the largest powers in Eq.~\eqref{eq:GG} are $\sim 1/Q$.
So the $G(l_1)  G(l_2)$ terms are subleading.  Notice that the validity of the argument relies on the use of 
deformed $l_1^-$ and $l_2^-$ contours.

\subsection{$K(l_2)$ Terms}
Next we write the general expression for the case where one gluon is a $K$-gluon.
Applying  Eqs.~(\ref{eq:k23},\ref{eq:k22}) to Eq.~\eqref{eq:rexpression} gives
\begin{align}
\left[ K(l_2)_{\mu_2^\prime \mu_2}  R^{\alpha \beta ; \mu_1 \mu_2} \right]_{\rm I.S. / F.S} = & \nonumber \\ 
\; \nonumber \\
 i e_q g_s^2 \, \underbrace{\bar{u}(k_4,S_4)}_\text{{\small $\sim  Q^{1/2}  $} }  \, & \underbrace{\left\{
\frac{\slashed{l}_2 ( \slashed{k}_4 - \slashed{l}_2 + m_q) \gamma^{\mu_1} (\slashed{k}_4 - \slashed{k}_2  + m_q) \gamma^{\sigma} t^\beta t^\alpha}{\left[ (k_4 - l_2)^2 - m_q^2 + i0 \right] \left[ (k_4 - k_2)^2 - m_q^2 + i0 \right]} \times \frac{1}{l_2^- - i0} \right. }_\text{{\small $\sim 1/Q^2$} }   & && \text{(a.)} \nonumber \\  
\; \nonumber \\
+ &
\underbrace{\left.
\frac{\gamma^{\mu_1} ( \slashed{k}_4 - \slashed{l}_1 + m_q) \slashed{l}_2 (\slashed{k}_4 - \slashed{k}_2 + m_q) \gamma^{\sigma}  t^\alpha t^\beta}{\left[ (k_4 - l_1)^2 - m_q^2 + i0 \right] \left[ (k_4 - k_2 )^2 - m_q^2 + i0 \right]} \times \frac{1}{l_2^-} \right.}_\text{{\small $\sim 1/Q^2$} } & && \text{(b.)} \nonumber \\    
\; \nonumber \\
+ & 
\underbrace{\left. 
\frac{\gamma^{\mu_1} ( \slashed{k}_4 - \slashed{l}_1 + m_q) \gamma^{\sigma} (\slashed{k}_1 + \slashed{l}_2 + m_q) \slashed{l}_2  t^\alpha t^\beta}{\left[ (k_4 - l_1)^2 - m_q^2 + i0 \right] \left[ (k_1 + l_2 )^2 - m_q^2 + i0 \right]} \times \frac{1}{l_2^- + i0} \right.}_\text{{\small $\sim 1/Q^2$} } & && \text{(c.)} \nonumber  \\  
\; \nonumber \\
 + & \underbrace{\left. 
\frac{\gamma^\sigma ( \slashed{k}_4 + \slashed{k}_3 + m_q) \slashed{l}_2 (\slashed{k}_1 + \slashed{l}_1  + m_q) \gamma^{\mu_1} t^\beta t^\alpha}{\left[ (k_4 + k_3)^2 - m_q^2 + i0 \right] \left[ (k_1 + l_1)^2 - m_q^2 + i0 \right]} \times \frac{1}{l_2^-} \right.}_\text{{\small $\sim 1/Q^2$} } & && \text{(d.)} \nonumber  \\   
\; \nonumber \\
+ &
\underbrace{\left. 
\frac{\gamma^\sigma ( \slashed{k}_4 + \slashed{k}_3 + m_q) \gamma^{\mu_1} (\slashed{k}_1 + \slashed{l}_2  + m_q) \slashed{l}_2  t^\alpha t^\beta}{\left[ (k_4 + k_3)^2 - m_q^2 + i0 \right] \left[ (k_1 + l_2)^2 - m_q^2 + i0 \right]} \times \frac{1}{l_2^- + i0} \right.}_\text{{\small $\sim 1/Q^2$} } & && \text{(e.)} \nonumber   \\ 
\; \nonumber \\
+ &
\underbrace{\left.
\frac{\slashed{l}_2 ( \slashed{k}_4 - \slashed{l}_2 + m_q) \gamma^{\sigma} (\slashed{k}_1 + \slashed{l}_1 + m_q) \gamma^{\mu_1} t^\beta t^\alpha}{\left[ (k_4 - l_2)^2 - m_q^2 + i0 \right] \left[ (k_1 + l_1)^2 - m_q^2 + i0 \right]} \times \frac{1}{l_2^- - i0} \right\}}_\text{{\small $\sim 1/Q^2$} } & && \text{(f.)} \nonumber   \\ & & \times
 n_{1\, \mu_2^\prime} \underbrace{\lblob(P_1;k_1)}_\text{{\small $\sim Q^{1/2}$} }  \, .   \label{eq:rexpressionk2}
\end{align} 
In Figs.~\ref{fig:twogluons}(b,d), $l_2$ attaches to an internal hard propagator so no intermediate line in the hard subgraph 
goes on shell when $l_2$ approaches the Glauber region;  there are no Glauber poles and so no need for replacements like Eqs.~(\ref{eq:g23}-\ref{eq:k22}).

\subsection{$G(l_1) K(l_2)$ Terms}
The ``${\mu_1}$" index in Eq.~\eqref{eq:rexpressionk2} is ultimately to be contracted with  
a $K(l_1)_{\mu_1^\prime \mu_1}$ and a $G(l_1)_{\mu_1^\prime \mu_1}$ in the amplitude in Eq.~\eqref{eq:gydecomptwo}.  We first consider the $G(l_1)_{\mu_1^\prime \mu_1}$ contraction:
 \begin{align}
\left[ M_{(2g)} \right]_{G(l_1) K(l_2)} & =  \left[ \ublob(P_2;l_1,l_2)^{\alpha \beta ; \mu_1^\prime \mu_2^\prime} 
 G(l_1)_{\mu_1^\prime \mu_1} K(l_2)_{\mu_2^\prime \mu_2}  R^{\alpha \beta ; \mu_1 \mu_2} \right]_{\rm I.S. / F.S.} \nonumber \\ 
\; \nonumber \\ & =  
 i e_q g_s^2 \, \ublob(P_2;l_1,l_2)^{\alpha \beta ; \mu_1^\prime \, , -}  \underbrace{\bar{u}(k_4,S_4)}_\text{{\small $\sim Q^{1/2}$} }  \times \nonumber \\ 
\; \nonumber \\  &  \times \underbrace{\left\{
\left( g_{\mu_1^\prime \mu_1} -  \frac{n_{1 \, , \mu_1^\prime} l_{1 \, , \mu_1}}{l_1^-} \right) \frac{\slashed{l}_2 ( \slashed{k}_4 - \slashed{l}_2 + m_q) \gamma^{\mu_1} (\slashed{k}_4 - \slashed{k}_2  + m_q) \gamma^{\sigma} t^\beta t^\alpha}{\left[ (k_4 - l_2)^2 - m_q^2 + i0 \right] \left[ (k_4 - k_2)^2 - m_q^2 + i0 \right]} \times \frac{1}{l_2^- - i0} \right. }_\text{{\small $ \sim 1/Q^2$} }   &  \text{(a.)} \nonumber \\  
\; \nonumber \\
& \; + 
\underbrace{\left.
\left( g_{\mu_1^\prime \mu_1} -  \frac{n_{1 \, , \mu_1^\prime} l_{1 \, , \mu_1}}{l_1^- - i0} \right) \times \frac{\gamma^{\mu_1} ( \slashed{k}_4 - \slashed{l}_1 + m_q) \slashed{l}_2 (\slashed{k}_4 - \slashed{k}_2 + m_q) \gamma^{\sigma}  t^\alpha t^\beta}{\left[ (k_4 - l_1)^2 - m_q^2 + i0 \right] \left[ (k_4 - k_2 )^2 - m_q^2 + i0 \right]} \times \frac{1}{l_2^-} \right.}_\text{{\small $\sim 1/Q^2$} } & \text{(b.)} \nonumber \\    
\; \nonumber \\
& \; + 
\underbrace{\left. 
\left( g_{\mu_1^\prime \mu_1} -  \frac{n_{1 \, , \mu_1^\prime} l_{1 \, , \mu_1}}{l_1^- - i0} \right) \times \frac{\gamma^{\mu_1} ( \slashed{k}_4 - \slashed{l}_1 + m_q) \gamma^{\sigma} (\slashed{k}_1 + \slashed{l}_2 + m_q) \slashed{l}_2  t^\alpha t^\beta}{\left[ (k_4 - l_1)^2 - m_q^2 + i0 \right] \left[ (k_1 + l_2 )^2 - m_q^2 + i0 \right]} \times \frac{1}{l_2^- + i0} \right.}_\text{{\small $\sim 1/Q^2$} } & \text{(c.)} \nonumber  \\  
\; \nonumber \\
& \; +  \underbrace{\left. 
\left( g_{\mu_1^\prime \mu_1} -  \frac{n_{1 \, , \mu_1^\prime} l_{1 \, , \mu_1}}{l_1^- + i0} \right) \times \frac{\gamma^\sigma ( \slashed{k}_4 + \slashed{k}_3 + m_q) \slashed{l}_2 (\slashed{k}_1 + \slashed{l}_1  + m_q) \gamma^{\mu_1} t^\beta t^\alpha}{\left[ (k_4 + k_3)^2 - m_q^2 + i0 \right] \left[ (k_1 + l_1)^2 - m_q^2 + i0 \right]} \times \frac{1}{l_2^-} \right.}_\text{{\small $\sim 1/Q^2$} } & \text{(d.)} \nonumber  \\   
\; \nonumber \\
& \; + 
\underbrace{\left. 
\left( g_{\mu_1^\prime \mu_1} -  \frac{n_{1 \, , \mu_1^\prime} l_{1 \, , \mu_1}}{l_1^-} \right) \times \frac{\gamma^\sigma ( \slashed{k}_4 + \slashed{k}_3 + m_q) \gamma^{\mu_1} (\slashed{k}_1 + \slashed{l}_2  + m_q) \slashed{l}_2  t^\alpha t^\beta}{\left[ (k_4 + k_3)^2 - m_q^2 + i0 \right] \left[ (k_1 + l_2)^2 - m_q^2 + i0 \right]} \times \frac{1}{l_2^- + i0} \right.}_\text{{\small $\sim 1/Q^2$} } & \text{(e.)} \nonumber   \\ 
\; \nonumber \\
& \; + 
\underbrace{\left.
\left( g_{\mu_1^\prime \mu_1} -  \frac{n_{1 \, , \mu_1^\prime} l_{1 \, , \mu_1}}{l_1^- + i0} \right) \times \frac{\slashed{l}_2 ( \slashed{k}_4 - \slashed{l}_2 + m_q) \gamma^{\sigma} (\slashed{k}_1 + \slashed{l}_1 + m_q) \gamma^{\mu_1} t^\beta t^\alpha}{\left[ (k_4 - l_2)^2 - m_q^2 + i0 \right] \left[ (k_1 + l_1)^2 - m_q^2 + i0 \right]} \times \frac{1}{l_2^- - i0} \right\}}_\text{{\small $\sim 1/Q^2$} } & \text{(f.)} \nonumber \\ & \times
\underbrace{\lblob(P_1;k_1)}_\text{{\small $\sim Q^{1/2}$} }  \, .   \label{eq:rexpressionk2g1}
\end{align}
Here we have completed the contraction with $\ublob(P_2;l_1,l_2)^{\alpha \beta ; \mu_1^\prime \mu_2^\prime}$ in Eq.~\eqref{eq:gydecomptwo} to get the full $G(l_1) K(l_2)$ contribution to the amplitude.

The next step is to exploit the cascade of cancellations that results when $\slashed{l}_2$ is substituted with the following Feynman replacement identities:
\begin{align}
& \slashed{l}_2 = - (\slashed{k}_4 - \slashed{l}_2 - m_q) + (\slashed{k}_4 - m_q)  & &&   &\text{In \, graphs \, (a.) \& (f.)} &   \label{eq:replacementaf1} \\
& \slashed{l}_2 =  (\slashed{l}_2 + \slashed{k}_1 - m_q) - (\slashed{k}_1 - m_q)  & && &\text{In \, graphs \, (c.) \& (e.)} & \label{eq:replacementaf2} \\
& \slashed{l}_2 =  -(\slashed{k}_4 - \slashed{k}_2 - m_q) + (\slashed{k}_4 - \slashed{l}_1 - m_q)  & && &\text{In \, graph  \, \, (b.)  \; \, \, \, } & \label{eq:replacementaf3} \\
& \slashed{l}_2 =  (\slashed{k}_3 + \slashed{k}_4 - m_q) - (\slashed{k}_1 + \slashed{l}_1 - m_q)\,.  & && &\text{In \, graph  \, \, (d.) \; \, \,  \, } & \label{eq:replacementaf4}
\end{align}
Furthermore, Eq.~\eqref{eq:rexpressionk2g1} may be separated into real and imaginary contributions by using the familiar substitution by distributions
convenient for evaluating integrals:
\begin{eqnarray}
\frac{1}{l_{1,2}^- + i0} \to {\rm P.V.} \frac{1}{l_{1,2}^-} - i \pi  \delta(l_{1,2}^-) \, , \label{eq:finalstate}  \\
\frac{1}{l_{1,2}^- - i0} \to {\rm P.V.} \frac{1}{l_{1,2}^-} + i \pi \delta(l_{1,2}^-)\, , \label{eq:initialstate} 
\end{eqnarray}
where ``${\rm P.V.}$" is the symbol for the principal value distribution.  
Thus, each $1/(l_{1,2}^- \pm i0)$ eikonal factor is rewritten as a sum or difference of two distributions, 
with the relative signs depending on the direction of the contour deformations. 

For this paper, the main goal is to demonstrate the appearance of a TMD-factorization breaking double spin asymmetry with $S_4$ and $\lambda_3$ in the final state.  
In the squared the amplitude, this arises at lowest order from the cross terms between the graphs of 
Fig.~\ref{fig:twogluons} and the lowest order amplitudes in Fig.~\ref{fig:onegluon}.  
With just one extra gluon, it is the real 
terms in Eqs.~(\ref{eq:finalstate},\ref{eq:initialstate}) which yield real terms in the double spin dependent cross 
section with a violation of the minimal TMD-factorization criteria.  Therefore, to get the double spin asymmetry we focus attention on the principal value 
contributions.\footnote{At this order the imaginary 
parts in Eqs.~(\ref{eq:finalstate},\ref{eq:initialstate}) give single spin asymmetries.  However, for the one-extra-gluon example there is no imaginary contribution
that violates \emph{maximally} generalized TMD-factorization.}

Using Eqs.~(\ref{eq:replacementaf1}-\ref{eq:replacementaf4}) in Eq.~\eqref{eq:rexpressionk2g1}, we obtain 
\begin{multline}
\left[ M_{(2g)} \right]_{G(l_1) K(l_2), \, {\rm P.V.}}  = 
 -e_q g_s^2 f^{\alpha \beta \kappa} \, \underbrace{\bar{u}(k_4,S_4)}_\text{{\small $\sim Q^{1/2} $}} \underbrace{\ublob(P_2;l_1,l_2)^{\alpha \beta ; \mu_1^\prime \, , -} \left( g_{\mu_1^\prime \rho} -  n_{1 \, , \mu_1^\prime} l_{1 \, , \rho} \; {\rm P.V.} \frac{1}{l_{1}^-} \right)}_\text{{\small $\sim Q $}} \underbrace{ {\rm P.V.} \frac{1}{l_2^-} }_\text{{\small $\sim1/Q  $}} \\ \\ \times  \underbrace{\left\{
 \frac{\gamma^{\rho} (\slashed{k}_4 - \slashed{k}_2  + m_q) \gamma^{\sigma} t^\kappa}{(k_4 - k_2)^2  - m_q^2 + i0} \right.  \left.  
  + \frac{\gamma^{\sigma} (\slashed{k}_4 + \slashed{k}_3  + m_q) \gamma^{\rho} t^\kappa}{(k_4 + k_3)^2  - m_q^2 + i0} \right\}}_\text{{\small $\sim 1/Q $}} 
\; \underbrace{\lblob(P_1;k_1)}_\text{{\small $\sim Q^{1/2} $}} 
\; + \; \mathcal{O}(1/Q) \, .   \label{eq:rexpressionfa2}
\end{multline}
Here we have combined terms using the non-Abelian commutation relation $[ t^\alpha , t^\beta ] = i f^{\alpha \beta \kappa} t^\kappa$.  In addition, we have 
used again the result from Eqs.~(\ref{eq:k2sub},~\ref{eq:k2subb}) that the action of $\slashed{k}_1 - m_q$  
on $\lblob(P_1;k_1)$ only gives 
terms suppressed by two powers of $Q$.
Equation~\eqref{eq:rexpressionfa2} is now almost in the form of the leading order result in Eq.~\eqref{eq:1gluonampresult}, with the leading order hard part, Eq.~\eqref{eq:treehardpart},
and $\lblob(P_1;k_1)$ factored away from each other on the second line.  The only difference now from Eq.~\eqref{eq:1gluonampresult} is in the details of the factor associated with the upper bubble in the first 
line of Eq.~\eqref{eq:rexpressionfa2}.

In the contraction with $\ublob(P_2;l_1,l_2)^{\alpha \beta ; \mu_1^\prime \, , -}$, only the transverse components of $\mu_1^\prime$ are non-suppressed.   The plus components cancel between the 
two terms in $G(l_1)_{\mu_1^\prime \mu_1}$, and the minus components are suppressed by an extra power of $Q$.  Therefore, we may make the replacements $l_{1 \, \mu_1^\prime} \to l_{1t \, \mu_1^\prime}$ 
and $g_{\mu_1^\prime \rho} \to g^t_{\mu_1^\prime \rho}$  in Eq.~\eqref{eq:rexpressionfa2} without introducing any leading-power errors and rewrite it as
\begin{multline}
\left[ M_{(2g)} \right]_{G(l_1) K(l_2), \, {\rm P.V.}}  = 
 -e_q g_s^2 f^{\alpha \beta \kappa} \, \underbrace{\bar{u}(k_4,S_4)}_\text{{\small $\sim Q^{1/2} $}} \underbrace{\ublob(P_2;l_1,l_2)^{\alpha \beta ; \mu_1^\prime \, , -} \left( g^t_{\mu_1^\prime \rho} -  n_{1 \, , \mu_1^\prime} l_{1t \, , \rho} {\rm P.V.} \frac{1}{l_{1}^-} \right)}_\text{{\small $\sim Q $}} \underbrace{ {\rm P.V.} \frac{1}{l_2^-} }_\text{{\small $\sim 1/Q  $}} \\ \\ \times  
 \underbrace{H_{\rm LO}(k_1,k_2,k_3,k_4)^{\kappa ; \rho \sigma}}_\text{{\small $\sim 1/Q $}} \;\;
\underbrace{\lblob(P_1;k_1)}_\text{{\small $\sim Q^{1/2} $}} 
\; + \; \mathcal{O}(1/Q) \, .   \label{eq:rexpressionfa2b}
\end{multline}
Hence, the $G(l_1) K(l_2)$ terms are leading power.

\subsection{$K(l_1) G(l_2)$ Terms}
The treatment of the $K(l_1) G(l_2)$ terms is an exact mirror of the previous subsection, but with the roles of $l_1$ and $l_2$ switched.  
Since this also involves an interchange of the $\alpha$ and $\beta$ color indices, there is a sign reversal 
relative to Eq.~\eqref{eq:rexpressionfa2} due to the anti-symmetry of $f^{\alpha \beta \kappa}$:
\begin{multline}
\left[ M_{(2g)} \right]_{K(l_1) G(l_2), \, {\rm P.V.}}  = 
 e_q g_s^2 f^{\alpha \beta \kappa} \, \underbrace{\bar{u}(k_4,S_4)}_\text{{\small $\sim Q^{1/2}$}} \underbrace{\ublob(P_2;l_1,l_2)^{\alpha \beta ; - \, , \mu_2^\prime} \left( g^t_{\mu_2^\prime \rho} -  n_{1 \, , \mu_2^\prime} l_{2t \, , \rho} {\rm P.V.} \frac{1}{l_{2}^-} \right)}_\text{{\small $\sim Q$}} \underbrace{ {\rm P.V.} \frac{1}{l_1^-} }_\text{{\small $\sim 1/Q$}} \\ \\ \times  \underbrace{H_{\rm LO}(k_1,k_2,k_3,k_4)^{\kappa ; \rho \sigma}}_\text{{\small $\sim 1/Q $}} \; \;
 \underbrace{\lblob(P_1;k_1)}_\text{{\small $\sim Q^{1/2}$}} \; 
+ \;  \mathcal{O}(1/Q) \, .   \label{eq:rexpressionfa3}
\end{multline}

 \subsection{$K(l_1) K(l_2)$ Terms}
 Extra caution is needed in treating the 
 $K(l_1) K(l_2)$ contributions because they project terms that are individually superleading.
 To get the $K(l_1) K(l_2)$ terms, we contract Eq.~\eqref{eq:rexpressionk2} with $K(l_1)_{\mu_1^\prime \mu_1}$  
 to obtain an explicit expression for the last term of Eq.~\eqref{eq:gydecomptwo}:
  \begin{align}
\left[ M_{(2g)} \right]_{K(l_1) K(l_2)} & = \left[  \ublob(P_2;l_1,l_2)^{\alpha \beta ; \mu_1^\prime \mu_2^\prime}  K(l_1)_{\mu_1^\prime \mu_1} K(l_2)_{\mu_2^\prime \mu_2}  R^{\alpha \beta ; \mu_1 \mu_2} \right]_{\rm I.S. / F.S.} \nonumber \\ 
\; \nonumber \\ & =  
 i e_q g_s^2 \, \underbrace{\ublob(P_2;l_1,l_2)^{\alpha \beta ; - \, , -}}_\text{{\small $\sim Q^2$}}  \underbrace{\bar{u}(k_4,S_4)}_\text{{\small $\sim Q^{1/2} $} }  \times \nonumber \\ 
\; \nonumber \\  &  \times \underbrace{\left\{
 \frac{1}{l_1^-} \times \frac{\slashed{l}_2 ( \slashed{k}_4 - \slashed{l}_2 + m_q) \slashed{l}_1 (\slashed{k}_4 - \slashed{k}_2  + m_q) \gamma^{\sigma} t^\beta t^\alpha}{\left[ (k_4 - l_2)^2 - m_q^2 + i0 \right] \left[ (k_4 - k_2)^2 - m_q^2 + i0 \right]} \times \frac{1}{l_2^- - i0} \right. }_\text{{\small $\sim 1/Q^2$} }   &  \text{(a.)} \nonumber \\  
\; \nonumber \\
 & \; +
\underbrace{\left.
 \frac{1}{l_1^- - i0} \times \frac{\slashed{l}_1 ( \slashed{k}_4 - \slashed{l}_1 + m_q) \slashed{l}_2 (\slashed{k}_4 - \slashed{k}_2 + m_q) \gamma^{\sigma}  t^\alpha t^\beta}{\left[ (k_4 - l_1)^2 - m_q^2 + i0 \right] \left[ (k_4 - k_2 )^2 - m_q^2 + i0 \right]} \times \frac{1}{l_2^-} \right.}_\text{{\small $\sim 1/Q^2$} } & \text{(b.)} \nonumber \\    
\; \nonumber \\
 & \; +
\underbrace{\left. 
\frac{1}{l_1^- - i0}  \times \frac{\slashed{l}_1 ( \slashed{k}_4 - \slashed{l}_1 + m_q) \gamma^{\sigma} (\slashed{k}_1 + \slashed{l}_2 + m_q) \slashed{l}_2  t^\alpha t^\beta}{\left[ (k_4 - l_1)^2 - m_q^2 + i0 \right] \left[ (k_1 + l_2 )^2 - m_q^2 + i0 \right]} \times \frac{1}{l_2^- + i0} \right.}_\text{{\small $\sim 1/Q^2$} } & \text{(c.)} \nonumber  \\  
\; \nonumber \\
  & \; + \underbrace{\left. 
 \frac{1}{l_1^- + i0} \times \frac{\gamma^\sigma ( \slashed{k}_4 + \slashed{k}_3 + m_q) \slashed{l}_2 (\slashed{k}_1 + \slashed{l}_1  + m_q) \slashed{l}_1  t^\beta t^\alpha}{\left[ (k_4 + k_3)^2 - m_q^2 + i0 \right] \left[ (k_1 + l_1)^2 - m_q^2 + i0 \right]} \times \frac{1}{l_2^-} \right.}_\text{{\small $\sim 1/Q^2$} } & \text{(d.)} \nonumber  \\   
\; \nonumber \\
 & \; +
\underbrace{\left. 
\frac{1}{l_1^-}  \times \frac{\gamma^\sigma ( \slashed{k}_4 + \slashed{k}_3 + m_q) \slashed{l}_1 (\slashed{k}_1 + \slashed{l}_2  + m_q) \slashed{l}_2  t^\alpha t^\beta}{\left[ (k_4 + k_3)^2 - m_q^2 + i0 \right] \left[ (k_1 + l_2)^2 - m_q^2 + i0 \right]} \times \frac{1}{l_2^- + i0} \right.}_\text{{\small $\sim 1/Q^2$} } & \text{(e.)} \nonumber   \\ 
\; \nonumber \\
 & \; +
\underbrace{\left.
 \frac{1}{l_1^- + i0}  \times \frac{\slashed{l}_2 ( \slashed{k}_4 - \slashed{l}_2 + m_q) \gamma^{\sigma} (\slashed{k}_1 + \slashed{l}_1 + m_q) \slashed{l}_1 t^\beta t^\alpha}{\left[ (k_4 - l_2)^2 - m_q^2 + i0 \right] \left[ (k_1 + l_1)^2 - m_q^2 + i0 \right]} \times \frac{1}{l_2^- - i0} \right\}}_\text{{\small $\sim 1/Q^2$} } & \text{(f.)} \nonumber \\ & \times
\underbrace{\lblob(P_1;k_1)}_\text{{\small $\sim Q^{1/2}$} }  \, .   \label{eq:rexpressionkk0}
\end{align}
Counting the powers of $Q$ written underneath Eq.~\eqref{eq:rexpressionkk0} shows the maximum powers to be superleading.
Next, we repeat the Feynman identity substitutions of Eqs.~(\ref{eq:replacementaf1}-\ref{eq:replacementaf4}), again using the result from Eqs.~(\ref{eq:k2sub},~\ref{eq:k2subb}) 
 that the action of $\slashed{k}_1 - m_q$  
on $\lblob(P_1;k_1)$ yields 
terms suppressed by two powers of $Q$.  In the $K(l_1) K(l_2)$ case above, it is critical that dropped terms have \emph{two} powers of suppression because of the potential for 
extra leading powers of $Q$ in the cross section coming from the multiplication with a superleading $\ublob(P_2;l_1,l_2)^{\alpha \beta ; -,-}$.

The principal value contributions are again obtained using Eqs.~(\ref{eq:finalstate},\ref{eq:initialstate}).
 Thus, after the cascade of cancelations, and neglecting power suppressed terms, the principle value $K(l_1) K(l_2)$ term from Eq.~\eqref{eq:m2gluon} becomes
 \begin{multline}
\left[ M_{(2g)} \right]_{K(l_1) K(l_2), \, {\rm P.V.}}  = 
 - e_q g_s^2 f^{\alpha \beta \kappa} \, \underbrace{\bar{u}(k_4,S_4)}_\text{{\small $\sim Q^{1/2}$}} \underbrace{\ublob(P_2;l_1,l_2)^{\alpha \beta ; - \, , -}}_\text{{\small $\sim Q^2$}} 
\; \; \underbrace{\left( {\rm P.V.} \frac{1}{l_1^-} \right)  \left( {\rm P.V.} \frac{1}{l_2^-} \right)  l_{1}^{\rho^\prime}  g_{\rho^\prime \rho} }_\text{{\small $\sim 1/Q $}} \\ \\ \times  \underbrace{\left\{
 \frac{\gamma^{\rho} (\slashed{k}_4 - \slashed{k}_2  + m_q) \gamma^{\sigma} t^\kappa}{(k_4 - k_2)^2  - m_q^2 + i0} \right.  \left.  
  + \frac{\gamma^{\sigma} (\slashed{k}_4 + \slashed{k}_3  + m_q) \gamma^{\rho} t^\kappa}{(k_4 + k_3)^2  - m_q^2 + i0} \right\}}_\text{{\small $\sim1/Q $}} \;
 \underbrace{\lblob(P_1;k_1)}_\text{{\small $\sim Q^{1/2} $}} \; + \; \mathcal{O}(1/Q) \, .   \label{eq:rexpressionkk}
\end{multline}  
The factors of $Q$ in Eq.~\eqref{eq:rexpressionkk} still combine to give superleading powers, and the equation 
does not appear to be symmetric in $l_1$ and $l_2$ in spite of it being the $K(l_1) K(l_2)$ contribution.  
However, Eq.~\eqref{eq:rexpressionkk} now has the basic structure of a leading order result like Eq.~\eqref{eq:1gluonampresult}.  
Therefore, it may be further simplified
with another iteration of the $G$/$K$ Grammer-Yennie decomposition, but now on $g_{\rho^\prime \rho}$: 
\begin{equation}
g^{\rho^\prime \rho} = G(k_2)^{\rho^\prime \rho} + K(k_2)^{\rho^\prime \rho} \, .
\end{equation}   
The $K(k_2)$ term then vanishes up to power suppressed corrections by an argument identical to that of the one-gluon case 
in Eqs.~(\ref{eq:oneKdecomp}--\ref{eq:simplecancel}).  
The  $G(k_2)$ term is  then the only remaining contribution and it is
 \begin{align}
\left[ M_{(2g)} \right]_{K(l_1) K(l_2), \,  {\rm P.V.}}  & = 
 -e_q g_s^2  f^{\alpha \beta \kappa} \, \underbrace{\bar{u}(k_4,S_4)}_\text{{\small $\sim Q^{1/2} $}} \underbrace{\ublob(P_2;l_1,l_2)^{\alpha \beta ; - \, , -}}_\text{{\small $\sim Q^2 $}} 
\; \; \underbrace{\left( {\rm P.V.} \frac{1}{l_1^-} \right)  \left( {\rm P.V.} \frac{1}{l_2^-} \right)  
l_{1, \rho^\prime}  \left(g^{\rho^\prime \rho} - \frac{n_{1}^{\rho^\prime} k_{2}^{\rho} }{k_2^-} \right) }_\text{{\small $\sim 1/Q^2$}} \nonumber \\ \nonumber \\ & \qquad \times  \underbrace{\left\{
 \frac{\gamma_{\rho} (\slashed{k}_4 - \slashed{k}_2  + m_q) \gamma^{\sigma} t^\kappa}{(k_4 - k_2)^2  - m_q^2 + i0} \right.  \left.  
  + \frac{\gamma^{\sigma} (\slashed{k}_4 + \slashed{k}_3  + m_q) \gamma_{\rho} t^\kappa}{(k_4 + k_3)^2  - m_q^2 + i0} \right\}}_\text{{\small $\sim 1/Q $}} \;
 \underbrace{\lblob(P_1;k_1)}_\text{{\small $\sim Q^{1/2} $}} 
\; + \; \mathcal{O}(1/Q) \,  
\nonumber  \\ \nonumber \\
 & = 
 -e_q g_s^2  f^{\alpha \beta \kappa} \, \underbrace{\bar{u}(k_4,S_4)}_\text{{\small $\sim  Q^{1/2} $}} \underbrace{\ublob(P_2;l_1,l_2)^{\alpha \beta ; - \, , -}}_\text{{\small $\sim Q^2 $}} 
\; \; \underbrace{\left( {\rm P.V.} \frac{1}{l_1^-} \right)  \left( {\rm P.V.} \frac{1}{l_2^-} \right)   \left(\frac{l_{1}^{\rho} k_2^-}{k_2^-}  - \frac{l_1^- k_{2}^{\rho} }{k_2^-} \right) }_\text{{\small $\sim 1/Q^2 $}} \nonumber \\ \nonumber \\ & \qquad \times  \underbrace{\left\{
 \frac{\gamma_{\rho} (\slashed{k}_4 - \slashed{k}_2  + m_q) \gamma^{\sigma} t^\kappa}{(k_4 - k_2)^2  - m_q^2 + i0} \right.  \left.  
  + \frac{\gamma^{\sigma} (\slashed{k}_4 + \slashed{k}_3  + m_q) \gamma_{\rho} t^\kappa}{(k_4 + k_3)^2  - m_q^2 + i0} \right\}}_\text{{\small $\sim 1/Q  $}} \;
 \underbrace{\lblob(P_1;k_1)}_\text{{\small $\sim Q^{1/2} $}} 
\; + \; \mathcal{O}(1/Q) \,  
\nonumber  \\ \nonumber \\
 & = 
 -e_q g_s^2  f^{\alpha \beta \kappa} \, \underbrace{\bar{u}(k_4,S_4)}_\text{{\small $\sim Q^{1/2}$}} \underbrace{\ublob(P_2;l_1,l_2)^{\alpha \beta ; - \, , -}}_\text{{\small $\sim Q^2 $}} 
\; \; \underbrace{\left( {\rm P.V.} \frac{1}{l_1^-} \right)  \left( {\rm P.V.} \frac{1}{l_2^-} \right)   \left(\frac{l_{1}^{\rho} l_2^-}{k_2^-}  - \frac{l_1^- l_{2}^{\rho} }{k_2^-} \right) }_\text{{\small $\sim 1/Q^2$}} \nonumber \\ \nonumber \\ &\qquad  \times  \underbrace{\left\{
 \frac{\gamma_{\rho} (\slashed{k}_4 - \slashed{k}_2  + m_q) \gamma^{\sigma} t^\kappa}{(k_4 - k_2)^2  - m_q^2 + i0} \right.  \left.  
  + \frac{\gamma^{\sigma} (\slashed{k}_4 + \slashed{k}_3  + m_q) \gamma_{\rho} t^\kappa}{(k_4 + k_3)^2  - m_q^2 + i0} \right\}}_\text{{\small $\sim 1/Q $}} 
 \underbrace{\lblob(P_1;k_1)}_\text{{\small $\sim Q^{1/2} $}} 
\; + \; \mathcal{O}(1/Q) \,  .
 \label{eq:rexpressionkk2}
\end{align}     
After the second equality, we have written the result of contracting $l_{1, \rho^\prime}$ with the $G(k_2)^{\mu_2^\prime \mu_2}$-factor.
To write the last equality, we have used that $k_2^- \equiv l_1^- + l_2^-$ and $k_{2, \rho} \equiv l_{1, \rho} + l_{2, \rho}$. 
By construction, $G(k_2)^{+-} = 0$ so the super leading powers of $Q$ have canceled  --- 
notice after the last equality of Eq.~\eqref{eq:rexpressionkk2} that the result vanishes exactly for $\rho = ``-"$.  
We are left after the last equality with 
a leading-power contribution that is symmetric in a relabeling of $l_1$ and $l_2$.   As a final step we may also make 
the leading-power replacement $l_{1}^{\rho} \to l_{1t}^{\rho}$, $l_{2}^{\rho} \to l_{2t}^{\rho}$:  
 \begin{multline}
\left[ M_{(2g)} \right]_{K(l_1) K(l_2), \,  {\rm P.V.}}   = \\ 
 -e_q g_s^2  f^{\alpha \beta \kappa} \, \underbrace{\bar{u}(k_4,S_4)}_\text{{\small $\sim Q^{1/2}  $}} \underbrace{\ublob(P_2;l_1,l_2)^{\alpha \beta ; - \, , -}}_\text{{\small $\sim Q^2 $}} 
\; \; \underbrace{\left( {\rm P.V.} \frac{1}{l_1^-} \right)  \left( {\rm P.V.} \frac{1}{l_2^-} \right)   \left(\frac{l_{1t, \rho} l_2^-}{k_2^-}  - \frac{l_1^- l_{2t, \rho} }{k_2^-} \right) }_\text{{\small $\sim 1/Q^2$}}  \underbrace{H_{\rm LO}(k_1,k_2,k_3,k_4)^{\kappa ; \rho\sigma}}_\text{{\small $\sim 1/Q $}} \; 
 \underbrace{\lblob(P_1;k_1)}_\text{{\small $\sim Q^{1/2} $}}  \\
\; + \; \mathcal{O}(1/Q) \,  .
 \label{eq:rexpressionkk3}
\end{multline}   
Here we have used the notation from Eq.~\eqref{eq:treehardpart} for the leading order hard part.
 
\subsection{Complete Amplitude}

Adding Eqs.~(\ref{eq:rexpressionfa2b},~\ref{eq:rexpressionfa3},~\ref{eq:rexpressionkk3}) gives the complete leading-power principal value contribution to the amplitude:
\begin{multline}
\label{eq:fullpv}
\left[ M_{(2g)} \right]_{ {\rm P.V.}}  = 
 -e_q g_s^2  f^{\alpha \beta \kappa} \, \bar{u}(k_4,S_4)  \;  \ublob(P_2;l_1,l_2)^{\alpha \beta ; \mu_1 \mu_2}
\; \left\{ 
g^t_{\mu_1 \rho} n_{1 \, \mu_2} \left( {\rm P.V.} \frac{1}{l_2^-} \right) - g^t_{\mu_2  \rho} n_{1 \, \mu_1} \left( {\rm P.V.} \frac{1}{l_1^-} \right) + \right. \\ \qquad + \left. 
n_{1 \, \mu_1} n_{1 \, \mu_2} \left( {\rm P.V.} \frac{1}{l_1^-} \right)  \left( {\rm P.V.} \frac{1}{l_2^-} \right)   \left(\frac{l_{2t , \rho} \, l_2^-}{k_2^-}  - \frac{ l_{1t, \rho} \,  l_1^-}{k_2^-} \right)
 \right.  \left. \vphantom{{\rm P.V.} \frac{1}{l_2^-}} \right\} \times
\\ \times  H_{\rm LO}(k_1,k_2,k_3,k_4)^{\kappa ; \rho \sigma} \; \lblob(P_1;k_1)
+ \mathcal{O}(1/Q) \,  .
\end{multline}
This is the same result found in Eq.~(81) of Ref.~\cite{Collins:2008sg}.  It also forms part of the gauge link factor for the 
ordinary collinear gluon PDF.

The next step is to isolate any terms that can violate 
the maximally general TMD-factorization criteria of Sect.~\ref{sec:partonmodel}.
For this it will be convenient to rewrite Eq.~\eqref{eq:fullpv} in terms of $l_{1t,\rho}$ and $k_{2t,\rho}$ using $l_{2t,\rho} = k_{2t,\rho} - l_{1t,\rho}$:
\begin{multline}
\label{eq:fullpvb}
\left[ M_{(2g)} \right]_{ {\rm P.V.}}  = 
 -e_q g_s^2  f^{\alpha \beta \kappa} \, \bar{u}(k_4,S_4) \;  \ublob(P_2;l_1,l_2)^{\alpha \beta ; \mu_1 \mu_2}
\; \left\{ 
g^t_{\mu_1 \rho} n_{1 \, \mu_2} \left( {\rm P.V.} \frac{1}{l_2^-} \right) - g^t_{\mu_2  \rho} n_{1 \, \mu_1} \left( {\rm P.V.} \frac{1}{l_1^-} \right) + \right. \\ + \left. 
n_{1 \, \mu_1} n_{1 \, \mu_2} \left( {\rm P.V.} \frac{1}{l_1^-} \right)  \left( {\rm P.V.} \frac{1}{l_2^-} \right)   \left(\frac{k_{2t, \rho} \, l_2^-}{k_2^-}  -  l_{1t, \rho} \right)  \right\} \times \\ \times  H_{\rm LO}(k_1,k_2,k_3,k_4)^{\kappa ; \rho \sigma} \; \lblob(P_1;k_1)
+ \mathcal{O}(1/Q) \,  .
\end{multline}
We now organize the terms in Eq.~\eqref{eq:fullpvb} according to their compatibility or non-compatibility with the 
maximally general TMD-factorization criteria established in Sects.~\ref{sec:genfact} and Sect.~\ref{sec:maxgen}.  We separate 
Eq.~\eqref{eq:fullpvb} into the following terms:
\begin{equation}
\label{eq:mdecomp}
\left[ M_{(2g)} \right]_{ {\rm P.V.}} = \left[ M_{(2g)} \right]^{\rm Fact.,1} + \left[ M_{(2g)} \right]^{\rm Fact.,2} + \left[ M_{(2g)} \right]^{\rm Fact.,3}
+  \left[ M_{(2g)} \right]^{\rm Fact. Viol.}  \;  + \; \mathcal{O}(1/Q) \,  ,
\end{equation}
where we have defined
\begin{align}
\left[ M_{(2g)} \right]^{\rm Fact.,1} \equiv -e_q g_s^2 f^{\alpha \beta \kappa}\; \sum_i 
\underbrace{\mathcal{J}^1(P_2;k_2,l_1)^{\alpha \beta ; i}}_\text{{\small $\sim Q^0 $}} \; \underbrace{\bar{u}(k_4,S_4)}_\text{{\small $\sim Q^{1/2} $}} \; \underbrace{H_{\rm LO}(k_1,k_3,k_4,k_2)^{\kappa ; i \sigma}}_\text{{\small $\sim 1/Q $}} 
\; \underbrace{\lblob(P_1;k_1)}_\text{{\small $\sim Q^{1/2} $}} \, ,  \label{eq:vgfactamps1} \\ 
\nonumber \\
\left[ M_{(2g)} \right]^{\rm Fact.,2} \equiv -e_q g_s^2 f^{\alpha \beta \kappa}\; \sum_i \underbrace{\mathcal{J}^2(P_2;k_2,l_1)^{\alpha \beta ; i}}_\text{{\small $\sim Q^0 $}} \; 
\underbrace{\bar{u}(k_4,S_4)}_\text{{\small $\sim Q^{1/2}$}} \; \underbrace{H_{\rm LO}(k_1,k_3,k_4,k_2)^{\kappa ; i \sigma}}_\text{{\small $\sim 1/Q$}} \; \underbrace{\lblob(P_1;k_1)}_\text{{\small $\sim Q^{1/2} $}} \, ,  \label{eq:vgfactamps2} \\
\nonumber \\
\left[ M_{(2g)} \right]^{\rm Fact.,3} \equiv -e_q g_s^2 f^{\alpha \beta \kappa}\; \sum_i 
\underbrace{\mathcal{J}^3(P_2;k_2,l_1)^{\alpha \beta ; i}}_\text{{\small $\sim Q^0$}} \; \underbrace{\bar{u}(k_4,S_4)}_\text{{\small $\sim Q^{1/2}$}} \; 
\underbrace{H_{\rm LO}(k_1,k_3,k_4,k_2)^{\kappa ; i \sigma}}_\text{{\small $\sim 1/Q$}} \; \underbrace{\lblob(P_1;k_1)}_\text{{\small $\sim Q^{1/2}$}} \, ,  \label{eq:vgfactamps3} \\ 
\nonumber \\
\left[ M_{(2g)} \right]^{\rm Fact. Viol.} \equiv -e_q g_s^2 f^{\alpha \beta \kappa}\; \sum_i \underbrace{\mathcal{J}^{\rm F.V.}(P_2;k_2,l_1)^{\alpha \beta ; i}}_\text{{\small $\sim Q^0$}} \; 
\underbrace{\bar{u}(k_4,S_4)}_\text{{\small $\sim Q^{1/2}$}} \; \underbrace{H_{\rm LO}(k_1,k_3,k_4,k_2)^{\kappa ; i \sigma}}_\text{{\small $\sim 1/Q$}} \; \underbrace{\lblob(P_1;k_1)}_\text{{\small $\sim Q^{1/2}$}} \, .  \label{eq:vgfactamps4}
\end{align}
The $\mathcal{J}$ factors are defined as
\begin{align}
\mathcal{J}^1(P_2;k_2,l_1)^{\alpha \beta ; i} & \equiv  \left( \ublob(P_2;l_1,l_2)_{ \mu_1 \mu_2}^{\alpha \beta} \, \left( {\rm P.V.} \frac{1}{l_2^-} \right)  \right) \; \mathcal{P}_{\ublob, 1}^{ \mu_1 \mu_2, i} \, \sim Q^0 \, , \label{eq:ueffposses1} \\
\mathcal{J}^2(P_2;k_2,l_1)^{\alpha \beta ; i} & \equiv  \left( \ublob(P_2;l_1,l_2)_{ \mu_1 \mu_2}^{\alpha \beta} \, \left( {\rm P.V.} \frac{1}{l_1^-} \right)  \right) \; \mathcal{P}_{\ublob, 2}^{ \mu_1 \mu_2, i} \, \sim Q^0 \, , \label{eq:ueffposses2} \\
\mathcal{J}^3(P_2;k_2,l_1)^{\alpha \beta ; i} & \equiv  \left( \ublob(P_2;l_1,l_2)_{ \mu_1 \mu_2}^{\alpha \beta} \, \left( \frac{l_2^-}{k_2^-} \right)  
\left( {\rm P.V.} \frac{1}{l_1^-} \right)  \left( {\rm P.V.} \frac{1}{l_2^-} \right)   \right) \; \mathcal{P}_{\ublob, 3}({\bf k}_{2t})^{\mu_1 \mu_2, i} \, \sim Q^0 \, , \label{eq:ueffposses3} \\
\mathcal{J}^{\rm F.V.}(P_2;k_2,l_1)^{\alpha \beta ; i} & \equiv  \left( \ublob(P_2;l_1,l_2)_{ \mu_1 \mu_2}^{\alpha \beta} \, 
\left( {\rm P.V.} \frac{1}{l_1^-} \right)  \left( {\rm P.V.} \frac{1}{l_2^-} \right)   \right) \; \mathcal{P}_{\ublob, {\rm F.V.}}({\bf l}_{1t})^{\mu_1 \mu_2, i} \, \sim Q^0 \, . \label{eq:ueffposses4}
\end{align}
The gluon polarization projections are
\begin{align}
\mathcal{P}_{\ublob, 1}^{\mu_1 \mu_2, i} & \equiv  n_{1}^{\mu_2} g_t^{\mu_1 i} \, , \label{eq:polar1} \\
\nonumber \\
\mathcal{P}_{\ublob, 2}^{\mu_1 \mu_2, i} & \equiv  - n_{1}^{\mu_1} g_t^{\mu_2 i}\, , \label{eq:polar2} \\
\nonumber \\
\mathcal{P}_{\ublob, 3}({\bf k}_{2t})^{\mu_1 \mu_2, i} & \equiv  n_1^{\mu_1} n_1^{\mu_2} k_{2t}^i\, , \label{eq:polar3} \\
\nonumber \\
 \mathcal{P}_{\ublob, {\rm F.V.}}({\bf l}_{1t})^{\mu_1 \mu_2, i} & \equiv  -n_1^{\mu_1} n_1^{\mu_2} l_{1t}^i \, . \label{eq:polar4}
\end{align}
Substituting Eqs.(\ref{eq:vgfactamps1}-\ref{eq:vgfactamps4}) into Eq.~\eqref{eq:mdecomp} reproduces Eq.~\eqref{eq:fullpvb}.  
Note that Eqs.~(\ref{eq:polar1},~\ref{eq:polar2}) only project $\ublob(P_2;l_1,l_2)^{- j} \sim Q$ components, giving an overall $Q^0$ power
when combined with the single eikonal denominators of Eqs.~(\ref{eq:ueffposses1},\ref{eq:ueffposses2}).  Equations~(\ref{eq:polar3},~\ref{eq:polar4}) project 
the $\ublob(P_2;l_1,l_2)^{\alpha \beta , \; --} \sim Q^2$ components, giving an overall power of $Q^0$  when combined with the 
double eikonal denominators in Eqs.~(\ref{eq:ueffposses3},~\ref{eq:ueffposses4}).

Now comparing Eqs.~(\ref{eq:vgfactamps1}--\ref{eq:polar4}) with Eqs.~(\ref{eq:vgfactamp}--\ref{eq:ueffpossviol}) from Sect.~\ref{sec:maxgen} verifies 
that Eqs.~(\ref{eq:vgfactamps1},~\ref{eq:vgfactamps2},~\ref{eq:vgfactamps3}) are consistent with TMD-factorization with the maximally general criteria.
However, Eq.~\eqref{eq:vgfactamps4} is a candidate TMD-factorization violating term in the form of Eq.~\eqref{eq:ueffpossviol}.  Specifically, the 
transverse momentum, ${\bf l}_{1t}$, in the gluon polarization projection of Eq.~\eqref{eq:polar4} differs from the transverse momentum entering the hard subprocess.  
Therefore, we have used ``${\rm F.V.}$" (super)subscripts in Eqs.~(\ref{eq:vgfactamps4},~\ref{eq:ueffposses4},~\ref{eq:polar4}) to label them as 
potential TMD-factorization violating contributions.

\subsection{Candidate TMD-Factorization Violating Terms in the Amplitude}

In Eqs.~(\ref{eq:vgfactamps1}--\ref{eq:vgfactamps4}) we have categorized each term in the amplitude according 
to its factorizability, and we have isolated the term in Eq.~\eqref{eq:vgfactamps4} that, according to Sect.~\ref{sec:maxgen}, can violate the
TMD-factorization criteria of Sect.~\ref{sec:genfact}.  Writing out this contribution explicitly in terms of 
its basic factors gives
\begin{align}
\left[ M_{(2g)} \right]^{\rm Fact. Viol.} & =  \nonumber \\ \nonumber \\
  = & -e_q g_s^2 f^{\alpha \beta \kappa}\; \sum_i \mathcal{J}^{\rm F.V.}(P_2;k_2,l_1)^{\alpha \beta ; i} \; \bar{u}(k_4,S_4) \; H_{\rm LO}(k_1,k_3,k_4,k_2)^{\kappa ; i \sigma} \; \lblob(P_1;k_1) \nonumber \\ 
\nonumber \\
  = & e_q g_s^2 f^{\alpha \beta \kappa}\; \sum_i \;  \underbrace{\ublob(P_2;l_1,l_2)^{\alpha \beta ; -,-}}_\text{{\small $\sim Q^2$}} \, 
\underbrace{\left( {\rm P.V.} \frac{1}{l_1^-} \right) \left( {\rm P.V.} \frac{1}{l_2^-} \right) l_{1t}^i}_\text{{\small $\sim 1/Q^2$}}  \; 
\underbrace{\bar{u}(k_4,S_4)}_\text{{\small $\sim Q^{1/2}$}} \; \underbrace{H_{\rm LO}(k_1,k_3,k_4,k_2)^{\kappa ; i \sigma}}_\text{{\small $\sim 1/Q$}} \; \underbrace{\lblob(P_1;k_1)}_\text{{\small $\sim Q^{1/2}$}} \nonumber \\ 
& \qquad \qquad {} + \mathcal{O}(1/Q) \, 
\nonumber \\
 = & - e_q g_s^2 f^{\alpha \beta \kappa}\; \sum_i \;  
\underbrace{\ublob(P_2;l_1,l_2)^{\alpha \beta ; \mu_1,\mu_2} \, \left( {\rm P.V.} \frac{1}{l_1^-} \right)  \left( {\rm P.V.} \frac{1}{l_2^-} \right) \mathcal{P}_{\ublob, {\rm F.V.}}({\bf l}_{1t})_{\mu_1 \mu_2}^{ i}}_\text{{\small $\sim Q^0$}}  \; 
\underbrace{\bar{u}(k_4,S_4)}_\text{{\small $\sim Q^{1/2}$}} \; \nonumber \\
& \qquad \qquad \times \underbrace{H_{\rm LO}(k_1,k_3,k_4,k_2)^{\kappa ; i \sigma}}_\text{{\small $\sim 1/Q$}} \; \underbrace{\lblob(P_1;k_1)}_\text{{\small $\sim Q^{1/2}$}} \, \; + \; \mathcal{O}(1/Q) \, .  \label{eq:factviolamp}
\end{align}
In the last equality, the amplitude has been written as in Eq.~\eqref{eq:vgfactamp}.
The order $\sim Q^0$ factor on the first line becomes a $\ublob_{eff}$, but now it is exactly in the 
form of a candidate TMD-factorization breaking contribution like Eq.~\eqref{eq:ueffpossviol}.  
The gluon transverse momentum, ${\bf l}_{1t}$, which determines the polarization projection in Eq.~\eqref{eq:polar4}, differs from the actual total transverse gluon momentum, ${\bf k}_{2t}$, that 
enters into the calculation of the leading-order hard part.  And it is not possible to shift $ \mathcal{P}_{\ublob, {\rm F.V.}}({\bf l}_{1t})^i$ into another factor without $H_L$ or $\lblob$ then acquiring dependence on ${\bf l}_{1t}$. 

For an Abelian theory, $f^{\alpha \beta \kappa} \to 0$ in which case contributions like Eq.~\eqref{eq:factviolamp} vanish.
Thus the appearance of structures like Eq.~\eqref{eq:factviolamp} is a consequence 
of the interplay of color, spin and transverse momentum degrees of freedom 
in a non-Abelian gauge theory.

\subsection{Reduction To Factorization in Collinear Case}

From the form of Eq.~\eqref{eq:factviolamp}, it is evident that any violation of the criteria of Sect.~\ref{sec:genfact}
is specific to observables that are sensitive to intrinsic transverse momentum.  The violation of generalized 
TMD-factorization is due to sensitivity to the difference between the two intrinsic transverse momenta, ${\bf l}_{1t}$ and ${\bf k}_{2t}$, 
outside of $\ublob$.  To have this, the observable must be sensitive to intrinsic transverse momenta to begin with.

More specifically, the ``${\rm F.V.}$" transverse polarization projection in  Eq.~\eqref{eq:polar4} appears at leading order in 
the cross section paired with the tree-level projection, $G(k_2)^{\mu^\prime j}$, from Eqs.~(\ref{eq:Gdefk2},~\ref{eq:Greplace},~\ref{eq:1gluonampresult}):
\begin{equation}
\mathcal{P}_{\ublob, {\rm F.V.}}({\bf l}_{1t})^{i} G(k_2)^{j} \, \label{eq:polcombo}
\end{equation}
where to simplify notation here we have only written the transverse $i,j$ indices that couple to the hard part.

If we were to restrict consideration to observables that are insensitive to a small ${\bf q}_t$ 
in Fig.~\ref{fig:processdiagram}, i.e. that are azimuthally symmetric, then the sum over the $i$ and $j$ components 
is diagonal.  Then we could follow Sect.~\RNum{3} D of Ref.~\cite{Collins:2008sg}, averaging over transverse gluon polarizations in 
the hard part while separately tracing over the transverse components in Eq.~\eqref{eq:polcombo} in the upper subgraph, $\ublob$, for hadron-2.  That is, the combination 
in Eq.~\eqref{eq:polcombo} could be replaced with 
\begin{equation}
\sum_i \mathcal{P}_{\ublob, {\rm F.V.}}({\bf l}_{1t})^{i} G(k_2)^{i} \, , \label{eq:polcombo2}
\end{equation}
in which case the sums over transverse polarization components, as well as the integrations over ${\bf l}_{1t}$ and ${\bf k}_{2t}$, 
take place independently of the rest of the graph.  The result is a contribution to the normal integrated collinear 
gluon PDF identified in Sect.~\RNum{5} G of Ref.~\cite{Collins:2008sg}.  
Hence, factorization of the gluon distribution is maintained so long as we restrict to collinear observables.  

In the next section, however,
we are specifically interested in observables that are sensitive to the direction of ${\bf q}_t$, and therefore we must account for the 
off-diagonal parts of Eq.~\eqref{eq:polcombo}.

\section{Final State Spin Dependence in the Presence of One Extra Gluon}
\label{sec:extra}

The aim of this section is to
verify, by directly counting powers of $Q$, that terms like Eq.~\eqref{eq:factviolamp} result in a breakdown of TMD-factorization with an extra 
TMD-factorization-violating spin correlation in the 
final state.  Recall the strategy outlined in Sect.~\ref{eq:symmetries} for identifying spin correlations that are specific 
to TMD-factorization breaking effects.  There it was noted that a leading-power correlation between the spin $S_4$ of the final state quark and 
the helicity $\lambda_3$ of the prompt photon does not fit into the standard classification scheme.  
Thus, the strategy of this section is to isolate terms that give contributions from these specific
spin correlations in the cross section.

Such extra asymmetries can only arise from contributions that violate the most general version of the
TMD-factorization hypothesis, stated in Sect.~\ref{sec:partonmodel}.  Therefore, we may
focus attention on the only term, isolated in Eq.~\eqref{eq:factviolamp}, that violates those criteria.

\subsection{TMD-Factorization Breaking Cross Section}
\label{sec:extrapart1}

The lowest order contribution to the cross section from the term in Eq.~\eqref{eq:factviolamp} 
is obtained by adding it to the tree level contribution in Eq.~\eqref{eq:1gluonampresult}, taking the square modulus, 
and keeping the order $g_s^6$ terms.  The steps are exactly analogous to those of 
Sect.~\ref{sec:tmdfactorization} and we write the details here.  Using Eq.~\eqref{eq:crosssection}, we find
\begin{align}
d \sigma^{\rm F.V.} & =  \mathcal{C} Q^2 \sum_{X_1}  \sum_{X_2}  \sum_s \int \frac{d^4 k_1}{(2 \pi)^4} \frac{d^4 1_1}{(2 \pi)^4} \frac{d^4 1_2}{(2 \pi)^4}\,  \left[ M_{(2g)} \right]^{\rm Fact. Viol.} M_{(1g)}^{G \, \dagger}   \, (2 \pi)^4 \delta^{(4)} (k_1 + 1_1 + l_2 - k_3 - k_4)  
\nonumber \\  & = 
\mathcal{C} Q^2 \sum_{X_1}  \sum_{X_2}  \sum_{s} \sum_{i j} \int \frac{d^4 k_1}{(2 \pi)^4} \frac{d^4 1_1}{(2 \pi)^4} \frac{d^4 1_2}{(2 \pi)^4}   \nonumber \\ & \qquad \qquad {}
\times \left(i e_q^2 g_s^3 f^{\alpha \beta \kappa} 
\ublob(P_2;l_1,l_2)^{\alpha \beta}_{\mu_1^\prime ,\mu_2^\prime } \ublob^{\dagger}(P_2,k_2)^{\kappa^\prime}_{ \nu^\prime} \left( {\rm P.V.} \frac{1}{l_1^-} \right)  \left( {\rm P.V.} \frac{1}{l_2^-} \right)  \mathcal{P}_{\ublob, {\rm F.V.}}({\bf l}_{1t})^{\mu_1^\prime \mu_2^\prime, i} G(k_2)^{\nu^\prime j} \right) 
\nonumber \\ &  \qquad \qquad  
\, \times
\tr{\bar{u}(k_4,S_4) \; H^{\kappa ; i}_{{\rm LO}\, \sigma} \;  \lblob \bar{\lblob}  \; H^{\dagger \; \kappa^\prime ; j}_{{\rm LO}\, \sigma^\prime} \; u(k_4,S_4)} \, \epsilon_{\lambda_3}^\sigma \epsilon_{\lambda_3}^{\ast \, \sigma^\prime} (2 \pi)^4 \delta^{(4)} (k_1 + l_1+ l_2  - k_3 - k_4) \, \nonumber \\ 
& \qquad \qquad {} + {\rm H.C.} \nonumber \\ \nonumber \\ 
& = i \mathcal{C} Q^2 e_q^2 g_s^3 f^{\alpha \beta \kappa} \sum_{X_1}  \sum_{X_2} \sum_{s} \sum_{i j} \int \frac{d^2 {\bf k}_{1t} \, d k_1^-}{(2 \pi)^3} \, \int \frac{d^2 {\bf l}_{1t} d l_1^+ d l_1^-}{(2 \pi)^4} \int \frac{d k_2^+}{2 \pi}  \nonumber \\ & \qquad \qquad {}
\times \left( 
\ublob(P_2;l_1,k_2 - l_2)^{\alpha \beta}_{\mu_1^\prime ,\mu_2^\prime } \ublob^{\dagger}(P_2,k_2)^{\kappa^\prime}_{ \nu^\prime} \left( {\rm P.V.} \frac{1}{l_1^-} \right)  \left( {\rm P.V.} \frac{1}{(k_2^- - l_1^-)} \right)  \mathcal{P}_{\ublob, {\rm F.V.}}({\bf l}_{1t})^{\mu_1^\prime \mu_2^\prime, i} G(k_2)^{\nu^\prime j} \right) \nonumber \\ &  \qquad \qquad  
\times \tr{\bar{u}(k_4,S_4) \; H^{\kappa ; i}_{{\rm LO} \, \sigma} \;  \lblob \bar{\lblob}  \; H^{\dagger \; \kappa^\prime ; j}_{{\rm LO}, \, \sigma^\prime} \; u(k_4,S_4)} \epsilon_{\lambda_3}^\sigma \epsilon_{\lambda_3}^{\ast \, \sigma^\prime}  + {\rm H.C.} \, \nonumber \\ \nonumber \\
& = i \mathcal{C} Q^2 e_q^2 g_s^3 f^{\alpha \beta \kappa} \sum_{s} \sum_{i j} \int \frac{d^2 {\bf k}_{1t} }{(2 \pi)^2} \, \int \frac{d^2 {\bf l}_{1t}}{(2 \pi)^2} \nonumber \\ 
&  \qquad \qquad \times  \left\{  \vphantom{{\rm P.V.} \frac{1}{l_2^-}}  \right. \left. \sum_{X_2} \underbrace{\int \frac{d l_1^+ d l_1^- d k_2^+}{(2 \pi)^3}}_\text{{\small $\sim 1/Q$}}
\right. \nonumber \\ &  \qquad \qquad  \times \left. 
\underbrace{\ublob(P_2;l_1,k_2 - l_2)^{\alpha \beta}_{\mu_1^\prime ,\mu_2^\prime } \ublob^{\dagger}(P_2,k_2)^{\kappa^\prime}_{ \nu^\prime} \left( {\rm P.V.} \frac{1}{l_1^-} \right)  \left( {\rm P.V.} \frac{1}{(k_2^- - l_1^-)} \right) 
\mathcal{P}_{\ublob, {\rm F.V.}}({\bf l}_{1t})^{\mu_1^\prime \mu_2^\prime, i} G(k_2)^{\nu^\prime j} }_\text{{\small $\sim Q^0$}} \right. \left. \vphantom{{\rm P.V.} \frac{1}{l_2^-}} \right\} 
\nonumber \\ &  \qquad \qquad  
\, \times \tr{ \vphantom{{\rm P.V.} \frac{1}{l_2^-}}  \right. \left. \underbrace{\bar{u}(k_4,S_4)}_\text{{\small $\sim Q^{1/2}$}} \right. \left. \; \underbrace{\hat{H}^{\kappa ; i}_{{\rm LO} \, \sigma}}_\text{{\small $\sim 1/Q$}} 
\; \; \underbrace{\left\{ \sum_{X_1} \int \frac{d k_1^-}{2 \pi}  \lblob \bar{\lblob}  \right\}}_\text{{\small $\sim Q^0$}} \; \; \underbrace{\hat{H}^{\dagger \; \kappa^\prime ; j}_{{\rm LO}, \, \sigma^\prime}}_\text{{\small $\sim 1/Q$}} 
\; \; \underbrace{u(k_4,S_4)}_\text{{\small $\sim Q^{1/2}$}} \right. \left. \vphantom{{\rm P.V.} \frac{1}{l_2^-}}   } \epsilon_{\lambda_3}^\sigma \epsilon_{\lambda_3}^{\ast \, \sigma^\prime} \; + \; {\rm H.C.} \, 
\; \sim Q^0 \, .\label{eq:facross}
\end{align}
In the second equality, we have written out the explicit factors.
In the the third equality, we have changed variables from $l_2$ to $k_2 - l_1$ and we have evaluated the momentum conserving $\delta$-functions using the $l_2$ integration. In the last equality we have applied the minimal 
TMD kinematic approximations of Eqs.~(\ref{eq:hat1}-\ref{eq:hardsub}), as indicated by the hats placed on $\hat{H}_{\rm LO}$, and we have 
shifted the positions of the $l_1^+$, $l_1^-$, $k_2^+$ and $k_1^-$ integrals into separate factors reminiscent of the step in Eq.~\eqref{eq:basicdecom}.  
Also, we have explicitly written the final state photon
polarization vectors $ \epsilon_{\lambda_3}^\sigma \epsilon_{\lambda_3}^{\ast \, \sigma^\prime}$ from Eq.~\eqref{eq:photonpol}.  

Eq.~\eqref{eq:facross} has almost exactly the same structure as the simple tree level result, especially if 
the factors in braces in the last equality are identified as special types of effective TMD PDFs.   Compare with Eq.~\eqref{eq:basicdecom};
the only modification is that we have isolated the TMD-factorization violating projection from Eq.~\eqref{eq:polcombo} that comes from a single 
extra soft/collinear gluon.

A natural strategy would be to attempt to rewrite this as an effective TMD-factorization formula.  
However, it is clear now from the structure of Eq.~\eqref{eq:facross} that this is impossible without the effective TMD PDFs 
violating at least one of the the minimal criteria for generalized TMD-factorization enumerated in Sect.~\ref{sec:genfact}:  
On one hand, if the $\mathcal{P}_{\ublob, {\rm F.V.}}({\bf l}_{1t})^{\mu_1^\prime \mu_2^\prime, i}$ projection is included inside the definition of 
an effective gluon TMD PDF then criterion 2.) is violated because the projection of Dirac components on the fermion line then depends on ${\bf l}_{1t}$. 
On the other hand, if $\mathcal{P}_{\ublob, {\rm F.V.}}({\bf l}_{1t})^{\mu_1^\prime \mu_2^\prime, i}$ is included in a redefinition of the hard part or in the definition of an effective quark 
TMD PDF then criteria 1.) or 3.) are violated respectively.  So there is no possible rearrangement of factors that allows for a factorization formula with 
the properties of Sect.~\ref{sec:genfact}.
Hence, the maximally general form of TMD-factorization is broken.   
 
\subsection{Effective Factorization Violating TMD Functions}

In spite of the problems noted above with TMD-factorization, let us push forward with an identification of effective TMD PDFs, but now allowing them to violate the 
maximally general criteria for TMD-factorization from Sect.~\ref{sec:genfact}.   Once we have organized factors into a set of effective (but TMD-factorization violating) 
TMD PDFs, we may then directly examine the consequences for power counting and the classification of final state spin dependence.  

We use the explicit expression for $\hat{H}^{i; \kappa}_{{\rm LO} \, \sigma}$ from Eq.~\eqref{eq:treehardpart} 
and directly apply the expression for the ``${\rm F.V.}$" projection in Eq.~\eqref{eq:polar4}.  Then, dropping power suppressed terms, the factor inside the 
Dirac trace on the last line of Eq.~\eqref{eq:facross} becomes
\begin{align}
& \frac{\mathcal{P}_{\ublob, {\rm F.V.}}({\bf l}_{1t})^{\mu_1^\prime \mu_2^\prime, i}}{\hat{s} \hat{t}} 
   \left( \bar{u}(k_4,S_4) \; \gamma^i \slashed{\hat{h}} \gamma_\sigma \;  \lblob(P_1;k_1) \bar{\lblob}(P_1;k_1)  \gamma^j \slashed{q} \gamma_{\sigma^\prime} \; u(k_4,S_4) 
   \right. \nonumber \\ &  \qquad \qquad + \left.
   \bar{u}(k_4,S_4) \; \gamma_\sigma \slashed{q} \gamma^i \;  \lblob(P_1;k_1) \bar{\lblob}(P_1;k_1)  \gamma_{\sigma^\prime} \slashed{\hat{h}} \gamma^j \; u(k_4,S_4) \right) 
   \epsilon_{\lambda_3}^\sigma \epsilon_{\lambda_3}^{\ast \, \sigma^\prime} t^\kappa t^{\kappa^\prime} \nonumber \\
& +  \frac{\mathcal{P}_{\ublob, {\rm F.V.}}({\bf l}_{1t})^{\mu_1^\prime \mu_2^\prime, i}}{\hat{s}^2} \left( \bar{u}(k_4,S_4) \; \gamma_\sigma \slashed{q} 
\gamma^i \;  \lblob(P_1;k_1) \bar{\lblob}(P_1;k_1)  \gamma^j \slashed{q} \gamma_{\sigma^\prime} \; u(k_4,S_4)  \vphantom{\hat{h}} \right) \epsilon_{\lambda_3}^\sigma \epsilon_{\lambda_3}^{\ast \, \sigma^\prime} t^\kappa t^{\kappa^\prime} 
\nonumber \\
& +  \frac{\mathcal{P}_{\ublob, {\rm F.V.}}({\bf l}_{1t})^{\mu_1^\prime \mu_2^\prime, i}}{\hat{t}^2} \left( \bar{u}(k_4,S_4) \; \gamma^i \slashed{\hat{h}} 
\gamma_\sigma \;  \lblob(P_1;k_1) \bar{\lblob}(P_1;k_1)  \gamma_{\sigma^\prime} \slashed{\hat{h}} \gamma^j \; u(k_4,S_4)  \vphantom{\hat{h}} \right) \epsilon_{\lambda_3}^\sigma \epsilon_{\lambda_3}^{\ast \, \sigma^\prime} 
t^\kappa t^{\kappa^\prime} 
\nonumber \\ 
\nonumber \\ 
 & =  \frac{-n_1^{\mu_1^\prime} n_1^{\mu_2^\prime}}{\hat{s} \hat{t}} \left( \underbracket{\bar{u}(k_4,S_4) \slashed{l}_{1t}}_{\text{F.S.I.}} \; \slashed{\hat{h}} 
\gamma_\sigma \;  \lblob(P_1;k_1) \bar{\lblob}(P_1;k_1)  \gamma^j \slashed{q} \gamma_{\sigma^\prime} \; u(k_4,S_4) + \right. \nonumber \\  & \qquad \qquad + \left. 
 \bar{u}(k_4,S_4) \; \gamma_\sigma \slashed{q} \; \underbracket{\slashed{l}_{1t} \lblob(P_1;k_1)}_{\text{I.S.I.}} \; \bar{\lblob}(P_1;k_1)  
\gamma_{\sigma^\prime} \slashed{\hat{h}} \gamma^j \; u(k_4,S_4) \right) \epsilon_{\lambda_3}^\sigma \epsilon_{\lambda_3}^{\ast \, \sigma^\prime}  
t^\kappa t^{\kappa^\prime} \nonumber \\
& - \frac{n_1^{\mu_1^\prime} n_1^{\mu_2^\prime}}{\hat{s}^2} \left( \bar{u}(k_4,S_4) \; \gamma_\sigma \slashed{q} 
\; \underbracket{\slashed{l}_{1t} \lblob(P_1;k_1)}_{\text{I.S.I.}} \;
\bar{\lblob}(P_1;k_1)  \gamma^j \slashed{q} \gamma_{\sigma^\prime} \; u(k_4,S_4)  \vphantom{\hat{h}} \right) \epsilon_{\lambda_3}^\sigma \epsilon_{\lambda_3}^{\ast \, \sigma^\prime}  t^\kappa t^{\kappa^\prime} \nonumber \\
& -  \frac{n_1^{\mu_1^\prime} n_1^{\mu_2^\prime}}{\hat{t}^2} \left( \underbracket{\bar{u}(k_4,S_4) \slashed{l}_{1t}}_{\text{F.S.I.}} \slashed{\hat{h}} \;
\gamma_\sigma \lblob(P_1;k_1) \bar{\lblob}(P_1;k_1)  \gamma_{\sigma^\prime} \slashed{\hat{h}} \gamma^j \; u(k_4,S_4)  \vphantom{\hat{h}} \right) \epsilon_{\lambda_3}^\sigma \epsilon_{\lambda_3}^{\ast \, \sigma^\prime}  
t^\kappa t^{\kappa^\prime} \, .
\label{eq:splittingup}
\end{align}
After the second equality, we have included horizontal brackets underneath 
the equation to indicate where the anomalous $\slashed{l}_{1t}$ transverse momentum contraction is an extra 
soft initial state attachment (labeled I.S.I.) and where it is an extra soft final state attachment (labeled F.S.I.). 

It is in principle possible for soft I.S.I. and F.S.I. attachments to cancel.  But
any such cancellations cannot be general because they depend on the separate non-perturbative 
details of $\slashed{l}_{1t} \lblob(P_1;k_1)$ in the initial state and $\bar{u}(k_4,S_4) \slashed{l}_{1t}$ in the final state, including sensitivity to 
the species of incoming and outgoing hadrons and the quark masses.

Therefore, without loss of generality, we may focus attention on the initial state interactions.   
These correspond to modified TMD-factorization breaking effective TMD PDFs.  
An analysis similar to what we give below may be applied to the extra final state 
interactions, but this will give modified effective TMD-factorization breaking TMD fragmentation or jet functions.  

It is the second and third terms in Eq.~\eqref{eq:splittingup} that contain extra initial state Dirac matrix contractions.
These correspond to a cross term and an $s$-channel term respectively in the squared amplitude. 
We rewrite them by first defining the effective TMD functions for the factors in braces in Eq.~\eqref{eq:facross},
\begin{align}
\Phi_{q / P_1}^{(2g)}\left( P_1;\hat{k}_{1}(x_1,{\bf k}_{1t}), q, {\bf l}_{1t} \right)  &  \equiv     \frac{Q}{m_q} \sum_{X_1} \int  \frac{d k_1^-}{2 \pi}  \slashed{q} \slashed{l}_{1t} \lblob \bar{\lblob}  \;  \; \sim Q^2 \, , \label{eq:NPVvioldef}  \\
\Delta_{k_4/q}\left(k_4, S_4 \right) & \equiv  \sum_s u(k_4,S_4) \bar{u}(k_4,S_4) 
= \left( \slashed{k}_4 + m_q \right) \left( 1 + \gamma_5 \frac{\slashed{S}_4}{m_q}   \right) \; \; \sim Q\, , \label{eq:OFF} 
\end{align}
and, \\
\begin{multline}
 \Phi_{g / P_2}^{(2g)}\left( P_2;\hat{k}_{2}(x_2,{\bf k}_{2t}), {\bf l}_{1t} \right)^{\alpha \beta \kappa^\prime ; j}  \\ \equiv  Q \sum_{X_2} 
 \int \frac{d l_1^+ d l_1^- d k_2^+}{(2 \pi)^3}
\ublob(P_2;l_1,k_2 - l_2)^{\alpha \beta ; -,-} \ublob^{\dagger}(P_2,k_2)^{\kappa^\prime}_{ \nu^\prime} G(k_2)^{\nu^\prime j}  \left( {\rm P.V.} 
\frac{1}{l_1^-} \right) \left( {\rm P.V.} \frac{1}{(k_2^- - l_1^-)} \right)  \; \sim Q^0 \, . \label{eq:gluonmoddef}
\end{multline}
Note the $\slashed{q} \slashed{l}_{1t}$ in the definition in Eq.~\eqref{eq:NPVvioldef} as compared with Eq.~\eqref{eq:1gluondef}.\footnote{An equivalent analysis may be performed with $\slashed{q}$ and/or $\slashed{l}_{1t}$ organized 
into different factors.  We find the above arrangement to be the most convenient for illustrating the basic result.}  
Equation~\eqref{eq:NPVvioldef} also has 
an extra power of $Q$ relative to Eq.~\eqref{eq:1gluondef} and is defined with 
an overall factor of $1/m_q$ to maintain the same units as Eq.~\eqref{eq:1gluondef}.
Equation~\eqref{eq:OFF} is just the spin sum for the final state quark, but represented in notation reminiscent of a jet or fragmentation function.

Using Eqs.~(\ref{eq:splittingup}--\ref{eq:gluonmoddef}) inside Eq.~\eqref{eq:facross} gives
\begin{multline}
d \sigma_{{\rm X-term}}^{\rm I.S.I.}  = 
 \underbrace{ -i \mathcal{C} \frac{e_q^2 g_s^3 f^{\alpha \beta \kappa}  t^\kappa t^{\kappa^\prime} m_q}{\hat{s} \hat{t}}}_\text{{\small $\sim 1/Q^4$}} \sum_{j} \int \frac{d^2 {\bf k}_{1t} }{(2 \pi)^2} \, \int \frac{d^2 {\bf l}_{1t}}{(2 \pi)^2} \times \\ \\
 \times \underbrace{\Phi_{g / P_2}^{(2g)}\left( P_2;\hat{k}_{2}(x_2,{\bf k}_{2t}), {\bf l}_{1t} \right)^{\alpha \beta \kappa^\prime ; j}}_\text{{\small $\sim Q^0$}}  \; 
 \tr{ \vphantom{{\rm P.V.} \frac{1}{l_2^-}} \right. \left. \underbrace{\Delta_{k_4/q}\left( k_4, S_4 \right)}_\text{{\small $\sim Q$}}  
\; \gamma_\sigma \;  \underbrace{\Phi_{q / P_1}^{(2g)}\left( P_1;\hat{k}_{1}(x_1,{\bf k}_{1t}), q,{\bf l}_{1t} \right)}_\text{{\small $\sim Q^2$}}  
\; \underbrace{\gamma_{\sigma^\prime} \slashed{\hat{h}} \gamma^j}_\text{{\small $\sim Q$}} \right. \left.  \vphantom{{\rm P.V.} \frac{1}{l_2^-}}  } \epsilon_{\lambda_3}^\sigma \epsilon_{\lambda_3}^{\ast \, \sigma^\prime} 
\left. \right.  + {\rm H.C.} \,  \label{eq:facross2}
\end{multline}
and
\begin{multline}
d \sigma_{\rm s-chan}^{\rm I.S.I.}  = 
 \underbrace{ -i \mathcal{C} \frac{e_q^2 g_s^3 f^{\alpha \beta \kappa}  t^\kappa t^{\kappa^\prime} m_q}{\hat{s}^2}}_\text{{\small $\sim 1/Q^4$}} \sum_{j} \int \frac{d^2 {\bf k}_{1t} }{(2 \pi)^2} \, \int \frac{d^2 {\bf l}_{1t}}{(2 \pi)^2} \times \\ \\
 \times \underbrace{\Phi_{g / P_2}^{(2g)}\left( P_2;\hat{k}_{2}(x_2,{\bf k}_{2t}), {\bf l}_{1t} \right)^{\alpha \beta \kappa^\prime ; j}}_\text{{\small $\sim Q^0$}}  \; 
 \tr{ \vphantom{{\rm P.V.} \frac{1}{l_2^-}} \right. \left. \underbrace{\Delta_{k_4/q}\left( k_4, S_4 \right)}_\text{{\small $\sim Q$}}  
\; \gamma_\sigma \;  \underbrace{\Phi_{q / P_1}^{(2g)}\left( P_1;\hat{k}_{1}(x_1,{\bf k}_{1t}), q,{\bf l}_{1t} \right)}_\text{{\small $\sim Q^2$}}  
\; \underbrace{\gamma^j \slashed{q} \gamma_{\sigma^\prime}}_\text{{\small $\sim Q$}} \right. \left.  \vphantom{{\rm P.V.} \frac{1}{l_2^-}}  } \epsilon_{\lambda_3}^\sigma \epsilon_{\lambda_3}^{\ast \, \sigma^\prime} \left. \right.  + {\rm H.C.} \, . \label{eq:facross22}
\end{multline}
In Eq.~\eqref{eq:facross2}, we have pulled the $1/\hat{t}$ outside the integrals using the leading-power approximation $\hat{t} \approx (k_3 - \hat{k}_1)^2 \approx - x_1 P_1^+ k_3^-$.  
The subscripts ``${\rm X-term}$" and ``${\rm s-chan}$" label the cross-term and $s$-channel contributions.

The above steps for separating the cross section into factors are very similar to those of Sect.~\ref{sec:tmdfactorization}.
However, Eqs.~(\ref{eq:facross2},~\ref{eq:facross22}) violate the minimal criteria for TMD-factorization from Sect.~\ref{sec:genfact} because of 
the extra momentum arguments of $\Phi_{g / P_2}^{(2g)}\left( P_2;\hat{k}_{2}(x_2,{\bf k}_{2t}), {\bf l}_{1t} \right)$ and 
$\Phi_{q / P_1}^{(2g)}\left( P_1;\hat{k}_{1}(x_1,{\bf k}_{1t}), q,{\bf l}_{1t} \right)$.  As explained in the last paragraph of Sect.~\ref{sec:extrapart1}, at 
least one such violation of the minimal criteria is required by the pattern of non-Abelian Ward identity cancellations in Sect.~\ref{sec:twogluons}.

For the sake of demonstrating with explicit color factors, let us now
take the target hadrons $P_1$ and $P_2$ to be on-shell color triplet quarks, and calculate the contribution for averaged initial and summed final 
state quark color.  Also, let us consider the component of $\Phi_{g / P_2}^{(2g)}\left( P_2;\hat{k}_{2}(x_2,{\bf k}_{2t}), {\bf l}_{1t} \right)^{\alpha \beta \kappa^\prime ; j}$ 
proportional to $-i f^{\alpha \beta \kappa^{\prime \prime}} t^{\kappa^{\prime \prime}} t^{\kappa^\prime}$ 
so that we may define an effective colorless gluon TMD function via
\begin{equation}
\Phi_{g / P_2}^{(2g)}\left( P_2;\hat{k}_{2}(x_2,{\bf k}_{2t}), {\bf l}_{1t} \right)^{\alpha \beta \kappa^\prime ; j} \to -i f^{\alpha \beta \kappa^{\prime \prime}} t^{\kappa^{\prime \prime}} t^{\kappa^\prime} \Phi_{g / P_2}^{(2g)}\left( P_2;\hat{k}_{2}(x_2,{\bf k}_{2t}), {\bf l}_{1t} \right)^{j}  \, . 
\end{equation}
Averaging over initial state triplet color and summing over final state triplet color then gives a specific overall color factor of $(N_c^2 - 1)/ 4 N_c^2$.  
Equations~(\ref{eq:facross2},~\ref{eq:facross22}) become:
\begin{multline}
d \sigma_{{\rm X-term}}^{\rm I.S.I.} =  \underbrace{-\mathcal{C} \frac{e_q^2 g_s^3 (N_c^2 - 1) m_q}{4 N_c^2 \hat{s} \hat{t}}}_\text{{\small $\sim 1/Q^4$}} \; \sum_{j} \int \frac{d^2 {\bf k}_{1t} }{(2 \pi)^2} \, \int \frac{d^2 {\bf l}_{1t}}{(2 \pi)^2} \times \\ \\
 \times \underbrace{\Phi_{g / P_2}^{(2g)}\left( P_2;\hat{k}_{2}(x_2,{\bf k}_{2t}), {\bf l}_{1t} \right)^{j}}_\text{{\small $\sim Q^0$}} \, 
 \tr{  \vphantom{{\rm P.V.} \frac{1}{l_2^-}} \right. \left.  \underbrace{\Delta_{k_4/q}\left( k_4, S_4 \right)}_\text{{\small $\sim Q$}} \gamma_\sigma  
 \underbrace{{\rm Tr}_C \left\{ \Phi_{q / P_1}^{(2g)}\left( P_1;\hat{k}_{1}(x_1,{\bf k}_{1t}),q, {\bf l}_{1t} \right) \right\}}_\text{{\small $\sim Q^2$}} 
 \; \underbrace{\gamma_{\sigma^\prime} \slashed{\hat{h}} \gamma^j}_\text{{\small $\sim Q$}}  \right. \left.  \vphantom{{\rm P.V.} \frac{1}{l_2^-}} }  \epsilon_{\lambda_3}^\sigma \epsilon_{\lambda_3}^{\ast \, \sigma^\prime} 
  + {\rm H.C.}  \label{eq:facross3a}
\end{multline}
and 
\begin{multline}
d \sigma_{{\rm s-chan}}^{\rm I.S.I.} =  \underbrace{-\mathcal{C} \frac{e_q^2 g_s^3 (N_c^2 - 1) m_q}{4 N_c^2 \hat{s}^2 }}_\text{{\small $\sim 1/Q^4$}} \; \sum_{j} \int \frac{d^2 {\bf k}_{1t} }{(2 \pi)^2} \, \int \frac{d^2 {\bf l}_{1t}}{(2 \pi)^2} \times \\ \\
 \times \underbrace{\Phi_{g / P_2}^{(2g)}\left( P_2;\hat{k}_{2}(x_2,{\bf k}_{2t}), {\bf l}_{1t} \right)^{j}}_\text{{\small $\sim Q^0$}} 
 \tr{ \vphantom{{\rm P.V.} \frac{1}{l_2^-}} \right. \left.  \underbrace{\Delta_{k_4/q}\left( k_4, S_4 \right)}_\text{{\small $\sim Q$}}  \gamma_\sigma  
 \underbrace{{\rm Tr}_C \left\{ \Phi_{q / P_1}^{(2g)}\left( P_1;\hat{k}_{1}(x_1,{\bf k}_{1t}),q, {\bf l}_{1t} \right) \right\}}_\text{{\small $\sim Q^2$}} 
 \;  \underbrace{\gamma^j \slashed{q} \gamma_{\sigma^\prime}}_\text{{\small $\sim Q$}}  \right. \left.  \vphantom{{\rm P.V.} \frac{1}{l_2^-}} }  \epsilon_{\lambda_3}^\sigma \epsilon_{\lambda_3}^{\ast \, \sigma^\prime}  + {\rm H.C.}  . \label{eq:facross3aschan}
\end{multline}
In these equations we have used that $\lblob \bar{\lblob}$ is 
diagonal in triplet color; the ${\rm Tr}_C \left\{ \right\}$ denotes a trace over color triplet indices.  

To obtain an analogue of Eq.~\eqref{eq:basicdecom2}, 
we first decompose $\Phi_{g / P_2}^{(2g)}\left( P_2;\hat{k}_{2}(x_2,{\bf k}_{2t}), {\bf l}_{1t} \right)^{j}$ into 
its possible transverse index structures.  The only dependence on transverse indices is from ${\bf l}_{1t}$ and ${\bf k}_{2 t}$ so the analogue of Eq.~\eqref{eq:gluondecomp} is 
\begin{equation}
\label{eq:2gluondecomp}
\Phi_{g / P_2}^{(2g)}\left( P_2;\hat{k}_{2}(x_2,{\bf k}_{2t}), {\bf l}_{1t} \right)^{j} 
\equiv \frac{k_{2t}^j}{m_q} \underbrace{\Phi_{g / P_2}^{(2g) \; [k_{2t}] }\left( P_2;\hat{k}_{2}(x_2,{\bf k}_{2t}), {\bf l}_{1t} \right)}_\text{{\small $\sim Q^0$}} 
+ \frac{l_{1t}^j}{m_q} \underbrace{\Phi_{g / P_2}^{(2g) \; [l_{1t}] }\left( P_2;\hat{k}_{2}(x_2,{\bf k}_{2t}), {\bf l}_{1t} \right)}_\text{{\small $\sim Q^0$}}  \sim Q^0\, .
\end{equation}
There may in general be a separate contribution from each effective gluon TMD PDF labeled with the $[k_{2t}]$ and $[l_{1t}]$ superscripts. 

We will focus first on the contribution from the $[k_{2t}]$-term in Eq.~\eqref{eq:2gluondecomp}.  
Using an arbitrary Dirac projection $\Gamma_{q}$ and following by analogy with Eq.~\eqref{eq:genproj2}, Eqs.~(\ref{eq:facross3a},~\ref{eq:facross3aschan}) then become
\begin{multline}
d \sigma_{{\rm X-term}}^{{\rm I.S.I.} \; \left[\Gamma_{q} \right]}
=  
- \underbrace{\mathcal{C} \frac{e_q^2 g_s^3 (N_c^2 - 1)}{4 N_c^2 \hat{s} \hat{t}}}_\text{{\small $\sim 1/Q^4$}} \int \frac{d^2 {\bf k}_{1t} }{(2 \pi)^2} \, \int \frac{d^2 {\bf l}_{1t}}{(2 \pi)^2} \times \\ \\
\times 
\underbrace{\Phi_{g / P_2}^{(2g) \; [k_{2t}] }\left( P_2;q - \hat{k}_{1}(x_1,{\bf k}_{1t}), {\bf l}_{1t} \right)}_\text{{\small $\sim Q^0$}} \; 
\underbrace{\Phi_{q / P_1}^{(2g) \; [\Gamma_{q}]}\left( P_1;\hat{k}_{1}(x_1,{\bf k}_{1t}),q, {\bf l}_{1t} \right)}_\text{{\small $\sim Q^2$}}   \times \\ \\
 \times \tr{ \vphantom{\slashed{\hat{k}}_1} \right. \left. \underbrace{\Delta_{k_4/q}\left( k_4, S_4 \right)}_\text{{\small $\sim Q$}} \, \slashed{\epsilon}_{\lambda_3} \;  {\rm P}_{\Gamma_q}
 \; \slashed{\epsilon}^{\ast}_{\lambda_3} \underbrace{\slashed{\hat{h}} ( \slashed{q}_{t} - \slashed{k}_{1t})}_\text{{\small $\sim Q$}}  
\right. \left.  \vphantom{\slashed{\hat{k}}_1}  } \, + {\rm H.C.} \, , \label{eq:facross4a}
\end{multline}
and
\begin{multline}
d \sigma_{{\rm s-chan}}^{{\rm I.S.I.} \; \left[\Gamma_{q} \right]}
=  
- \underbrace{\mathcal{C} \frac{e_q^2 g_s^3 (N_c^2 - 1)}{4 N_c^2 \hat{s}^2 }}_\text{{\small $\sim 1/Q^4$}} \int \frac{d^2 {\bf k}_{1t} }{(2 \pi)^2} \, \int \frac{d^2 {\bf l}_{1t}}{(2 \pi)^2} \times \\ \\
\times 
\underbrace{\Phi_{g / P_2}^{(2g) \; [k_{2t}] }\left( P_2;q - \hat{k}_{1}(x_1,{\bf k}_{1t}), {\bf l}_{1t} \right)}_\text{{\small $\sim Q^0$}} \; 
\underbrace{\Phi_{q / P_1}^{(2g) \; [\Gamma_{q}]}\left( P_1;\hat{k}_{1}(x_1,{\bf k}_{1t}),q, {\bf l}_{1t} \right)}_\text{{\small $\sim Q^2$}}   \times \\ \\
 \times \tr{ \vphantom{\slashed{\hat{k}}_1} \right. \left. \underbrace{\Delta_{k_4/q}\left( k_4, S_4 \right)}_\text{{\small $\sim Q$}} \, \slashed{\epsilon}_{\lambda_3} \;  {\rm P}_{\Gamma_q}
 \;  \underbrace{( \slashed{q}_{t} - \slashed{k}_{1t}) \slashed{q} }_\text{{\small $\sim Q$}} \slashed{\epsilon}^{\ast}_{\lambda_3}  
\right. \left.  \vphantom{\slashed{\hat{k}}_1}  } \, + {\rm H.C.} \, . \label{eq:facross4aschan}
\end{multline}
Above we have used
\begin{equation}
\label{eq:traceNPVproj}
\Phi_{q / P_1}^{(2g) \; [\Gamma_{q}]}\left( P_1;\hat{k}_{1}(x_1,{\bf k}_{1t}), q,{\bf l}_{1t} \right)
\equiv {\rm Tr}_C \left(  \frac{1}{4} \tr{ \Gamma_{q} \, \Phi_{q / P_1}^{(2g)}\left( P_1;\hat{k}_{1}(x_1,{\bf k}_{1t}),q, {\bf l}_{1t} \right)} \right) \, ,
\end{equation}
in accordance with Eq.~\eqref{eq:genproj}.  

In Eqs.~(\ref{eq:facross4a},~\ref{eq:facross4aschan}) we have reached the analogues of Eq.~\eqref{eq:genproj2}, but now in a form that accounts for the extra initial state interactions.  The effective TMD PDFs now depend 
on extra momenta external to their parent hadrons and therefore violate the minimal requirements for TMD-factorization.  
The organization of factors in Eqs.~\eqref{eq:splittingup} through~\eqref{eq:traceNPVproj} is 
represented graphically in Fig~\ref{fig:anomalygraphs}.  The special ``\textcircled{$\times$}" vertex denotes the TMD-factorization anomaly factor of $\slashed{q} \slashed{l}_{1t}$ in Eq.~\eqref{eq:NPVvioldef} that
enters into the trace of Dirac matrices along the lower active quark line.  The extra $\slashed{l}_{1t} \neq \slashed{k}_{2t}$ from hadron-2 is trapped by the $\slashed{q}$ on the hadron-1 side of the graph.   
\begin{figure*}
\centering
  \begin{tabular}{c@{\hspace*{25mm}}c}
    \includegraphics[scale=0.4]{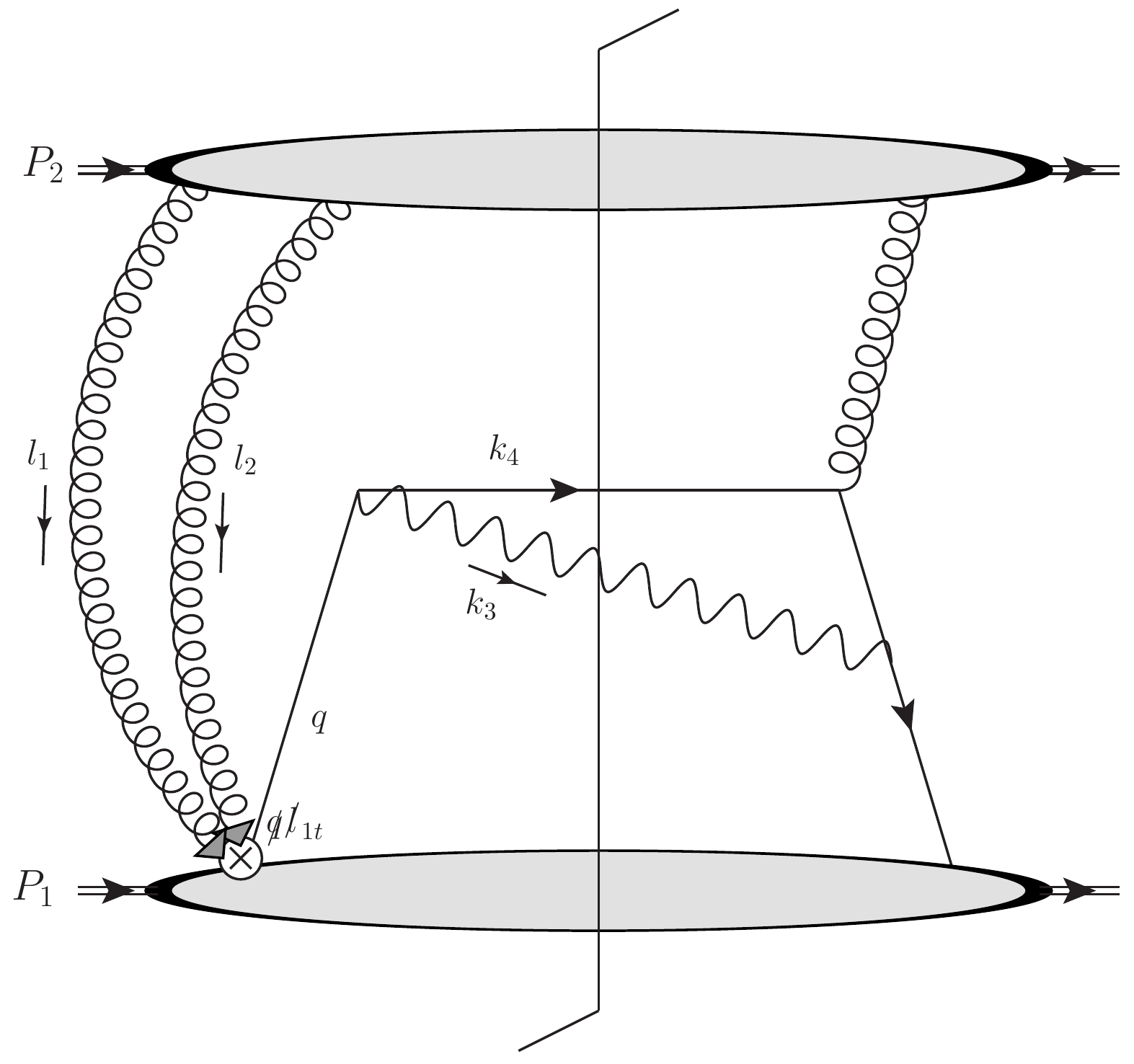}
    &
    \includegraphics[scale=0.4]{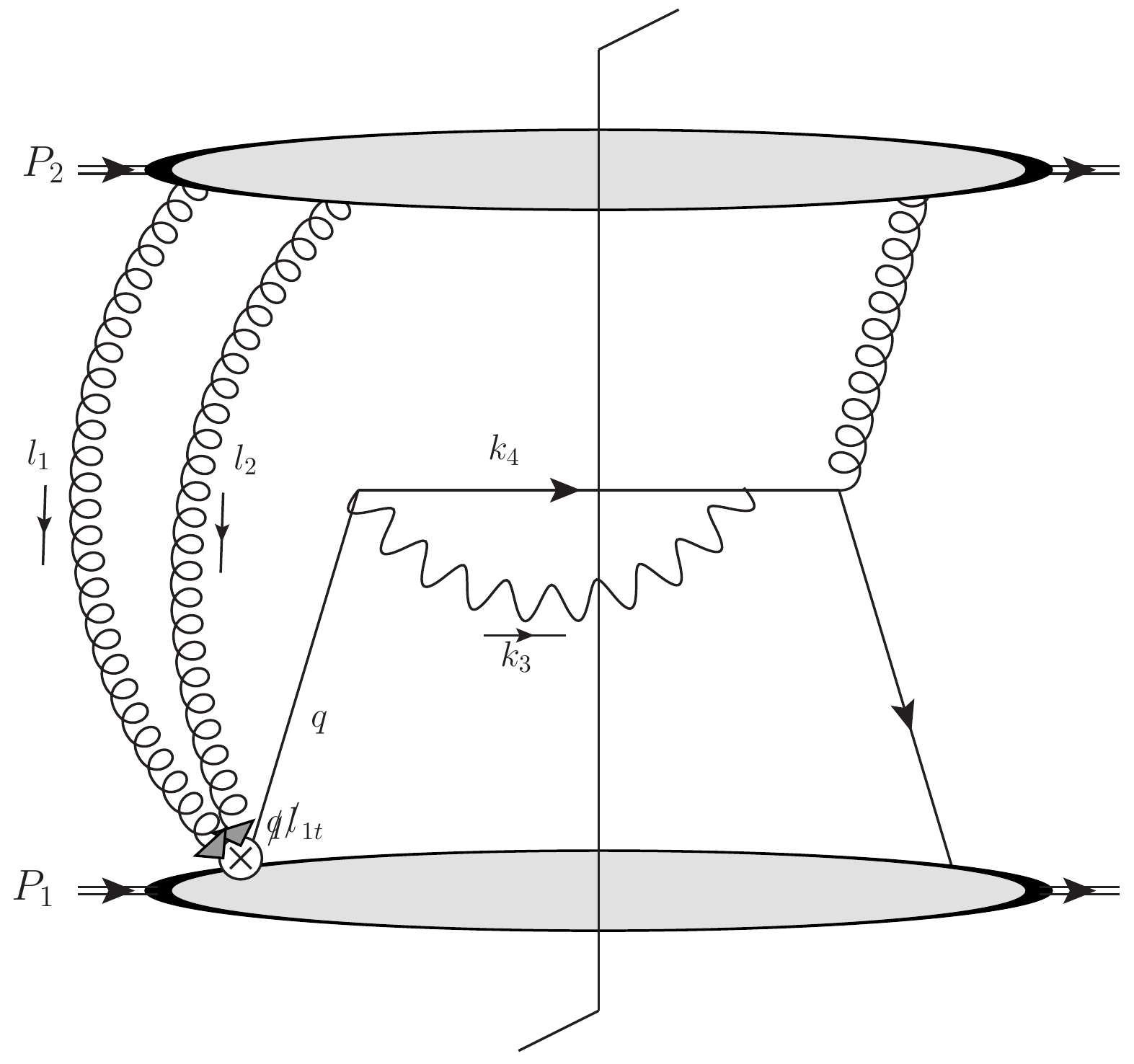}
    \\
    (a) & (b)
    \\[5mm]
   \end{tabular}
\caption{Graphs representing the organization of factors in a.) Eq.~\eqref{eq:facross4a} and b.) Eq.~\eqref{eq:facross4aschan}.  Hermitian conjugate 
graphs should also be included.  The ``\textcircled{$\times$}" vertex denotes the anomalous factor of $\slashed{q} \slashed{l}_{1t}$ in Eq.~\eqref{eq:NPVvioldef}, which 
cannot be associated with a gauge link operator in a TMD PDF for either hadron alone.}
\label{fig:anomalygraphs}
\end{figure*}

\subsection{Power Counting and TMD-Factorization Violating Spin Asymmetries}
\label{eq:npvasymetries}

In Eqs.~(\ref{eq:facross4a},\ref{eq:facross4aschan}), the initial state interaction contribution to the cross section from 
Eq.~\eqref{eq:vgfactamps4} has been written in a form analogous to Eq.~\eqref{eq:genproj2}, but with the 
effective TMD functions now explicitly violating the minimal TMD-factorization criteria of Sect.~\ref{sec:genfact}.  We can now
examine whether the exclusion of certain leading-power correlations on the basis of parity invariance, as discussed in Sect.~\ref{eq:symmetries}, 
still applies to the effective TMD PDFs in the TMD-factorization breaking scenario of Eqs.~(\ref{eq:facross4a},\ref{eq:facross4aschan}).   

In fact, a close inspection of the more general multi-gluon effective 
TMD objects like Eq.~\eqref{eq:traceNPVproj} hints at 
a large variety of extra angular and spin dependencies.   
It is beyond the scope of this paper to classify all of them in detail,
so we will instead illustrate the general existence of such effects by focusing on the specific case, already discussed in Sect.~\ref{eq:symmetries}, of a final 
state correlation between photon helicity $\lambda_3$ and transverse quark spin $S_4$.

\subsubsection{Initial State Effective TMD PDF}

Applying the projection in Eq.~\eqref{eq:longprojj} to 
Eqs.~(\ref{eq:facross4a},~\ref{eq:traceNPVproj}) gives the effective (but TMD-factorization violating) TMD PDF:
\begin{equation}
\label{eq:traceNPVproj2}
\Phi_{q / P_1}^{(2g) \; [ \gamma_5 ]}\left( P_1;\hat{k}_{1}(x_1,{\bf k}_{1t}),q, {\bf l}_{1t} \right)
= {\rm Tr}_C \left(  \frac{1}{4} \tr{ \gamma_5 \, \Phi_{q / P_1}^{(2g)}\left( P_1;\hat{k}_{1}(x_1,{\bf k}_{1t}), q,{\bf l}_{1t} \right)} \right) \, .
\end{equation}

Now the effective TMD PDF is a function of $q$ and ${\bf l}_{1t}$ as well as $P_1$ and $k_1$, in violation of the minimal criteria for TMD-factorization 
from Sect.~\ref{sec:genfact}.
So Eq.~\eqref{eq:traceNPVproj2} in general gives a non-vanishing leading-power trace,
\begin{align}
\Phi_{q / P_1}^{(2g) \; [ \gamma_5]}\left( P_1;\hat{k}_{1}(x_1,{\bf k}_{1t}), q,{\bf l}_{1t} \right) 
& \propto  \frac{1}{4} \tr{\gamma_5 \slashed{q} \slashed{l}_{1t} \slashed{k}_{1} \slashed{P}_1}   \nonumber \\
& \approx i P_1 \cdot q \; \left( k_{1t}^i l_{1t}^j \epsilon^{ij} \right)_{f_1} \sim Q^2 \, . \label{eq:npvtrace}
\end{align}
The approximation sign is used on the second line because we have neglected subleading terms in the evaluation of the trace.  

\subsubsection{Effective Hard Subprocess}

There is one more Dirac trace in each of Eq.~\eqref{eq:facross4a} and Eq.~\eqref{eq:facross4aschan}.  It is the trace
for the bottom quark line in Fig.~\ref{fig:anomalygraphs} involving the final state polarizations $\lambda_3$ and $S_4$.  Since our main
interest is in the dependence on the final state transverse polarization for the quark, we keep only the leading power 
transverse spin contribution from Eq.~\eqref{eq:OFF}:
\begin{equation}
\Delta_{k_4/q}\left(k_4, S_4 \right) \to \gamma_5 \slashed{k}_4 \slashed{S}_4 / m_q\, .
\end{equation} 
Applying Eq.~\eqref{eq:longprojjj}, the Dirac trace for the effective hard subprocess 
in Eq.~\eqref{eq:facross4a} for the cross-term contribution becomes
\begin{equation}
\label{eq:npvtrace3}
 \tr{\slashed{k}_4 \slashed{S}_4 \slashed{\epsilon}_{\lambda_3}  
\slashed{\epsilon}^{\ast}_{\lambda_3} \slashed{\hat{h}} ( \slashed{q}_{t} - \slashed{k}_{1t}) } \, .
\end{equation}
(Recall the definition of $h$ from Eq.~\eqref{eq:hdef}.)
To get a final state double spin dependence, we must extract the imaginary 
part of Eq.~\eqref{eq:npvtrace3} to multiply the $i$ in Eq.~\eqref{eq:npvtrace}.  For this it is useful to note that for 
arbitrary four-vectors $v_{1 \, \mu}$ and $v_{2 \, \nu}$,
\begin{equation}
v_{1 \, \mu} v_{2 \, \nu} {\rm Im} \left(\epsilon_{\lambda_3}^\mu  \epsilon^{\ast \, \nu}_{\lambda_3} \right) =
\lambda_3 \, v_{1 \, \mu} v_{2 \, \nu} {\rm Im} \left(\epsilon_{1}^\mu  \epsilon^{\ast \, \nu }_{1} \right) \, .
\end{equation}
Then, the imaginary terms in Eq.~\eqref{eq:npvtrace3} are
\begin{equation}
i \, {\rm Im} \left( 
\tr{   \slashed{k}_4 \slashed{S}_4 \slashed{\epsilon}_{\lambda_3} 
\slashed{\epsilon}^{\ast}_{\lambda_3} \slashed{\hat{h}} ( \slashed{q}_{t} - \slashed{k}_{1t})  } \right) 
=  i \lambda_3 \, {\rm Im} \left(  \tr{\slashed{k}_4 \slashed{S}_4 \slashed{\epsilon}_1 
\slashed{\epsilon}^{\ast}_1 \slashed{\hat{h}} ( \slashed{q}_{t} - \slashed{k}_{1t}) } \right) \, . \\ 
 \label{eq:npvhardtrace2}
\end{equation}
This trace may be straightforwardly evaluated and the leading-power terms kept.
It is helpful to note that
\begin{equation}
{\rm Im} \left( v_{1 \, \mu} v_{2 \, \nu} \epsilon_1^\mu \, \epsilon_1^{\ast \nu} \right) = \frac{1}{2} \left( \epsilon^{ij} v_{1}^i v_{2}^j \right)_{f_3}  \, , 
\end{equation}
where the $f_3$ subscript is a reminder that here the transverse indices $i,j$ are defined with respect to frame-3.
Evaluating the trace in Eq.~\eqref{eq:npvhardtrace2} then gives
\begin{align}
 i \lambda_3 \, {\rm Im} \left(  \tr{\slashed{k}_4 \slashed{S}_4 \slashed{\epsilon}_1
\slashed{\epsilon}^{\ast}_1 \slashed{\hat{h}} ( \slashed{q}_{t} - \slashed{k}_{1t}) } \right) & = 
4 i  \lambda_3 \; \underbrace{\left(S_4^i \hat{k}_1^j \epsilon^{ij} \right)_{f_3}}_\text{{\small $\sim Q$}} \;
 \underbrace{(k_4 \cdot (q_t - k_{1t}))}_\text{{\small $\sim Q$}} \nonumber \\ & + 4 i \lambda_3 \; \underbrace{\left( (q_t - k_{1t})^i S_4^j \epsilon^{ij} \right)_{f_3}}_\text{{\small $\sim Q^0$}} \;
 \underbrace{(k_4 \cdot \hat{h})}_\text{{\small $\sim Q^2$}} \, \sim Q^2
 . \label{eq:npvhardtrace4}
\end{align}
Here we have used the definition of $h$ in Eq.~\eqref{eq:hdef}.  On first line, note that
the transverse components of $\hat{k}_1$ are taken with respect to frame-3 and, 
recalling Eq.~\eqref{eq:largep2trans}, the frame-3 transverse components of the incoming hadrons and partons are of order $\sim Q$.

The same steps apply to 
the $s$-channel trace of Eq.~\eqref{eq:facross4aschan}.  The result is obtained by replacing
$\slashed{\hat{h}} ( \slashed{q}_{t} - \slashed{k}_{1t}) \to  ( \slashed{q}_{t} - \slashed{k}_{1t}) \slashed{q}$ in Eq.~\eqref{eq:npvhardtrace2} .
The trace is
\begin{align}
 i \lambda_3 \, {\rm Im} \left(  \tr{\slashed{k}_4 \slashed{S}_4 \slashed{\epsilon}_1
\slashed{\epsilon}^{\ast}_1 ( \slashed{q}_{t} - \slashed{k}_{1t}) \slashed{q} } \right) & = 
4 i  \lambda_3 \; \underbrace{\left(S_4^i (q_t - k_{1t})^j \epsilon^{ij} \right)_{f_3}}_\text{{\small $\sim Q^0$}} \;
 \underbrace{(k_3 \cdot k_4)}_\text{{\small $\sim Q^2$}} \sim Q^2
 . \label{eq:npvhardtrace5}
\end{align}

Analogous steps also apply if we use the $[l_{1t}]$ effective gluon TMD PDF from 
second term in Eq.~\eqref{eq:2gluondecomp} rather than the $[k_{2t}]$ one.  This contributions differ from the above only in 
that $q_t - k_{1t}$ gets replaced by $l_{1t}$ in Eqs.~(\ref{eq:npvhardtrace4},~\ref{eq:npvhardtrace5}) and 
$\Phi_{g / P_2}^{(2g) \; [k_{2t}]}$ gets replaced by $\Phi_{g / P_2}^{(2g) \; [l_{1t}]}$ in 
Eqs.~(\ref{eq:facross4a},~\ref{eq:facross4aschan}). 

\subsubsection{Double Final State Spin Asymmetry}
\label{sec:doubelfinal}

By counting powers of $Q$ in Eqs.~(\ref{eq:2gluondecomp},~\ref{eq:npvtrace},~\ref{eq:npvhardtrace4},~\ref{eq:npvhardtrace5}) inside Eqs.~(\ref{eq:facross4a},~\ref{eq:facross4aschan}) we 
are now able to tally the largest powers contributing to a TMD-factorization violating final state double spin asymmetry:
\begin{equation}
\label{eq:result}
d \sigma^{\rm I.S.I. \; [ \gamma_5]} \propto \lambda_3 \left(S_4^j \right)_{f_3} \times \frac{N_c^2 - 1}{N_c^2} \times \mathcal{O}(Q^0)^j \, .
\end{equation}
The imaginary $i$ in Eq.~\eqref{eq:npvhardtrace4} combines with the $i$ in Eq.~\eqref{eq:npvhardtrace2} to give a real contribution 
to the cross section so the graphs in Fig.~\ref{fig:anomalygraphs} and their Hermitian conjugates add.

To summarize, power counting implies a leading contribution to the cross section from a double 
final state $S_4$-$\lambda_3$ spin dependence.
In Eq.~\eqref{eq:result}, we have explicitly written the $\lambda_3 \left( S_4^j \right)_{f_3}$ to emphasize the final state spin-helicity correlation, and 
we have explicitly displayed the color factor to emphasize the non-Abelian nature of the effect --- the spin asymmetry exactly vanishes in the Abelian limit of $N_c \to 1$ while in 
QCD it comes with a color factor of $8/9$.  
All remaining factors from the substitution of Eqs.~(\ref{eq:2gluondecomp},~\ref{eq:npvtrace},~\ref{eq:npvhardtrace4},~\ref{eq:npvhardtrace5}) 
into Eqs.~(\ref{eq:facross4a},~~\ref{eq:facross4aschan}) have been abbreviated by the $\mathcal{O}(Q^0)^j$.  
This corresponds to a convolution integral of effective non-perturbative multiparton correlation functions.

Recall that the $[\gamma_5]$ superscript
in Eq.~\eqref{eq:result} means that we have applied the pseudo-scalar $\gamma_5$ projection from Eq.~\eqref{eq:longprojj} to the effective TMD PDF in Eq.~\eqref{eq:NPVvioldef}.
Without the breakdown in maximally generalized TMD-factorization, projections like 
these would be applied only to ordinary TMD PDFs with the minimal TMD-factorization properties from Sect.~\ref{sec:genfact}.   They would therefore be discarded
on the grounds that they violate parity. 

To totally characterize the final state angular and spin dependence for Eq.~\eqref{eq:theprocess}, all other 
leading-power projections should also be accounted for, including the axial vector and 
tensor terms from Eq.~\eqref{eq:clifford} for the quark subgraph.  We leave a complete classification of the 
types of general behavior possible at leading power to future work.  We also plan to obtain explicit expressions 
for contributions like Eq.~\eqref{eq:result}, for various initial and final state spin configurations, 
by using specific non-perturbative models for the incoming hadron wave functions.

\section{Summary and Discussion}
\label{sec:conclusion}
By directly counting powers of the hard scale for the process of Sect.~\ref{sec:theprocess}, we have
shown how extra spin asymmetries may arise at leading power due to Ward identity non-cancellations that 
break TMD-factorization.  This is sufficient to conclude that TMD-factorization-based classification schemes are
too limiting to account for all the qualitatively distinct types of non-perturbative correlations that are possible in the 
limit of fixed energy and large $Q \gg \Lambda$.  The $(N_c^2 - 1)/ N_c^2$ factor in Eq.~\eqref{eq:result} reflects 
the special role of non-Abelian gauge invariance in producing such extra spin dependence.  That the leftover terms 
arise from an incomplete Ward identity cancellation is indicative of the role of TMD-factorization breaking effects in 
preserving gauge invariance over large distances at leading power in the cross section.  The essential steps for 
analyzing Feynman diagrams were already presented in Ref.~\cite{Collins:2008sg} for the collinear case.  In this paper, 
we have argued that certain terms in the Wilson line factors correspond to the breakdown of factorization when 
transitioning from the collinear to TMD cases.

It should be emphasized that the ``extra" soft or collinear gluon attachments here are in the regime of 
large coupling, so we are not actually justified in stopping with a fixed number of gluons.  Hence, the failure of the extra 
gluons to factorize in perturbation theory corresponds to the influence of extra non-perturbative effects.  The types of interactions  
are similar to those normally associated with Wilson lines in the definitions of TMD PDFs.  However, for the process considered 
in this paper, they cannot be attributed to Wilson line operators for any separate external hadron alone.   Instead, they 
describe truly global non-perturbative physical properties of the process.

The specific example used in Sects.~\ref{sec:twogluons} and~\ref{sec:extra} was chosen to illustrate the existence of 
TMD-factorization breaking correlations in the simplest way possible and with only one extra soft gluon needing to be 
considered.  This was sufficient for the current purpose of demonstrating a non-trivial interplay between transverse 
momentum, spin, and color degrees of freedom.  However, other extra spin and azimuthal asymmetries are also likely to 
arise following similar mechanisms, especially when going beyond the single extra gluon case.  For example, the term 
in Eq.~\eqref{eq:vgfactamps4} may contribute to an azimuthal angular dependence in the totally unpolarized cross section.  
Also, if the imaginary parts in Eqs.~(\ref{eq:finalstate},~\ref{eq:initialstate}) are taken into consideration, then a single final 
state spin asymmetry may be generated at the single extra gluon level.  A full classification of spin and angular asymmetries
in TMD-factorization breaking scenarios is left for future work.  

The interactions discussed here and in earlier work on 
similar topics (e.g.,~\cite{Collins:2002kn,Brodsky:2002rv,Brodsky:2002cx,Bomhof:2004aw,Bomhof:2006dp,Bomhof:2007xt,Vogelsang:2007jk,Collins:2007nk,Collins:2007jp,Rogers:2010dm})
are novel because they tend to give rise to effects outside of the common classical intuition that originates in a parton model picture.
While the general appearance of quantum mechanical correlations is not surprising, these specific effects are distinguishable by 
the fact that they are induced by the kinds of large coupling, long-range interactions that are more commonly associated with 
hadronic binding and confinement in separate hadrons.  Even for processes where TMD-factorization is valid, physical effects 
from initial and final state soft and collinear interactions are already predicted in, for example, studies of T-odd effects such as the 
Sivers effect~\cite{Brodsky:2002rv,Brodsky:2002cx}.  Namely, the Sivers function~\cite{Sivers:1989cc,Sivers:1990fh} (a type of TMD PDF) 
is predicted to reverse sign between Drell-Yan and SIDIS due to a reversal in the direction of the gauge link between these two 
processes~\cite{Collins:2002kn}.  (For physical pictures of the sign reversal mechanism see, e.g., Refs.~\cite{Burkardt:2003uw,Collins:2008ht}.)  
Interesting connections between QCD process dependence and a QCD version of the Aharonov-Bohm effect have been suggested in 
Refs.~\cite{Pijlman:2006vm,Sivers:2011ci}.

The existence of TMD-factorization breaking implies generally that unexpected experimental phenomena are possible.
In this context, it is worth recalling again the production of transversely polarized $\Lambda$-hyperons in unpolarized hadron-hadron 
collisions (spontaneous $\Lambda$ polarization)~\cite{Bunce:1976yb}.  Past TMD-based explanations have focused on effects in the 
fragmentation process,  such as a polarizing fragmentation function.  The production of transversely polarized $\Lambda$-hyperons 
in $p p \to \Lambda^\uparrow + X$ collisions was described within a hybrid TMD/collinear-factorization formalism in 
Ref.~\cite{Anselmino:2000vs}, with the spontaneous transverse polarization being ascribed to the TMD polarizing fragmentation function.   
It was argued in Ref.~\cite{Kanazawa:2000cx} (assuming universality for TMD fragmentation functions) that this should imply similar $\Lambda$ 
transverse polarizations in the $e^+ e^-$ collisions in the ALEPH experiment at LEP~\cite{Buskulic:1996vb}, and that the failure to observe 
such effects suggests that spontaneous production of $\Lambda$ polarizations in hadron-hadron collisions might also require contributions 
from initial state interactions.  (Significant $\Lambda$ polarizations also have not been observed in SIDIS measurements~\cite{Chekanov:2006wz}.)

However, the semi-inclusive $p p \to \Lambda^\uparrow + X$ and $e^+ e^- \to \Lambda^\uparrow + X$ processes are not differential in a small 
$q_t$, and so are not strictly appropriate for treatment within a TMD formalism.  Moreover, for totally inclusive $e^+ e^-$ annihilation processes, 
twist-three formalisms~\cite{Lu:1995rp,Boer:1997mf} are so far consistent with measurements of spontaneous $\Lambda$ polarization.
At best, information about the TMDs must be extracted from these measurements indirectly through, for example, hybrid formalisms like 
that of Ref.~\cite{Anselmino:2000vs}.  Furthermore, the $e^+ e^-$ measurements have been performed with a hard scale near the $Z$-pole, 
whereas the measurements of spontaneous transverse $\Lambda$-hyperon polarization in the $p p \to \Lambda^\uparrow + X$ process has a
hard scale of only a few GeV.  QCD evolution likely leads to a large suppression in the $e^+ e^-$ measurements relative to the hadro-production 
measurements, and this may be sufficient to explain the discrepancy.   In addition, current SIDIS measurements may be small due to effects 
specific to small-$x$ dynamics.  Thus, the status of comparisons of spontaneous transverse $\Lambda$ polarization between different processes 
remains inconclusive, and new studies are needed.  Ideally, investigations of TMD universality should be made by comparing processes like 
$p p \to {\rm hadron (jet)} + \Lambda^\uparrow + X$ and $e^+ e^- \to {\rm hadron (jet)} + \Lambda^\uparrow + X$ where a TMD-style
description is most appropriate, as explained recently in Refs.~\cite{Boer:2007nh,Boer:2010ya}.  New SIDIS measurement may also be useful, as 
this is a process where TMD-factorization is expected to hold.  
(The universality of fragmentation functions between $e^+ e^- \to {\rm hadron (jet)} + \Lambda^\uparrow + X$ and SIDIS has been shown 
in Refs.~\cite{Metz:2002iz,Collins:2004nx}.)
Future measurements may be possible, and the above general observations suggest that spontaneous $\Lambda$ polarization studies 
are ideal for comparing TMD-factorization versus TMD-factorization breaking scenarios.  See also the discussion in Ref.~\cite{Goldstein:2012py}.
Alternative explanations for spontaneous $\Lambda$ polarization have proposed in Refs.~\cite{Kanazawa:2000cx,Zhou:2008fb} within the twist-three 
collinear factorization formalism.  It is also interesting to note that, in Ref.~\cite{Albino:2008fy}, a possible inconstancy was found with universal 
unpolarized TMD $\Lambda$-fragmentation functions in $pp \to \Lambda/\bar{\Lambda} + X$ measurements as compared with 
$e^+ e^- \to \Lambda/\bar{\Lambda}+X$.  (See also the discussion in Ref.~\cite{Boer:2010yp}.)

Other observables are also likely to be useful for probing the role of TMD-factorization breaking effects.  
The role of three-gluon correlators in a twist-three approach has been recently discussed in Ref.~\cite{Kanazawa:2012kt}.
In Ref.~\cite{Metz:2012ct}, a proposed solution to the Sivers sign discrepancy between Drell-Yan and SIDIS~\cite{Kang:2011hk} was also 
given in the twist-three formalism.  Including a treatment of fragmentation in the description of transverse single spin asymmetries 
was also found to be important within the twist-three formalism in Ref.~\cite{Kang:2010zzb}.  In principle, it should be possible to relate the 
TMD and higher twist collinear formalisms through integrations over transverse momentum.  In this way, it may be possible to relate the 
breakdown of TMD-factorization discussed hear to higher twist collinear correlation functions.  Studies of the detailed properties of final 
state jets may also help clarify the dynamics responsible for final state asymmetries.  Rather than the polarizations of final state hadrons, 
for instance, one may instead investigate the azimuthal distribution of hadrons around a 
jet axis~\cite{Kang:2010vd,Boer:2010zf,DAlesio:2010am,Fatemi:2012ry}.  The ``CMS ridge" of Ref.~\cite{Khachatryan:2010gv}
has also been explained in terms of effects that are beyond the scope of standard leading twist factorization.   These include both 
Wilson line descriptions~\cite{Cherednikov:2010tr} and color glass condensate 
descriptions~\cite{Dusling:2012cg,Dusling:2012wy,Dusling:2013oia}.  
Interestingly, both descriptions incorporate subtle quantum mechanical entanglement effects.

Although a thorough discussion of how TMD-factorization breaking effects might be calculated 
precisely from first principles QCD is beyond the scope of this paper, it is natural to speculate that a form of 
TMD perturbative QCD, with a partonic description of the hard scattering, might be recovered by a rearranged treatment of 
non-perturbative contributions.  Rather than having separate TMD PDFs and TMD fragmentation functions, for example, one might use a 
set of non-perturbative functions wherein the constituents of multiple external hadrons remain entangled, but a hard partonic factor is 
separated out.  We leave the question of whether such a separation is valid to future studies.  Already, some versions of 
soft-collinear-effective-theory (SCET) refrain from identifying operator definitions for separate hadrons~\cite{Becher:2010tm}.  Other approaches 
that are likely to be useful in describing TMD-factorization breaking effects are those that are now more commonly associated with the 
physics of high energy scattering by large nuclei, 
where partons are described as propagating through a large size and/or high density gluon 
field~\cite{Qiu:2004da,Frankfurt:2007rn,Arleo:2009ch,Frankfurt:2011cs,McLerran:1993ni,Blaizot:2004wv,Gelis:2008rw,Akcakaya:2012si}.

\section*{Acknowledgments}
I gratefully acknowledge many conversations with John~Collins  and George~Sterman that influenced the content of this article.
I also thank Christine Aidala, Dani\"{e}l Boer, Maarten Buffing, Aurore Courtoy, Renee Fatemi, Bryan Field, Leonard Gamberg, Simonetta Liuti, Andreas Metz, Piet Mulders, Alexei Prokudin,
Mark Strikman, Bowen Xiao, and Feng Yuan for useful discussions and
comments on the text.
Feynman diagrams were produced using Jaxodraw~\cite{jaxo,Binosi:2008ig}.
This work was supported by the National Science Foundation, grant PHY-0969739.

\end{widetext}

\bibliography{factviolpart2}

\end{document}